\documentclass[english,prd,onecolumn,superscriptaddress,nofootinbib,preprintnumbers]{revtex4}
\usepackage[latin1]{inputenc}
\usepackage{graphicx}
\usepackage{amsmath,amsthm,amssymb}

\makeatletter

\usepackage{amsfonts}
\usepackage{dcolumn}
\usepackage{hyperref}

\def\be{\begin{equation}}
\def\ee{\end{equation}}
\def\ba{\begin{eqnarray}}
\def\ea{\end{eqnarray}}
\def\bs{\begin{subequations}}
\def\es{\end{subequations}}

\newcommand{\vp}{\varphi}
\newcommand{\rd}{{\rm d}}
\newcommand{\gsim}{\mbox{\raisebox{-1.ex}{$\stackrel
{\textstyle>}{\textstyle\sim}$}}}
\newcommand{\lsim}{\mbox{\raisebox{-1.ex}{$\stackrel
{\textstyle<}{\textstyle \sim}$}}}

\makeatother

\usepackage{babel}
\makeatother
\begin{document}

\title{Dark energy: investigation and modeling}

\author{Shinji Tsujikawa}
\affiliation{Department of Physics, Faculty of Science, Tokyo University of Science, 
1-3, Kagurazaka, Shinjuku-ku, Tokyo 162-8601, Japan}

\begin{abstract}

Constantly accumulating observational data continue to confirm 
that about 70 \% of the energy density today consists of dark energy 
responsible for the accelerated expansion of the Universe.
We present recent observational bounds on dark energy
constrained by the type Ia supernovae, cosmic microwave 
background, and baryon acoustic oscillations.
We review a number of theoretical approaches that have been 
adopted so far to explain the origin of dark energy.
This includes the cosmological constant, modified matter models 
(such as quintessence, k-essence, coupled
dark energy, unified models of dark energy and dark matter), 
modified gravity models (such as $f(R)$ gravity, 
scalar-tensor theories, braneworlds), and inhomogeneous models.
We also discuss observational and experimental constraints
on those models and clarify which models are favored or
ruled out in current observations.

\end{abstract}

\date{\today}

\maketitle

\tableofcontents

\section{Introduction}	

The discovery of the late-time cosmic acceleration reported in 
1998 \cite{Riess,Perlmutter} based on the type Ia Supernovae (SN Ia) 
observations opened up a new field of research in cosmology.
The source for this acceleration,
dubbed dark energy \cite{Huterer}, 
has been still a mystery in spite of tremendous efforts 
to understand its origin over the last
decade \cite{Sahnireview,Carroll,Peebles,Paddy,SahniLecture,review,Durrerreview,Caldwell09,Tsujibook}.
Dark energy is distinguished from ordinary matter in that it has a negative pressure
whose equation of state $w_{\rm DE}$ is close to $-1$.
Independent observational data such as 
SN Ia \cite{SNLS,Gold1,Gold2,Essence1,Essence2,Kowalski}, 
Cosmic Microwave Background (CMB) \cite{WMAP1,WMAP3,WMAP5,WMAP7}, 
and Baryon acoustic oscillations (BAO) \cite{Eisenstein,Percival1,Percival2}
have continued to confirm that about 70 \% of the energy density of the 
present Universe consists of dark energy.

The simplest candidate for dark energy is the so-called cosmological constant 
$\Lambda$ whose equation of state is $w_{\rm DE}=-1$.
If the cosmological constant originates from a vacuum energy 
of particle physics, its energy scale is significantly larger than
the dark energy density today \cite{Weinberg}
($\rho_{\rm DE}^{(0)} \simeq 10^{-47}$~GeV$^4$).
Hence we need to find a mechanism to obtain the tiny value
of $\Lambda$ consistent with observations. A lot of efforts have
been made in this direction under the framework of particle physics.
For example, the recent development of string theory shows that it
is possible to construct de Sitter vacua
by compactifying extra dimensions in the presence of fluxes with an
account of non-perturbative corrections \cite{KKLT}. 

The first step toward understanding the property of dark energy
is to clarify whether it is a simple cosmological constant
or it originates from other sources that dynamically change in time.
The dynamical dark energy models can be distinguished
from the cosmological constant by considering 
the evolution of $w_{\rm DE}$.
The scalar field models of DE such as 
quintessence \cite{quin1,Ford,quin2,quin3,quin4,Ferreira1,Ferreira2,CLW,quin5,Zlatev,Paul99}
and k-essence \cite{kes1,kes2,kes3} predict a wide variety of variations
of $w_{\rm DE}$, but still the current observational data
are not sufficient to provide some preference of such models
over the $\Lambda$-Cold-Dark-Matter ($\Lambda$CDM) model.
Moreover, the field potentials need
to be sufficiently flat such that the field evolves slowly enough to
drive the present cosmic acceleration. This demands that the field
mass is extremely small ($m_{\phi}\simeq10^{-33}$\,eV) relative
to typical mass scales appearing in particle physics \cite{Carrollqui,Kolda}.
However it is not entirely hopeless to construct viable scalar-field dark 
energy models in the framework of particle physics.
We note that there is another class of modified matter models
based on perfect fluids--so-called (generalized) the 
Chaplygin gas model \cite{Kamen,Bento}. 
If these models are responsible for explaining the origin of 
dark matter as well as dark energy, 
then they are severely constrained from 
the matter power spectrum in galaxy clustering \cite{Waga}.

There exists another class of dynamical dark energy models that modify
General Relativity. The models that belong to this class are 
$f(R)$ gravity \cite{fR1,fR1d,fR2,fR3,Nojiri03} 
($f$ is a function of the Ricci scalar $R$), 
scalar-tensor theories \cite{st1,st2,st3,st4,st5}, and 
Dvali, Gabadadze and Porrati (DGP) braneworld model \cite{DGP}.
The attractive feature of these models is that the cosmic acceleration 
can be realized without recourse to a dark energy component.
If we modify gravity from General Relativity, however, there 
are stringent constraints coming from local gravity tests 
as well as a number of observational constraints such as large-scale 
structure (LSS) and CMB.
Hence the restriction on modified gravity models is in general 
very tight compared to modified matter models.
We shall construct viable modified gravity models and 
discuss their observational and experimental signatures.

In addition to the above mentioned models, there are attempts 
to explain the cosmic acceleration without dark energy.
One example is the void model in which an apparent accelerated expansion
is induced by a large spatial inhomogeneity \cite{Tomita1,Tomita2,Celerier,Iguchi,Alnes}.
Another example is the so-called backreaction model in which 
the backreaction of spatial inhomogeneities on the 
Friedmann-Lema\^{i}tre-Robertson-Walker (FLRW) background
is responsible for the real acceleration \cite{Rasanen,Kolb1,Kolb2}.
We shall discuss these models as well.

This review is organized as follows. In Sec.~\ref{obsec}
we provide recent observational constraints on dark energy
obtained by SN Ia, CMB, and BAO data.
In Sec.~\ref{cossec} we review theoretical attempts to 
explain the origin of the cosmological constant 
consistent with the low energy scale of dark energy.
In Sec.~\ref{mattersec} we discuss modified gravity models 
of dark energy--including quintessence, k-essence, 
coupled dark energy, and unified models of dark energy 
and dark matter.
In Sec.~\ref{mosec} we review modified gravity models
and provide a number of ways to distinguish those models
observationally from the $\Lambda$CDM model. 
Sec.~\ref{withoutsec} is devoted to the 
discussion about the cosmic acceleration without 
using dark energy.
We conclude in Sec.~\ref{consec}.

We use units such that $c=\hbar=1$,
where $c$ is the speed of light and $\hbar$
is reduced Planck's constant. 
The gravitational constant $G$ is related to the Planck mass 
$m_{{\rm pl}}=1.2211\times10^{19}$\,GeV
via $G=1/m_{{\rm pl}}^{2}$ and the reduced Planck mass
$M_{{\rm pl}}=2.4357\times10^{18}$\,GeV via 
$\kappa^{2}\equiv8\pi G=1/M_{{\rm pl}}^{2}$, respectively. 
We write the Hubble constant today as 
$H_0=100~h$\,km\,sec$^{-1}$\,Mpc$^{-1}$, where 
$h$ describes the uncertainty on the value $H_0$. 
We use the metric signature $(-,+,+,+)$.

%%%%%%%%%%%%%%%%%%%%%%%%%%%%
\section{Observational constraints on dark energy}	
\label{obsec}
%%%%%%%%%%%%%%%%%%%%%%%%%%%%

The late-time cosmic acceleration is supported by 
a number of independent observations--such as 
(i) supernovae observations, (ii) Cosmic Microwave Background (CMB),
and (iii) Baryon acoustic oscillations (BAO).
In this section we discuss observational constraints on the 
property of dark energy. 

\subsection{Supernovae Ia observations}
\label{SNIasec}

In 1998 Riess {\it et al.} \cite{Riess} and Perlmutter et al. {\it et al.} \cite{Perlmutter} 
independently reported the late-time cosmic acceleration 
by observing distant supernovae of type Ia (SN Ia). 
The line-element describing a 4-dimensional homogeneous and isotropic Universe, 
which is called the FLRW space-time, is given by \cite{Weinbergbook}
\begin{equation}
\rd s^2=g_{\mu \nu} \rd x^{\mu} \rd x^{\nu}
=-\rd t^2+a^2(t) \left[ \frac{\rd r^2}{1-Kr^2}+r^2 ({\rd} \theta^2+
\sin^2 \theta\,\rd \phi^2) \right]\,,
\label{metric}
\end{equation}
where $a(t)$ is the scale factor with cosmic time $t$, and 
$K = +1,-1,0$ correspond to closed, open and 
flat geometries, respectively. 
The redshift $z$ is defined by $z=a_0/a-1$, where $a_0=1$
is the scale factor today.

In order to discuss the cosmological evolution in the low-redshift regime
($z< {\cal O}(1)$), let us consider non-relativistic matter with 
energy density $\rho_m$ and dark energy with energy density 
$\rho_{\rm DE}$ and pressure $P_{\rm DE}$, 
satisfying the continuity equations
\begin{eqnarray}
\label{rhomeq}
& &\dot{\rho}_m+3H\rho_m=0\,,\\
\label{rhoDEeq}
& &\dot{\rho}_{\rm DE}+3H(\rho_{\rm DE}+P_{\rm DE})=0\,,
\end{eqnarray}
which correspond to the conservation of the energy-momentum tensor $T_{\mu \nu}$
for each component ($\nabla_{\mu} T^{\mu \nu}$=0, where $\nabla$ represents
a covariant derivative). Note that a dot represents a derivative with 
respect to $t$.
The cosmological dynamics is known by solving the Einstein equations 
\begin{equation}
G_{\mu \nu}=8\pi G T_{\mu \nu}\,,
\label{Einsteineq}
\end{equation}
where $G_{\mu \nu}$ is the Einstein tensor.
For the metric (\ref{metric}) the (00) component of the Einstein equations 
gives \cite{Weinbergbook}
\begin{equation}
H^2=\frac{8\pi G}{3} (\rho_m+\rho_{\rm DE})-\frac{K}{a^2}\,,
\label{Hubbleeq}
\end{equation}
where $H \equiv \dot{a}/a$ is the Hubble parameter.
We define the density parameters
\begin{equation}
\Omega_m \equiv \frac{8\pi G \rho_m}{3H^2}\,,\qquad
\Omega_{\rm DE} \equiv \frac{8\pi G \rho_{\rm DE}}{3H^2}\,,\qquad
\Omega_K \equiv -\frac{K}{(aH)^2}\,,
\end{equation}
which satisfy the relation $\Omega_m+\Omega_{\rm DE}+\Omega_K=1$ from 
Eq.~(\ref{Hubbleeq}). Integrating Eqs.~(\ref{rhomeq}) and (\ref{rhoDEeq}),
we obtain 
\begin{equation}
\rho_m=\rho_m^{(0)}(1+z)^3\,,\qquad 
\rho_{{\rm DE}}=\rho_{{\rm DE}}^{(0)}\,\exp\left[\int_{0}^{z}
\frac{3(1+w_{{\rm DE}})}{1+\tilde{z}}{\rm d}\tilde{z}\right]\,,\label{rhoDE}
\end{equation}
where ``0'' represents the values today and 
$w_{\rm DE}=P_{\rm DE}/\rho_{\rm DE}$ is the equation of 
state of dark energy. 
Plugging these relations into Eq.~(\ref{Hubbleeq}), it follows that 
\begin{equation}
H^{2}(z)=H_{0}^{2}\left[ \Omega_{m}^{(0)}(1+z)^{3}+
\Omega_{{\rm DE}}^{(0)}\exp\left\{ \int_{0}^{z}
\frac{3(1+w_{{\rm DE}})}{1+\tilde{z}}{\rm d}\tilde{z}\right\}
+\Omega_{K}^{(0)}(1+z)^{2}\right].
\label{Heqpa}
\end{equation}

The expansion rate $H(z)$ can be known observationally by measuring the 
luminosity distance $d_L(z)$ of SN Ia.
The luminosity distance is defined by $d_L^2 \equiv L_s/(4\pi {\cal F})$, 
where $L_s$ is the absolute luminosity of a source and ${\cal F}$ is 
an observed flux. It is a textbook exercise \cite{review,Tsujibook,Weinbergbook} 
to derive $d_L (z)$ for the FLRW metric (\ref{metric}):
\begin{equation}
d_{L} (z)=\frac{1+z}{H_{0}\sqrt{\Omega_{K}^{(0)}}}
\sinh\left(\sqrt{\Omega_{K}^{(0)}}\int_{0}^{z}
\frac{{\rm d}\tilde{z}}{E(\tilde{z})}\right)\,,
\label{dLex}
\end{equation}
where $E(z) \equiv H(z)/H_0$.
The function $f_K (\chi) \equiv 1/\sqrt{\Omega_{K}^{(0)}}\,
\sinh (\sqrt{\Omega_{K}^{(0)}}\chi)$ 
can be understood as $f_K (\chi)=\sin \chi$ (for $K=+1$), 
$f_K (\chi)=\chi$ (for $K=0$), and 
$f_K (\chi)=\sinh \chi$ (for $K=-1$).
For the flat case ($K=0$), Eq.~(\ref{dLex}) reduces to 
$d_{L} (z)=(1+z) \int_0^z \rd \tilde{z}/H(\tilde{z})$, i.e.
\begin{equation}
H(z)=\left[ \frac{\rd}{\rd z} \left(
\frac{d_L (z)}{1+z} \right) \right]^{-1}\,.
\end{equation}
Hence the measurement of the luminosity distance $d_L (z)$ of 
SN Ia allows us to find the expansion history of the Universe
for $z < {\cal O}(1)$.

The luminosity distance $d_L$ is expressed in terms of an apparent magnitude 
$m$ and an absolute magnitude $M$ of an object, as
\begin{equation}
\label{mMre}
m-M=5 \log_{10} \left( \frac{d_L}{10~{\rm pc}} \right)\,.
\end{equation}
The absolute magnitude $M$ at the peak of brightness is the same 
for any SN Ia under the assumption 
of standard candles, which is around $M \simeq -19$ \cite{Riess,Perlmutter}.
The luminosity distance $d_L(z)$ is known from Eq.~(\ref{mMre}) 
by observing the apparent magnitude $m$.
The redshift $z$ of an object is known by measuring
the wavelength $\lambda_0$ of light relative to its 
wavelength $\lambda$ in the rest frame, i.e. $z=\lambda_0/\lambda-1$.  
The observations of many SN Ia provide the dependence of 
the luminosity distance $d_L$ in terms of $z$. 

Expanding the function (\ref{dLex}) around $z=0$, it follows that 
\begin{eqnarray}
d_L(z) &=& \frac{1}{H_0} \left[ z+\left\{ 1-\frac{E'(0)}{2}
\right\} z^2+{\cal O}(z^3) \right] \nonumber \\
&=& \frac{1}{H_0} \left[ z+\frac14 
\left( 1-3w_{\rm DE}\Omega_{\rm DE}^{(0)}+\Omega_K^{(0)}
\right)z^2+{\cal O} (z^3) \right]\,,
\label{dLap}
\end{eqnarray}
where a prime represents a derivative with respect to $z$.
Note that, in the second line, we have used Eq.~(\ref{Heqpa}).
In the presence of dark energy ($w_{\rm DE}<0$ and $\Omega_{\rm DE}^{(0)}>0$) 
the luminosity distance gets larger than that in the flat Universe 
without dark energy. For smaller (negative) $w_{\rm DE}$ and for larger 
$\Omega_{\rm DE}^{(0)}$ this tendency becomes more significant. 
The open Universe without dark energy can also give rise to a larger
value of $d_{L}(z)$, but the density parameter $\Omega_K^{(0)}$ 
is constrained to be close to 0 from the WMAP data 
(more precisely, $-0.0175<\Omega_K^{(0)}<0.0085$ \cite{WMAP5}).
Hence, in the low redshift regime ($z < 1$),  
the luminosity distance in the open Universe is hardly 
different from that in the flat Universe without dark energy.

As we see from Eq.~(\ref{dLap}), the observational data in the high redshift regime
($z>0.5$) allows us to confirm the presence of dark energy.
The SN Ia data released by Riess {\it et al.} \cite{Riess} and 
Perlmutter {\it et al.} \cite{Perlmutter} in 1998
in the redshift regime $0.2 < z < 0.8$ showed that 
the luminosity distances of observed SN Ia tend to be larger than 
those predicted in the flat Universe without dark energy.
Assuming a flat Universe with a dark energy equation of state $w_{\rm DE}=-1$
(i.e. the cosmological constant), 
Perlmutter {\it et al.} \cite{Perlmutter} found that the cosmological 
constant is present at the 99 \% confidence level.
According to their analysis the density parameter of non-relativistic matter today
was constrained to be $\Omega_m^{(0)}=0.28^{+0.09}_{-0.08}$ 
(68 \% confidence level) in the flat universe with the cosmological constant.

%%%%%%%%%%%%%%%%%%%%%%%%%%%%%
\begin{figure}
\begin{centering}
\includegraphics[width=3.0in,height=3.0in]{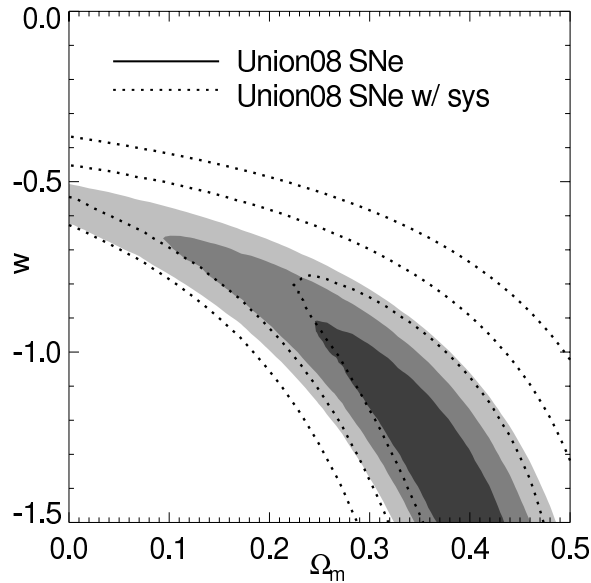} 
\par\end{centering}
\caption{68.3 \%, 95.4 \%, and 99.7 \% confidence level contours 
on $w_{\rm DE}$ and $\Omega_m^{(0)}$ (denoted as 
$w$ and $\Omega_m$ in the figure) constrained by the 
Union08 SN Ia datasets. The equation of state $w_{\rm DE}$ is
assumed to be constant.
{}From Ref.~\cite{Kowalski}.}
\centering{}\label{wcon} 
\end{figure}
%%%%%%%%%%%%%%%%%%%%%%%%%%%%

Over the past decade, more SN Ia data have been collected by a number of 
high-redshift surveys--such as SuperNova Legacy Survey (SNLS) \cite{SNLS}, 
Hubble Space Telescope (HST) \cite{Gold1,Gold2}, 
and ``Equation of State: SupErNovae trace Cosmic Expansion''
(ESSENCE) \cite{Essence1,Essence2} survey.
These data also confirmed that the Universe entered the epoch of 
cosmic acceleration after the matter-dominated epoch.
If we allow the case in which dark energy is different from 
the cosmological constant (i.e. $w_{\rm DE} \neq -1$), then 
observational constraints on $w_{\rm DE}$ and $\Omega_{\rm DE}^{(0)}$
(or $\Omega_m^{(0)}$) are not so stringent.
In Fig.~\ref{wcon} we show the observational contours on 
$(w_{\rm DE}, \Omega_m^{(0)})$ for constant $w_{\rm DE}$ 
obtained from the ``Union08'' SN Ia data by Kowalski {\it et al.} \cite{Kowalski}.
Clearly the SN Ia data alone are not yet sufficient to place tight bounds
on $w_{\rm DE}$. 

In the flat Universe dominated by dark energy with constant $w_{\rm DE}$, 
it follows from Eq.~(\ref{Heqpa}) that $H^2 \simeq H_0^2 \Omega_{\rm DE}^{(0)}
(1+z)^{3(1+w_{\rm DE})} \propto a^{-3(1+w_{\rm DE})}$.
Integrating this equation, we find that the scale factor evolves as
$a \propto t^{2/(3(1+w_{\rm DE}))}$ for $w_{\rm DE}>-1$ and 
$a \propto e^{Ht}$ for $w_{\rm DE}=-1$. 
The cosmic acceleration occurs for $-1 \le w_{\rm DE}<-1/3$.
In fact, Fig.~\ref{wcon} shows that $w_{\rm DE}$ is constrained 
to be smaller than $-1/3$.
If $w_{\rm DE}<-1$, which is called phantoms or ghosts \cite{phantom},
the solution corresponding to the expanding Universe is given by  
$a \propto (t_s-t)^{2/(3(1+w_{\rm DE}))}$, where 
$t_s$ is a constant. In this case the Universe ends at $t=t_s$ 
with a so-called big rip singularity \cite{Star00,CKW} at which the curvature 
grows toward infinity.\footnote{There are other classes of finite-time singularities
studied in Refs.~\cite{Barrow86,Barrow04,Barrow04d,Ste05,NOT05,Brevik,Dab05,Sami05,Marium}. 
In some cases quantum effects can  
moderate such singularities \cite{Nojiri04,NOT05,Singh06,Samart}.}
The current observations allow the possibility 
of the phantom equation of state. We note, however, that 
the dark energy equation of state smaller than $-1$ does not 
necessarily imply the appearance of the big rip singularity.
In fact, in some of modified gravity models such as $f(R)$ gravity, 
it is possible to realize $w_{\rm DE}<-1$ without having 
a future big rip singularity \cite{AmenTsuji}.

If the dark equation of state is not constant, we need to 
parametrize $w_{\rm DE}$ as a function of the redshift $z$.
This smoothing process is required because the actual observational data
have discrete values of redshifts with systematic and statistical errors.
There are several ways of parametrizations proposed
so far. In general one can write such parametrizations in the form 
\begin{equation}
w_{{\rm DE}}(z)=\sum_{n=0}w_{n}x_{n}(z)\,,
\label{Taylor}
\end{equation}
where $n$'s are integers.
We show a number of examples for the expansions:
\begin{eqnarray}
 &  & {\rm (i)~Redshift}:\quad\quad~~~~x_{n}(z)=z^{n}\,,\\
 &  & {\rm (ii)~Scale~factor}:\quad~~x_{n}(z)=\left(1-a \right)^{n}=\left(\frac{z}{1+z}\right)^{n}\,,\\
 &  & {\rm (iii)~Logarithmic}:\quad~x_{n}(z)=\left[\ln\,(1+z)\right]^{n}\,.
\end{eqnarray}
The parametrization (i) was introduced by Huterer and Turner \cite{Huterer}
and Weller and Albrecht \cite{Weller02} with $n\le1$, i.e. $w_{{\rm DE}}=w_{0}+w_{1}z$.
Chevalier and Polarski \cite{ChePo} and Linder \cite{Linderpara}
proposed the parametrization (ii) with $n\le1$, i.e.
\begin{equation}
w_{{\rm DE}}(z)=w_{0}+w_{1}(1-a)=w_{0}+w_{1}\frac{z}{1+z}\,.
\label{Polapara}
\end{equation}
This has a dependence $w_{{\rm DE}}(z)=w_{0}+w_{1}$ for $z\to\infty$
and $w_{{\rm DE}}(z)\to w_{0}$ for $z\to0$. 
A more general form, $w_{{\rm DE}}(z)=w_{0}+w_{1}z/(1+z)^{p}$,
was proposed by Jassal {\it et al.} \cite{Jassal}.
The parametrization (iii) with $n\le1$ was introduced by 
Efstathiou \cite{Efstathiou}. A functional form that can be used for a fast transition
of $w_{{\rm DE}}(z)$ was also proposed \cite{Bassett1,Bassett2}.
In addition to the parametrization of $w_{\rm DE}$, a number of authors 
assumed parametric forms of $d_{L}(z)$ \cite{Star98},
or $H(z)$ \cite{Saini,Alam,Alam03}.
Many works placed observational constraints on the property of dark energy 
by using such parametrizations \cite{Maor,Cora,Wang,Nesseris1,Nesseris2,Ishak,GuoOhta05,Wang2,Nesseris3,Crittenden,Ichikawa,Zhao1,Zhao2,Marina,Ichikawa2}.

%%%%%%%%%%%%%%%%%%%%%%%%%%%%%
\begin{figure}
\begin{centering}
\includegraphics[width=3.2in,height=3.0in]{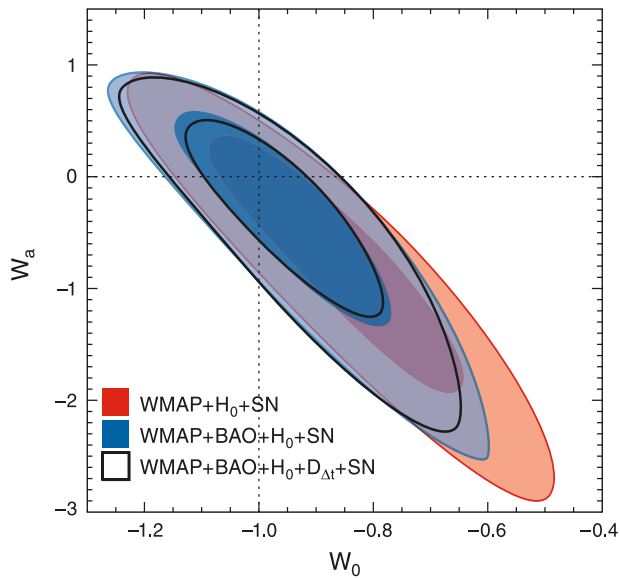} 
\par\end{centering}
\caption{Observational constraints on the parameters 
$w_0$ and $w_1$ (denoted as $w_a$ in the figure) 
for the parametrization (\ref{Polapara}).
The contours show the 68 \% and 95 \% confidence level from 
WMAP+$H_0$+SN (red), WMAP+BAO+$H_0$+SN (blue), 
and WMAP+BAO+$H_0$+$D_{\Delta t}$+SN (black), 
for a flat universe. {}From Ref.~\cite{WMAP7}.}
\centering{}\label{wvary} 
\end{figure}
%%%%%%%%%%%%%%%%%%%%%%%%%%%%

In Fig.~\ref{wvary} we show the SN Ia constraints combined with other 
measurements such as the WMAP 7-year \cite{WMAP7} and 
the BAO data \cite{Percival2}. 
The parametrization (\ref{Polapara}) is used in this analysis.
The Gaussian prior on the present-day Hubble constant \cite{Riess09}, 
$H_0=74.2 \pm 3.6$~km~sec$^{-1}$~Mpc$^{-1}$ (68 \% confidence level),
is also included in the analysis (obtained from the magnitude-redshift 
relation of 240 low-$z$ SN Ia at $z<0.1$).
In Fig.~\ref{wvary}, ``$D_{\Delta t}$'' means a constraint coming from 
the measurement of gravitational lensing time delays \cite{Suyu}.
The joint constraint from WMAP+BAO+$H_0$+$D_{\Delta t}$+SN 
gives the bound
\begin{equation}
w_0=-0.93 \pm 0.13\,,\qquad
w_1=-0.41^{+0.72}_{-0.71}\,,
\end{equation}
at the 68 \% confidence level.
Hence the current observational data are consistent with the 
flat Universe in the presence of the cosmological constant 
($w_0=-1, w_1=0$).

\subsection{CMB}

The temperature anisotropies in CMB are affected by the presence of dark energy.
The position of the acoustic peaks in CMB anisotropies depends on the 
expansion history from the decoupling epoch to the present.
Hence the presence of dark energy leads to the shift for the positions 
of acoustic peaks. There is also another effect called 
the Integrated-Sachs-Wolfe (ISW) effect \cite{ISW} induced by the variation 
of the gravitational potential during the epoch of the cosmic acceleration.
Since the ISW effect is limited to large-scale perturbations, the former
effect is typically more important.

The cosmic inflation in the early Universe \cite{Sta79,Kazanas,Sato,Guth} 
predicts nearly scale-invariant 
spectra of density perturbations through the quantum fluctuation of a scalar field.
This is consistent with the CMB temperature anisotropies observed by 
COBE \cite{COBE} and WMAP \cite{WMAP1}. 
The perturbations are ``frozen'' 
after the scale $\lambda=(2\pi/k)a$ ($k$ is a comoving wavenumber)
leaves the Hubble radius $H^{-1}$ during 
inflation ($\lambda>H^{-1}$) \cite{Liddlebook,BTW}.
After inflation, the perturbations cross inside the Hubble radius again 
($\lambda<H^{-1}$) and they start to oscillate as sound waves.
This second horizon crossing occurs earlier for larger $k$
(i.e. for smaller scale perturbations).

We define the sound horizon as $r_s (\eta)
=\int^{\eta}_0 {\rm d}\tilde{\eta} c_s (\tilde{\eta})$, 
where $c_s$ is the sound speed and 
${\rm d}\eta=a^{-1} {\rm d}t$.
The sound speed squared is given by 
\begin{equation}
c_s^2=1/[3(1+R_s)]\,,\qquad
R_s=3\rho_b/(4\rho_\gamma)\,,
\end{equation}
where $\rho_b$ and $\rho_\gamma$
are the energy densities of baryons and photons, respectively.
The characteristic angle for the location of CMB acoustic 
peaks is \cite{Page}
\begin{equation}
\theta_A \equiv \frac{r_s (z_{\rm dec})}{d_A^{(c)}(z_{\rm dec})}\,,
\label{thetaA}
\end{equation}
where $d_A^{(c)}$ is the comoving angular diameter distance 
related with the luminosity distance $d_L$ via the duality relation 
$d_A^{(c)}=d_L/ (1+z)$ \cite{Weinbergbook,Tsujibook}, 
and $z_{\rm dec} \simeq 1090$
is the redshift at the decoupling epoch.
The CMB multipole $\ell_A$ that corresponds to 
the angle (\ref{thetaA}) is 
\begin{equation}
\ell_A=\frac{\pi}{\theta_A}=\pi \frac{d_A^{(c)}(z_{\rm dec})}
{r_s (z_{\rm dec})}\,.
\label{ell}
\end{equation}
Using Eq.~(\ref{dLex}) and the background equation
$3H^2=8\pi G (\rho_m+\rho_r)$ for the redshift $z>z_{\rm dec}$
(where $\rho_m$ and $\rho_r$ are the energy density of non-relativistic
matter and radiation, respectively), 
we obtain \cite{HuSugi1,HuSugi2}
\begin{equation}
\ell_{A}=\frac{3\pi}{4}\sqrt{\frac{\omega_{b}}{\omega_{\gamma}}}
\left[{\rm ln}\,\left(\frac{\sqrt{R_{s} (a_{\rm dec})+R_{s} (a_{\rm eq})}
+\sqrt{1+R_{s}(a_{\rm dec})}}{1+\sqrt{R_{s} (a_{\rm eq})}}\right)\right]^{-1}
{\cal R}\,,
\label{ell2}
\end{equation}
where $\omega_b \equiv \Omega_b^{(0)}h^2$ and 
$\omega_{\gamma} \equiv \Omega_{\gamma}^{(0)}h^2$, and
${\cal R}$ is the so-called CMB shift parameter defined by \cite{Efs99}
\begin{equation}
{\cal R} \equiv \sqrt{\frac{\Omega_{m}^{(0)}}{\Omega_{K}^{(0)}}}
\sinh\left(\sqrt{\Omega_{K}^{(0)}}\int_{0}^{z_{{\rm dec}}}\frac{\rd z}{E(z)}\right)\,.
\label{CMBshift}
\end{equation}
The quantity $R_s=3\rho_b/(4\rho_\gamma)$ can be expressed as
\begin{equation}
R_s(a)=(3\omega_b/4\omega_{\gamma})a\,.
\end{equation}
In Eq.~(\ref{ell2}), $a_{\rm dec}$ and $a_{\rm eq}$ correspond to the 
scale factor at the decoupling epoch and at the radiation-matter equality, 
respectively.

The change of cosmic expansion history from the decoupling epoch to the present 
affects the CMB shift parameter, which gives rise to the shift for the multipole $\ell_A$.
The general relation for all peaks and troughs of observed CMB anisotropies 
is given by \cite{Doran}
\begin{equation}
\ell_{m}=\ell_{A}(m-\phi_{m})\,,
\label{Doran}
\end{equation}
where $m$ represents peak numbers ($m=1$ for the first peak, $m=1.5$
for the first trough,...) and $\phi_{m}$ is the shift of multipoles.
For a given cosmic curvature $\Omega_{K}^{(0)}$, 
the quantity $\phi_{m}$ depends weakly on $\omega_{b}$ and 
$\omega_{m} \equiv \Omega_m^{(0)}h^2$.
The shift of the first peak can be fitted as $\phi_{1}=0.265$ \cite{Doran}.
The WMAP 5-year bound on the CMB shift parameter is given by \cite{WMAP5}
\begin{equation}
{\cal R}=1.710 \pm 0.019\,, 
\label{Rbound}
\end{equation}
at the 68 \% confidence level.
Taking ${\cal R}=1.710$ together with other values $\omega_b=0.02265$, 
$\omega_m=0.1369$, and $\omega_\gamma=2.469 \times 10^{-5}$
constrained by the WMAP 5-year data, we obtain $\ell_A \simeq 300$ from
Eq.~(\ref{ell2}). Using the relation (\ref{Doran}) with $\phi_{1}=0.265$ 
we find that the first acoustic peak corresponds to $\ell_1 \simeq 220$,
as observed in CMB anisotropies.

%%%%%%%%%%%%%%%%%%%%%%%%%%%%%
\begin{figure}
\begin{centering}
\includegraphics[width=4.5in,height=3.0in]{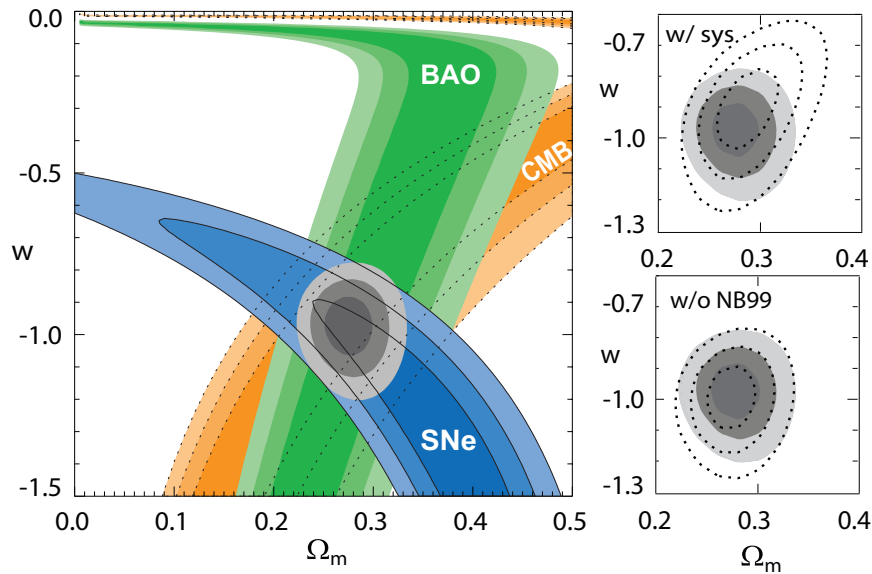} 
\par\end{centering}
\caption{68.3 \%, 95.4 \% and 99.7 \% confidence level contours 
on $w_{{\rm DE}}$ and $\Omega_m^{(0)}$ (denoted as $w$ and $\Omega_m$
in the figure, respectively) for a flat Universe. 
The left panel illustrates the individual constraints from SN Ia, CMB, and BAO, 
as well as the combined constraints (filled gray contours, statistical errors only). 
The upper right panel shows the effect of including systematic errors. 
The lower right panel illustrates the impact of the
Supernova Cosmology Project (SCP)
Nearby 1999 data. {}From Ref.~\cite{Kowalski}.}
\centering{}\label{joint} 
\end{figure}
%%%%%%%%%%%%%%%%%%%%%%%%%%%%

In the flat Universe ($K=0$) the CMB shift parameter is simply given by 
${\cal R}=\sqrt{\Omega_m^{(0)}} \int_0^{z_{\rm dec}} \rd z/E(z)$.
For smaller $\Omega_{m}^{(0)}$ (i.e. for larger $\Omega_{\rm DE}^{(0)}$), 
${\cal R}$ tends to be smaller. For the cosmological constant ($w_{\rm DE}=-1$)
the normalized Hubble expansion rate is given by 
$E(z)=[\Omega_m^{(0)}(1+z)^3+\Omega_{\rm DE}^{(0)}]^{1/2}$.
Under the bound (\ref{Rbound}) the density parameter is constrained to be 
$0.72<\Omega_{\rm DE}^{(0)}<0.77$.
This is consistent with the bound coming from the SN Ia data.
One can also show that, for increasing $w_{\rm DE}$, the observationally 
allowed values of $\Omega_m^{(0)}$ gets larger.
However, ${\cal R}$ depends weakly on the $w_{\rm DE}$.
Hence the CMB data alone do not provide a tight constraint on $w_{\rm DE}$.
In Fig.~\ref{joint} we show the joint observational constraints
on $w_{\rm DE}$ and $\Omega_m^{(0)}$ 
(for constant $w_{\rm DE}$) obtained from the WMAP 
5-year data and the Union08 SN Ia data \cite{Kowalski}.
The joint observational constraints provide much tighter bounds 
compared to the individual constraint from CMB and SN Ia.
For the flat Universe Kowalski {\it et al.} \cite{Kowalski} obtained the bounds 
$w_{\rm DE}=-0.955^{+0.060+0.059}_{-0.066-0.060}$ and 
$\Omega_m^{(0)}=0.265^{+0.022+0.018}_{-0.021-0.016}$
(with statistical and systematic errors) from the combined 
data analysis of CMB and SN Ia. See also 
Refs.~\cite{MMOT03,Hannestad02,Maor,WL03,Cora05,LeeNg,WT04,Rape,Pogo,Xia06} 
for related observational constraints.

\subsection{BAO}

The detection of baryon acoustic oscillations first reported in 2005 
by Eisenstein {\it et. al.} \cite{Eisenstein} in a spectroscopic sample 
of 46,748 luminous red galaxies observed by the Sloan Digital Sky Survey (SDSS)
has provided another test for probing the property of dark energy.
Since baryons are strongly coupled to photons prior to the decoupling epoch, 
the oscillation of sound waves is imprinted in baryon perturbations as well as
CMB anisotropies.

The sound horizon at which baryons were released from the Compton
drag of photons determines the location of baryon acoustic oscillations. 
This epoch, called the drag epoch, occurs at the redshift $z_{d}$. 
The sound horizon at $z=z_{d}$ is given by 
$r_{s}(z_d)=\int_{0}^{\eta_d}{\rm d}\eta\, c_{s}(\eta)$,
where $c_{s}$ is the sound speed.
According to the fitting formula of $z_d$ by Eisenstein and 
Hu \cite{EiHu}, $z_d$ and $r_{s}(z_d)$ are constrained to be 
around $z_d \approx 1020$ and $r_s (z_d) \approx 150$~Mpc.

We observe the angular and redshift distributions of galaxies as a
power spectrum $P(k_{\perp},k_{\parallel})$ in the redshift space,
where $k_{\perp}$ and $k_{\parallel}$ are the wavenumbers perpendicular
and parallel to the direction of light respectively. 
In principle we can measure the following two ratios \cite{Shoji} 
\begin{equation}
\theta_{s}(z)=\frac{r_{s}(z_{d})}{d_{A}^{(c)}(z)}\,,\qquad
\delta z_{s}(z)=\frac{r_{s}(z_{d})H(z)}{c}\,,
\label{thetasz}
\end{equation}
where the speed of light $c$ is recovered for clarity. 
In the first equation
$d_A^{(c)}$ is the comoving angular diameter distance related with 
the proper angular diameter distance $d_A$ via the relation 
$d_A^{(c)}=d_A/a=d_A (1+z)$.
The quantity $\theta_{s}(z)$ characterizes the angle orthogonal to the line of sight, 
whereas the quantity $\delta z_{s}$ corresponds to the 
oscillations along the line of sight.

The current BAO observations are not sufficient to measure both 
$\theta_{s}(z)$ and $\delta z_{s}(z)$ independently.
From the spherically averaged spectrum one can find 
a combined distance scale ratio given by \cite{Shoji} 
\begin{equation}
\left[\theta_{s}(z)^{2}\delta z_{s}(z)\right]^{1/3}\equiv
\frac{r_{s}(z_{d})}{[(1+z)^{2}d_{A}^{2}(z)c/H(z)]^{1/3}}\,,
\label{baoave}
\end{equation}
or, alternatively, the effective distance ratio \cite{Eisenstein} 
\begin{equation}
D_V (z) \equiv \left[ (1+z)^2 d_A^2 (z)cz/H(z) \right]^{1/3}\,.
\end{equation}
In 2005 Eisenstein {\it et al.} \cite{Eisenstein} obtained the constraint 
$D_V(z)=1370 \pm 64$~Mpc at the redshift $z=0.35$.
In 2007 Percival {\it et al.} \cite{Percival1} measured 
the effective distance ratio defined by 
\begin{equation}
r_{\rm BAO}(z) \equiv r_s (z_d)/D_V(z)\,,
\end{equation}
at the two redshifts: $r_{{\rm BAO}}(z=0.2)=0.1980 \pm 0.0058$ 
and $r_{{\rm BAO}}(z=0.35)=0.1094 \pm 0.0033$.
This is based on the data from the 2-degree Field (2dF) Galaxy 
Redshift Survey. These data provide the observational contour
of BAO plotted in Fig.~\ref{joint}.
{}From the joint data analysis of SN Ia \cite{Kowalski}, 
WMAP 5-year \cite{WMAP5}, and BAO data \cite{Percival1},
Kowalski {\it et al.} \cite{Kowalski}
placed the constraints $w_{\rm DE}=-0.969^{+0.059}_{-0.063}({\rm stat})
^{+0.063}_{-0.066}({\rm sys})$ 
and $\Omega_m^{(0)}=0.274^{+0.016}_{-0.016}
({\rm stat})^{+0.013}_{-0.012}({\rm sys})$
for the constant equation of state of dark energy.

The recent measurement of the 2dF as well as the SDSS data provided
the effective distance ratio to be $r_{{\rm BAO}}(z=0.2)=0.1905\pm0.0061$ 
and $r_{{\rm BAO}}(z=0.35)=0.1097\pm0.0036$ \cite{Percival2}.
Using these data together with the WMAP 7-year data \cite{WMAP7} 
and the Gaussian prior on the Hubble constant 
$H_0=74.2 \pm 3.6$~km~sec$^{-1}$~Mpc$^{-1}$ \cite{Riess09}, 
Komatsu {\it et al.} \cite{WMAP7} derived the constraint 
$w_{\rm DE}=-1.10 \pm 0.14$ (68 \% confidence level) 
for the constant equation of state in the flat Universe.
Adding the high-$z$ SN Ia in their analysis they found the most  
stringent bound: $w_{\rm DE}=-0.980 \pm 0.053$ 
(68 \% confidence level).
Hence the $\Lambda$CDM model is well consistent with
a number of independent observational data.
 
\vspace{0.2cm} 
Finally we should mention that there are other constraints coming from 
the cosmic age \cite{Feng05}, large-scale 
clustering \cite{LSS1,LSS2,LSS3}, gamma ray 
bursts \cite{Hooper,Oguri,Bai08,Wang08,Tsutsui}, and 
weak lensing \cite{Jain:2003tb,Takada04,Ishak05,Schimd,Bridle,Hollen}. 
So far we have not found strong evidence for supporting
dynamical dark energy models over the $\Lambda$CDM model, but 
future high-precision observations may break this degeneracy.

%%%%%%%%%%%%%%%%%%
\section{Cosmological constant}
\label{cossec}
%%%%%%%%%%%%%%%%%%

The cosmological constant $\Lambda$ is one of the simplest candidates
of dark energy, and as we have seen in the previous section, it is favored 
by a number of observations.
However, if the origin of the cosmological constant is
a vacuum energy, it suffers from a serious problem of its 
energy scale relative to the dark energy density
today \cite{Weinberg}.
The zero-point energy of some field of mass
$m$ with momentum $k$ and frequency $\omega$ is given by 
$E=\omega/2=\sqrt{k^{2}+m^{2}}/2$.
Summing over the zero-point energies
of this field up to a cut-off scale $k_{{\rm max}}$~$(\gg m)$,
we obtain the vacuum energy density 
\begin{equation}
\rho_{{\rm vac}}=\int_{0}^{k_{{\rm max}}}\frac{{\rm d}^{3}k}
{(2\pi)^{3}}\frac{1}{2}\sqrt{k^{2}+m^{2}}\,.
\label{venergy1}
\end{equation}
Since the integral is dominated by the mode with large $k$ ($\gg m$), 
we find that 
\begin{equation}
\rho_{{\rm vac}} \approx \int_{0}^{k_{{\rm max}}}\frac{4\pi k^{2} {\rm d} k}
{(2\pi)^{3}}\,\frac{1}{2}k=
\frac{k_{{\rm max}}^{4}}{16\pi^{2}}\,.
\label{venergy2}
\end{equation}
Taking the cut-off scale $k_{{\rm max}}$ to be the Planck mass $m_{{\rm pl}}$,
the vacuum energy density can be estimated as 
$\rho_{{\rm vac}}\simeq10^{74}~{\rm GeV}^{4}$.
This is about $10^{121}$ times larger than the observed value 
$\rho_{\rm DE}^{(0)} \simeq 10^{-47}\,{\rm GeV}^4$.

Before the observational discovery of dark energy in 1998, 
most people believed that the cosmological constant is 
exactly zero and tried to explain why it is so. 
The vanishing of a constant may imply the existence
of some symmetry. In supersymmetric theories the bosonic
degree of freedom has its Fermi counter part which contributes to the
zero point energy with an opposite sign\footnote{The readers 
who are not familiar with supersymmetric theories may consult the 
books \cite{Bailin,Green}.}. If supersymmetry
is unbroken, an equal number of bosonic and fermionic
degrees of freedom is present such that the total vacuum energy vanishes.
However it is known that supersymmetry is broken at sufficient high
energies (for the typical scale $M_{{\rm SUSY}} \approx 10^{3}$~GeV).
Therefore the vacuum energy is generally
non-zero in the world of broken supersymmetry. 

Even if supersymmetry is broken there is a hope to obtain a vanishing 
$\Lambda$ or a tiny amount of $\Lambda$.
In supergravity theory the effective cosmological constant is given 
by an expectation value of the potential $V$ for chiral scalar 
fields $\vp^{i}$ \cite{Bailin}:
\begin{equation}
V(\vp,\vp^{*})=e^{\kappa^{2}K}
\left[D_{i}W(K^{ij^{*}})(D_{j}W)^{*}-3\kappa^{2}|W|^{2}\right]\,,
\label{superpo}
\end{equation}
where $K$ and $W$ are the so-called K\"{a}hler
potential and the superpotential, respectively, which
are the functions of $\vp^{i}$ and its complex conjugate $\vp^{i*}$.
The quantity $K^{ij^{*}}$ is an inverse of the derivative
$K_{ij^{*}} \equiv \partial^{2}K/\partial\vp^{i}\partial\vp^{j^{*}}$, 
whereas the derivative $D_{i}W$ is defined by 
$D_{i}W\equiv \partial W/\partial\vp^{i}+
\kappa^{2}W(\partial K/\partial\vp^{i})$.

The condition $D_{i}W \neq 0$ corresponds to
the breaking of supersymmetry. In this case it is possible to find
scalar field values leading to the vanishing potential ($V=0$), but this
is not in general an equilibrium point of the potential $V$. Nevertheless
there is a class of  K\"{a}hler potentials and superpotentials giving 
a stationary scalar-field configuration at $V=0$.
The gluino condensation model in $E_{8}\times E_{8}$ superstring theory
proposed by Dine \cite{Dine} belongs to this class. The reduction
of the 10-dimensional action to the 4-dimensional action 
gives rise to a so-called modulus field $T$. This field characterizes the scale of the compactified
6-dimensional manifold. Generally one has another complex scalar field
$S$ corresponding to 4-dimensional dilaton/axion fields. 
The fields $T$ and $S$ are governed by the  K\"{a}hler potential 
\begin{equation}
K(T,S)=-(3/\kappa^{2})\,{\rm ln}\,(T+T^{*})
-(1/\kappa^{2}){\rm ln}\,(S+S^{*})\,,
\label{KTS}
\end{equation}
where $(T+T^{*})$ and $(S+S^{*})$ are positive definite. 
The field $S$ couples to the gauge
fields, while $T$ does not. An effective superpotential for $S$
can be obtained by integrating out the gauge fields under the use
of the $R$-invariance \cite{Affleck}: 
\begin{equation}
W(S)=M_{{\rm pl}}^{3}\left[c_{1}+c_{2}\exp(-3S/2c_{3})\right]\,,
\label{WS}
\end{equation}
where $c_{1},c_{2}$, and $c_{3}$ are constants.

Substituting Eqs.~(\ref{KTS}) and (\ref{WS}) into Eqs.~(\ref{superpo}), 
we obtain the field potential 
\begin{eqnarray}
V &=& \frac{1}{(T+T^{*})^{3}(S+S^{*})}
(D_S W)K^{SS^*} (D_S W)^* \nonumber \\
&=& \frac{M_{{\rm pl}}^{4}}{(T+T^{*})^{3}(S+S^{*})}\left|c_{1}
+c_{2}\exp(-3S/2c_3)\left\{ 1+\frac{3}{2c_{3}}(S+S^{*})\right\} \right|^{2}\,,
 \end{eqnarray}
where, in the first line, we have used the property 
$(D_T W)K^{TT^*} (D_T W)^*=3\kappa^2 |W|^2$
for the modulus term.
This potential is positive because of the cancellation of the last term in Eq.~(\ref{superpo}).
The stationary field configuration with $V=0$ is realized under the
condition $D_{S}W=\partial W/\partial S-W/(S+S^{*})=0$.
The derivative, 
$D_{T}W=\kappa^{2}W \partial K/\partial T=-3W/(T+T^{*})$, 
does not necessarily vanish. When $D_{T}W\neq0$ the supersymmetry
is broken with a vanishing potential energy. Therefore it is possible
to obtain a stationary field configuration with $V=0$ 
even if supersymmetry is broken. 

The discussion above is based on the lowest-order
perturbation theory. This picture is not necessarily valid
to all finite orders of perturbation theory because the non-supersymmetric
field configuration is not protected by any symmetry. Moreover some
non-perturbative effect can provide a large contribution to the effective
cosmological constant \cite{Kolda}. The so-called flux compactification
in type IIB string theory allows us to realize a metastable de Sitter (dS)
vacuum by taking into account a non-perturbative correction to the
superpotential (coming from brane instantons) as well as a number
of anti D3-branes in a warped geometry \cite{KKLT}. Hence it is not
hopeless to obtain a small value of $\Lambda$ or a vanishing $\Lambda$
even in the presence of some non-perturbative corrections.

Kachru, Kallosh, Linde and Trivedi (KKLT) \cite{KKLT} constructed
dS solutions in type II string theory compactified on a Calabi-Yau
manifold in the presence of flux. 
The construction of the dS vacua in the KKLT scenario
consists of two steps. The first step is to freeze all moduli fields
in the flux compactification at a supersymmetric Anti de Sitter (AdS)
vacuum. Then a small number of the anti D3-brane is added in a warped
geometry with a throat, so that the AdS minimum is uplifted 
to yield a dS vacuum with
broken supersymmetry. If we want to use the KKLT dS minimum 
derived above for the present cosmic acceleration,
we require that the potential energy $V_{{\rm dS}}$ at the minimum
is of the order of $V_{{\rm dS}}\simeq10^{-47}$\,GeV$^{4}$. 
Depending on the number of fluxes there are a vast of dS vacua, 
which opened up a notion called string landscape \cite{Susskind}.

The question why the vacuum we live in has a very small energy density 
among many possible vacua has been sometimes answered 
with the anthropic principle \cite{ant1,ant2}.
{}Using the anthropic arguments, Weinberg put the bound on the 
vacuum energy density \cite{Weinbergbound}
\begin{equation}
-10^{-123}\,m_{\rm pl}^4~\lsim~ \rho_{\Lambda}
~\lsim~3 \times 10^{-121}m_{\rm pl}^4\,.
\label{Webound}
\end{equation}
The upper bound comes from the requirement that the vacuum energy 
does not dominate over the matter density 
for the redshift $z~\gsim~1$.
Meanwhile the lower bound comes from the condition that 
$\rho_{\Lambda}$ does not cancel the present cosmological density. 
Some people have studied landscape statistics
by considering the relative abundance of long-lived low-energy 
vacua satisfying the bound (\ref{Webound}) \cite{Garriga00,Denef,GLV,Blum}. 
These statistical approaches are still under study, but it will be
interesting to pursue the possibility to obtain high probabilities
for the appearance of low-energy vacua.

Even in 1980s there were some pioneering works for finding a mechanism 
to make the effective cosmological constant small.
For example, let us consider a 4-form field $F^{\mu \nu \lambda \sigma}$
expressed by a unit totally anti-symmetric tensor $\epsilon^{\mu \nu \lambda \sigma}$, 
as $F^{\mu \nu \lambda \sigma}=c\epsilon^{\mu \nu \lambda \sigma}$
($c$ is a constant). Then the energy density of the 4-form field is given by 
$F^{\mu \nu \lambda \sigma}F_{\mu \nu \lambda \sigma}/(2 \cdot 4 !)=c^2/2$.
Taking into account a scalar field $\phi$ with a potential energy $V(\phi)$,
the total energy density is $\Lambda=V(\phi)+c^2/2$.
In 1984 Linde \cite{Linde84} considered the quantum creation of the Universe and 
claimed that the final value of $\Lambda$ can appear 
with approximately the same probability because $V(\phi)$ can take 
any initial value such that $\Lambda \approx m_{\rm pl}^4$.

In 1987-1988 Brown and Teltelboim \cite{Brown1,Brown2} studied the quantum 
creation of closed membranes by totally antisymmetric tensor and gravitational fields 
to neutralize the effective cosmological constant with small values.
The constant $c$ appearing in the energy density of the 4-form field 
can be quantized in integer multiples of the membrane charge $q$, 
i.e. $c=nq$. If we consider a negative bare cosmological constant $-\Lambda_b$
(as in the KKLT model) in the presence of the flux energy density $n^2 q^2/2$, 
then the effective gravitational constant is given by $\Lambda=-\Lambda_b+
n^2 q^2/2$. The field strength of the 4-form field is slowly discharged by 
a quantum Schwinger pair creation of field sources [$nq \to (n-1)q$].
However, in order to get a tiny value of $\Lambda$ consistent with 
the dark energy density today, the membrane change $q$ is constrained to be 
very small (for natural choices of  $\Lambda_b$) \cite{Boussoreview}.

In 2000 Bousso and Polchinski \cite{BoussoPol} considered multiple 4-form fields that arise in 
M-theory compactifications and showed that the small value of $\Lambda$
can be explained for natural choices of $q$.
More precisely, if we consider $J (> 1)$ 4-form fields as well as
$J$ membrane species with charges $q_1, q_2, \cdots , q_J $
and the quantized flux $F_i^{\mu \nu \lambda \sigma}=n_iq_i 
\epsilon^{\mu \nu \lambda \sigma}$, 
the effective cosmological constant is given by 
\begin{equation}
\Lambda=-\Lambda_b+\sum_{i=1}^J n_i^2 q_i^2/2\,.
\end{equation}
Bousso and Polchinski \cite{BoussoPol} showed that, for natural values of charges 
($q_i<{\cal O}(0.1)$) there exists integers $n_i$ such that 
$2\Lambda_b<\sum_{i=1}^{J }n_i^2 q_i^2<2(\Lambda_b+\Delta \Lambda)$
with $\Delta \Lambda \approx 10^{-47}\,{\rm GeV}^4$.
This can be realized for $J>100$ and $\Lambda_b \approx m_{\rm pl}^4$.

There are some interesting works for decoupling $\Lambda$ from 
gravity. In the cascading gravity scenario proposed in Ref.~\cite{Rham} the cosmological 
constant can be made gravitationally inactive by shutting off large-scale 
density perturbations. In Ref.~\cite{Afshordi} an incompressible 
gravitational Aether fluid was introduced to degraviate the vacuum.
In Refs.~\cite{Paddy1,Paddy2} Padmanabhan showed an example 
to gauge away the cosmological 
constant from gravity according to the variational principle different from 
the standard method.
See also Refs.~\cite{Hawking84,Kachru00,Kaloper00,Feng01,Tye,Garriga01,Yokoyama02,Burgess03,Burgess04,Mukoh04,Sorkin,Kane,Dolgov08} 
for other possibilities to solve 
the cosmological constant problem.
If the cosmological constant is completely decoupled from gravity, 
it is required to find alternative models of dark energy consistent with observations.

In the subsequent sections we shall consider alternative models of
dark energy, under the assumption that the cosmological 
constant problem is solved in such a way that it vanishes completely.

%%%%%%%%%%%%%%%%%%
\section{Modified matter models}
\label{mattersec}
%%%%%%%%%%%%%%%%%%

In this section we discuss ``modified matter models''
in which the energy-momentum tensor $T_{\mu \nu}$ on the r.h.s. of the Einstein 
equations contains an exotic matter source with a negative pressure.
The models that belong to this class are quintessence, k-essence, 
coupled dark energy, and  generalized Chaplygin gas.

\subsection{Quintessence}
\label{quinsec}

A canonical scalar field $\phi$ responsible for dark energy is 
dubbed quintessence \cite{quin5,Zlatev}
(see also Refs.~\cite{quin1,Ford,quin2,quin3,quin4,Ferreira1,Ferreira2} 
for earlier works).
The action of quintessence is described by 
\begin{equation}
S=\int {\rm d}^{4}x\sqrt{-g}\,\left[\frac{1}{2\kappa^{2}}R
-\frac{1}{2}g^{\mu\nu}\partial_{\mu}\phi\partial_{\nu}\phi-V(\phi)
\right]+S_{M}\,,
\label{quin}
\end{equation}
where $R$ is a Ricci scalar, and $\phi$ is a scalar field with 
a potential $V(\phi)$. 
As a matter action $S_{M}$, we consider perfect fluids
of radiation (energy density $\rho_r$, equation of state $w_r=1/3$) 
and non-relativistic matter (energy density $\rho_m$, 
equation of state $w_m=0$).

In the flat FLRW background radiation and non-relativistic matter 
satisfy the continuity equations $\dot{\rho}_r+4H\rho_r=0$
and $\dot{\rho}_m+3H\rho_m=0$, respectively.
The energy density $\rho_{\phi}$
and the pressure $P_{\phi}$ of the field are 
$\rho_{\phi}=\dot{\phi}^{2}/2+V(\phi)$ and 
$P_{\phi}=\dot{\phi}^{2}/2-V(\phi)$, respectively.
The continuity equation, 
$\dot{\rho}_{\phi}+3H (\rho_{\phi}+P_{\phi})=0$, 
translates to
\begin{equation}
\ddot{\phi}+3H\dot{\phi}+V_{,\phi}=0\,,
\label{Klein}
\end{equation}
where $V_{,\phi} \equiv {\rm d}V/{\rm d}\phi$.
The field equation of state is given by 
\begin{equation}
w_{\phi}\equiv\frac{P_{\phi}}{\rho_{\phi}}
=\frac{\dot{\phi}^{2}-2V(\phi)}{\dot{\phi}^{2}+2V(\phi)}\,.
\label{wphiqui}
\end{equation}
{}From the Einstein equations (\ref{Einsteineq}) 
we obtain the following equations
\begin{eqnarray}
H^{2} &=& \frac{\kappa^{2}}{3}\left[\frac{1}{2}\dot{\phi}^{2}
+V(\phi)+\rho_{m}+\rho_r \right]\,,
\label{Heq}
\\
\dot{H} &=&-\frac{\kappa^2}{2} \left( \dot{\phi}^2+\rho_m+
\frac43 \rho_r \right)\,.
\label{Heq1}
\end{eqnarray}

Although $\{ \rho_r, \rho_m \} \gg \rho_{\phi}$ during radiation and 
matter eras, the field energy density needs to dominate at late times
to be responsible for dark energy.
The condition to realize the late-time cosmic acceleration
corresponds to $w_{\phi}<-1/3$, i.e. 
$\dot{\phi}^{2}<V(\phi)$ from Eq.~(\ref{wphiqui}).
This means that the scalar potential needs to be flat 
enough for the field to evolve slowly.
If the dominant contribution to the energy density of the Universe is
the slowly rolling scalar field satisfying the condition $\dot{\phi}^{2} \ll V(\phi)$, 
we obtain the approximate relations $3H \dot{\phi}+V_{,\phi} \simeq 0$
and $3H^2 \simeq \kappa^2 V(\phi)$ from Eqs.~(\ref{Klein}) and 
(\ref{Heq}), respectively.
Hence the field equation of state in Eq.~(\ref{wphiqui}) is 
approximately given by 
\begin{equation}
w_{\phi} \simeq -1+2\epsilon_s/3\,,
\end{equation}
where $\epsilon_{s}\equiv \left(V_{,\phi}/V\right)^{2}/(2\kappa^2)$
is the so-called slow-roll parameter \cite{Liddlebook}.
During the accelerated expansion of the Universe, $\epsilon_s$ is 
much smaller than 1 because the potential is sufficiently flat.
Unlike the cosmological constant, the field equation of state 
deviates from $-1$ ($w_{\phi}>-1$).

Introducing the dimensionless variables $x_1 \equiv \kappa \dot{\phi}/(\sqrt{6}H)$, 
$x_2 \equiv \kappa \sqrt{V}/(\sqrt{3}H)$, and $x_3 \equiv \kappa \sqrt{\rho_r}/(\sqrt{3}H)$, 
we obtain the following equations from Eqs.~(\ref{Klein}), (\ref{Heq}), 
and (\ref{Heq1}) \cite{CLW,Macorra,Nunes,review}:
\begin{eqnarray}
 &  & x_1'=-3x_{1}+\frac{\sqrt{6}}{2}\lambda x_{2}^{2}+
 \frac{1}{2}x_{1}(3+3x_{1}^{2}-3x_{2}^{2}+x_{3}^{2})\,,\label{x1eq}\\
 &  & x_2'=-\frac{\sqrt{6}}{2}\lambda x_{1}x_{2}+\frac{1}{2}x_{2}
 (3+3x_{1}^{2}-3x_{2}^{2}+x_{3}^{2})\,,
 \label{x2eq}\\
 &  & x_3'=-2x_{3}+\frac{1}{2}x_{3}(3+3x_{1}^{2}-3x_{2}^{2}+x_{3}^{2})\,,
 \label{x3eq}
\end{eqnarray}
where a prime represents a derivative with respect to $N=\ln a$, and 
$\lambda$ is defined by $\lambda \equiv -V_{,\phi}/(\kappa V)$.
The density parameters of the field, radiation, and non-relativistic matter
are given by $\Omega_{\phi}=x_1^2+x_2^2$, $\Omega_r=x_3^2$, 
and $\Omega_m=1-x_1^2-x_2^2-x_3^2$, respectively.
One has constant $\lambda$ for the exponential 
potential \cite{CLW}
\begin{equation}
V(\phi)=V_0 e^{-\kappa \lambda \phi}\,,
\label{exppo}
\end{equation}
in which case the fixed points of the system (\ref{x1eq})-(\ref{x3eq}) can be
derived by setting $x_i'=0$ ($i=1, 2, 3$).
The fixed point that can be used for dark energy is given by 
\begin{equation}
(x_1,x_2,x_3)=\left(\lambda/\sqrt{6}, \sqrt{1-\lambda^2/6}, 0 \right)\,,
\quad w_{\phi}=-1+\lambda^2/3\,,\quad \Omega_{\phi}=1\,.
\label{fixedac}
\end{equation}
The cosmic acceleration can be realized for $w_{\phi}<-1/3$, i.e. 
$\lambda^2<2$. One can show that in this case the accelerated 
fixed point is a stable attractor \cite{CLW}.
Hence the solutions finally approach the fixed point (\ref{fixedac})
after the matter era [characterized by the fixed point $(x_1,x_2,x_3)=(0,0,0)$].

If $\lambda$ varies with time, we have the following relation
\begin{equation}
\lambda'=-\sqrt{6}\lambda^{2}(\Gamma-1)x_{1}\,,
\label{lambdaeq}
\end{equation}
where $\Gamma \equiv VV_{,\phi\phi}/V_{,\phi}^{2}$.
For monotonically decreasing potentials one has 
$\lambda>0$ and $x_1>0$ for $V_{,\phi}<0$ and 
$\lambda<0$ and $x_1<0$ for $V_{,\phi}>0$.
If the condition 
\begin{equation}
\Gamma=\frac{VV_{,\phi \phi}}
{V_{,\phi}^2}>1\,,
\label{Gamma}
\end{equation}
is satisfied, the absolute value of $\lambda$ decreases 
toward 0 irrespective of the signs of 
$V_{,\phi}$ \cite{Paul99}. 
Then the solutions finally approach the accelerated
``instantaneous'' fixed point (\ref{fixedac}) even if $\lambda^2$
is larger than 2 during radiation and matter eras \cite{Macorra,Nunes}.
In this case the field equation of state gradually decreases to $-1$, 
so the models showing this behavior are called 
``freezing'' models \cite{Caldwell}.
The condition (\ref{Gamma}) is the so-called tracking condition 
under which the field density eventually catches up that of 
the background fluid.

A representative potential of the freezing model is the inverse 
power-law potential $V(\phi)=M^{4+n}\phi^{-n}$ 
($n>0$) \cite{quin3,Paul99}, which can appear 
in the fermion condensate model as a dynamical
supersymmetry breaking \cite{Binetruy}. 
In this case one has $\Gamma=(n+1)/n>1$ and hence 
the tracking condition is satisfied.
Unlike the cosmological constant, even if the field energy density 
is not negligible relative to the background fluid density around 
the beginning of the radiation era, the field eventually 
enters the tracking regime to lead to the late-time 
cosmic acceleration \cite{Paul99}.
Another example of freezing models is $V(\phi)=M^{4+n}\phi^{-n}
\exp(\alpha\phi^{2}/m_{{\rm pl}}^{2})$, 
which has a minimum with a positive energy density 
at which the field is eventually trapped. 
This potential is motivated in the framework
of supergravity \cite{Brax}.

There is another class of quintessence potentials called 
``thawing'' models \cite{Caldwell}.
In thawing models the field with mass $m_{\phi}$ has been frozen
by the Hubble friction (i.e. the term $H\dot{\phi}$) until recently and
then it begins to evolve after $H$ drops below $m_{\phi}$. 
At early times the equation of state of dark energy is 
$w_{\phi}\simeq-1$, but it begins to grow for $H<m_{\phi}$. 
The representative potentials that belong to this class are
(a) $V(\phi)=V_{0}+M^{4-n}\phi^{n}$ ($n>0$) 
and (b) $V(\phi)=M^{4}\cos^{2}(\phi/f)$.
The potential (a) with $n=1$ was originally proposed by Linde \cite{Linde87} 
to replace the cosmological constant by a slowly evolving scalar field.
In Ref.~\cite{Kallosh03} this was revised to allow for negative 
values of $V(\phi)$. The universe will collapse in the
future if the system enters the region with $V(\phi)<0$. 
The potential (b) is motivated by the Pseudo-Nambu-Goldstone Boson
(PNGB), which was introduced in Ref.~\cite{Frieman}
in response to the first tentative suggestions for the existence of 
the cosmological constant. 
The small mass of the PNGB model required for dark energy 
is protected against radiative corrections, 
so this model is favored theoretically.
In fact there are a number of interesting works to explain the small energy scale
$M \approx 10^{-3}$\,eV required for the PNGB quintessence in supersymmetric
theories \cite{Nomura,Choi,Kim,Hall}. 
See Refs.~\cite{CNR,Townsend,Heller,Kallosh1,Kallosh2,Fre,Piazza1,Piazza2,Albrecht02,Burgess03d,Burgess} 
for the construction of quintessence potentials
in the framework of supersymmetric theories.

%%%%%%%%%%%%%%%%%%%%%%%%%%%%%
\begin{figure}
\begin{centering}
\includegraphics[width=3.2in,height=3.0in]{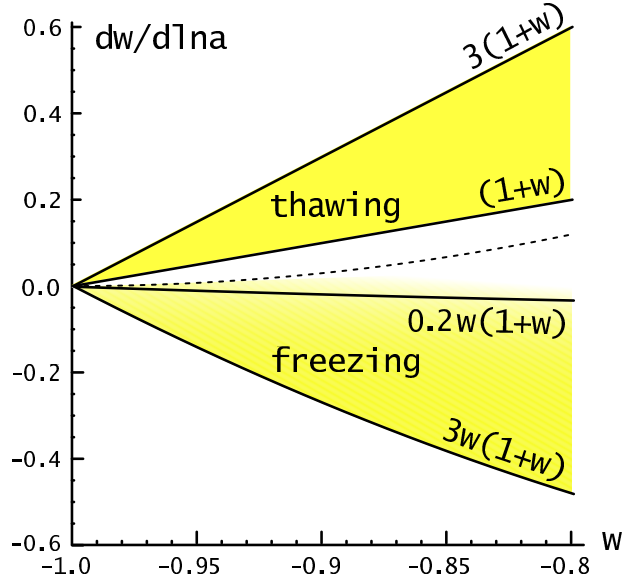} 
\par\end{centering}
\caption{The allowed region in the $(w_{\phi},w_{\phi}')$ plane for thawing
and freezing models of quintessence
($w_{\phi}$ is denoted as $w$ in the figure).
The thawing models correspond to the region between two curves: (a)
$w_{\phi}'=3(1+w_{\phi})$ and (b) $w_{\phi}'=1+w_{\phi}$, whereas
the freezing models are characterized by the region between two curves:
(c) $w_{\phi}'=0.2w_{\phi}(1+w_{\phi})$ and (d) $w_{\phi}'=3w_{\phi}(1+w_{\phi})$.
The dotted line shows the border between the acceleration and deceleration
of the field ($\ddot{\phi}=0$), which corresponds to 
$w_{\phi}'=3(1+w_{\phi})^{2}$. {}From Ref.~\cite{Caldwell}.}
\centering{}\label{wwd} 
\end{figure}
%%%%%%%%%%%%%%%%%%%%%%%%%%%%

The freezing models and the thawing
models are characterized by the conditions $w_{\phi}'\equiv 
{\rm d}w_{\phi}/{\rm d}N<0$
and $w_{\phi}'>0$, respectively. 
More precisely the allowed regions for the freezing and thawing 
models are given by 
$3w_{\phi}(1+w_{\phi})~\lsim~w_{\phi}'~\lsim~
0.2w_{\phi}(1+w_{\phi})$ and 
$1+w_{\phi}~\lsim~w_{\phi}'~\lsim~3(1+w_{\phi})$, 
respectively \cite{Caldwell} (see Ref.~\cite{Linder06} for details).
These regions are illustrated in Fig.~\ref{wwd}. 
While the observational data
up to now are not sufficient to distinguish freezing and thawing models
by the variation of $w_{\phi}$, we may be able to do so with the
next decade high-precision observations.

There is also another useful measure called ``statefinder'' \cite{state} by which 
the quintessence models can be distinguished from the $\Lambda$CDM 
model (see also Ref.~\cite{state2}). The statefinder parameters are defined by 
\begin{equation}
r \equiv \frac{1}{aH^3}\frac{{\rm d}^3 a}{{\rm d}t^3}\,,\qquad
s \equiv \frac{r-1}{3(q-1/2)}\,,
\label{rsdef}
\end{equation}
where $q=-\ddot{a}/(aH^2)$ is the deceleration parameter.
Let us consider the case in which the dark energy density 
$\rho_{\rm DE}$ satisfies the continuity equation 
$\dot{\rho}_{\rm DE}+3H(1+w_{\rm DE})\rho_{\rm DE}=0$.
In Einstein gravity in which the Friedmann equation 
$3H^2=\kappa^2 (\rho_{\rm DE}+\rho_m)$ holds, 
we obtain
\begin{equation}
r=1+\frac{9w_{\rm DE}\Omega_{\rm DE}}{2}s\,,\qquad
s=1+w_{\rm DE}-\frac{\dot{w}_{\rm DE}}{3H w_{\rm DE}}\,,
\end{equation}
where $\Omega_{\rm DE} \equiv \kappa^2 \rho_{\rm DE}/(3H^2)$.
The $\Lambda$CDM model ($w_{\rm DE}=-1$) corresponds to a
point $(r,s)=(1,0)$ in the $(r,s)$ plane, but the quintessence models
are characterized by the curves in the region $r<1$ and $s>0$ \cite{Alam}.
The Chaplygin gas model we discuss in Sec.~\ref{unifiedsec}
gives rise to the curves in the region $r>1$ and $s<0$.
Hence one can distinguish between dark energy models
by using the statefinders defined in Eq.~(\ref{rsdef}). 

It is possible to reconstruct quintessence potentials from the 
observational data of SN Ia. Neglecting the contribution of 
radiation and using the relation $\rho_m=\rho_m^{(0)}(1+z)^3$
for non-relativistic matter, we get the following equations 
from Eqs.~(\ref{Heq}) and (\ref{Heq1}) \cite{Star98,Saini,Nakamura,Chiba00}:
\begin{eqnarray}
 &  & \frac{\kappa^{2}}{2}\left(\frac{\rd\phi}{\rd z}\right)^{2}=
 \frac{1}{1+z}\frac{\rd\ln\, E(z)}{\rd z}-\frac{3\Omega_{m}^{(0)}}{2}
 \frac{1+z}{E(z)^2}\,,\label{reconst1}\\
 &  & \frac{\kappa^{2}V}{3H_{0}^{2}}=E(z)
 -\frac{1+z}{6}\,\frac{\rd\, E(z)^2}{\rd z}
 -\frac{1}{2}\Omega_{m}^{(0)}(1+z)^{3}\,,\label{reconst2}
 \end{eqnarray}
where $E(z)=H(z)/H_0$.
Note that we have changed the time derivative to the derivative 
with respect to the redshift $z$, by using the relation 
$\rd t=-\rd z/[H(1+z)]$.
Integrating Eq.~(\ref{reconst1}), the field $\phi$ is known as
a function of $z$. Inverting $\phi (z)$ to $z(\phi)$ and plugging 
it into Eq.~(\ref{reconst2}), one can reconstruct the potential $V$ 
with respect to $\phi$ by using the information of the observationally 
known values of $H(z)$ and $H'(z)$ as well as $\Omega_m^{(0)}$.
We caution, however, that the actual observational data (such as
the luminosity distance) are obtained at discrete values of redshifts.
Hence we need some smoothing process for reconstructing the
potential $V(\phi)$ and the field equation of state $w_{\phi}(z)$.
This smoothing was already discussed in Sec.~\ref{SNIasec}.

{}From the viewpoint of particle physics, the quintessence energy density 
can be comparable to the background fluid density in the early Universe.
It is possible to construct quintessence models in which the field energy 
density is proportional to the fluid density during radiation and matter eras.
For the exponential potential (\ref{exppo}) there is a fixed point 
giving a constant field density parameter 
$\Omega_{\phi}=3(1+w_M)/\lambda^2$ with $w_{\phi}=w_M$, 
where $w_M$ is the fluid equation of state \cite{CLW}.
This is called a scaling solution, which is stable for $\lambda^2>3(1+w_M)$.
During radiation and matter eras one has $\Omega_{\phi}=4/\lambda^2$
and $\Omega_{\phi}=3/\lambda^2$, respectively.
The Big Bang Nucleosynthesis (BBN) places the bound $\Omega_\phi<0.045$ 
(95 \% confidence level) around the temperature $T =1$\,MeV \cite{BBN}, 
which gives the bound $\lambda>9.4$.
In this case, however, the scalar field does not exit to the accelerated
fixed point given by Eq.~(\ref{fixedac}).

There are several ways to allow a transition from the scaling regime to 
the epoch of cosmic acceleration. 
One of them is to introduce a field potential that becomes shallow at late times, 
$V(\phi)=c_1 e^{-\kappa \lambda \phi}+c_2 e^{-\kappa \mu \phi}$
with $\lambda^2>3(1+w_M)$ and $\mu^2<2$ \cite{BCN}
(see Refs.~\cite{SahniWang,Albrecht,Stewart,Matos,BBN,Sethi,Blais} for related works).
For this double exponential potential the field equation of state 
of the final attractor is given by $w_{\phi}=-1+\mu^2/3$.
Another way is to consider multiple scalar fields with exponential
potentials,	$V(\phi_1, \phi_2)=c_1e^{-\kappa \lambda_1 \phi_1}+
c_2 e^{-\kappa \lambda_2 \phi_2}$ \cite{Coley,KLT}.
In this case the phenomenon called assisted inflation \cite{assisted} 
occurs for the multi-field exponential potential, even if the individual field 
has too steep a potential to lead to cosmic acceleration. 
The scalar field equation of state finally 
approaches the value $w_{\phi}=-1+\lambda_{\rm eff}^2/3$, 
where $\lambda_{\rm eff}=(1/\lambda_1^2+1/\lambda_2^2)^{-1/2}$
is smaller than each $\lambda_i$ ($i=1,2$).
In the presence of three assisting scalar fields it is possible to 
realize the observational bound $w_{\phi}(z=0)<-0.8$ today \cite{Ohashi}. 

There is a class of models dubbed quintessential inflation \cite{PVilenkin}
in which a single scalar field $\phi$ is responsible 
for both inflation and dark energy. 
One example of quintessential inflation is given by 
$V(\phi)=\lambda(\phi^{4}+M^{4})$ for $\phi<0$ and 
$V(\phi)=\lambda M^{4}/\left[1+(\phi/M)^{n}\right]$ ($n>0$)
for $\phi \ge 0$. In the regime $\phi<0$ with $|\phi| \gg M$
the potential behaves as $V(\phi) \simeq \lambda \phi^4$, which 
leads to inflation. In the regime $\phi>0$ with $\phi \gg M$ one has 
$V(\phi) \simeq \lambda M^{4+n} \phi^{-n}$, which leads the 
late-time cosmic acceleration.
Since the potential does not have a minimum, the reheating 
after inflation proceeds with a gravitational particle production.
Although this process is not efficient in general, 
it may be possible to make the reheating more efficient under 
the instant preheating mechanism proposed in Ref.~\cite{Felder}.
See Refs.~\cite{Giovan,Peloso,Kaga,Yahiro,Dimo,Nunes02,Dimo2,Tashiro,Sami04,Rosen05,Liddle06,Neupane08} 
about related works on quintessential inflation.

\subsection{k-essence}

Scalar fields with non-canonical kinetic terms often appear
in particle physics.
In general the action for such theories can be expressed as 
\begin{equation}
S=\int {\rm d}^{4}x\sqrt{-g}\left[\frac{1}{2\kappa^{2}}R
+P(\phi,X)\right]+S_{M}\,,
\label{kessence}
\end{equation}
where $P(\phi,X)$ is a function in terms of a scalar field
$\phi$ and its kinetic energy 
$X=-(1/2)g^{\mu\nu}\partial_{\mu}\phi\partial_{\nu}\phi$,
and $S_M$ is a matter action.
Even in the absence of the field potential $V(\phi)$ it is possible to 
realize the cosmic acceleration due to the kinetic energy $X$ \cite{kinf}.
The application of these theories to dark
energy was first carried out by Chiba {\it et al.} \cite{kes1}. 
In Ref.~\cite{kes2} this was extended to more general cases and the models based on the action
(\ref{kessence}) were named {}``k-essence''.
The action (\ref{kessence}) includes a wide variety of theories listed below.
\begin{itemize}
\item (A) Low energy effective string theory

The action of low energy effective string theory in the presence of a 
higher-order derivative term $(\tilde{\nabla}\phi)^4$ is 
given by \cite{PBB,Gasreview}
\begin{equation}
S=\frac{1}{2\kappa^2} \int {\rm d}^{4}\tilde{x}\sqrt{-\tilde{g}}
\left[ F(\phi) \tilde{R}
+\omega (\phi) (\tilde{\nabla} \phi)^2
+\alpha'B(\phi)(\tilde{\nabla} \phi)^4+{\cal O}(\alpha'^2) \right]\,,
\label{lowene}
\end{equation}
which is derived by the expansion in terms of the Regge 
slope parameter $\alpha'$ (this is related to the string mass scale $M_s$
via the relation $M_s=\sqrt{2/\alpha'}$).
The scalar field $\phi$, dubbed a dilaton field, is coupled to the Ricci scalar $R$
with the strength $F(\phi)$. This frame is called the Jordan frame, in which 
the tilde is used in the action (\ref{lowene}).
Under a conformal transformation, 
$g_{\mu \nu}=F(\phi)\tilde{g}_{\mu \nu}$, 
we obtain the action in the Einstein frame \cite{kinf}
\begin{equation}
S_E=\int {\rm d}^{4}x \sqrt{-g}\left[ 
\frac{1}{2\kappa^2}R+K(\phi)X
+L(\phi) X^2+\cdots \right]\,,
\end{equation}
where $K (\phi)=3(F_{,\phi}/F)^2-2\omega/F$ and 
$L(\phi)=2\alpha' B(\phi)/\kappa^2$.

\vspace{0.3cm}
\item (B) Ghost condensate

The theories with a negative kinetic energy $-X$ generally suffers from 
the vacuum instability \cite{Trod,Moore}, but the presence of the quadratic term
$X^2$ can evade this problem. The model constructed in this vein 
is the ghost condensate model characterized by the Lagrangian \cite{Arkani}
\begin{equation}
P=-X+X^2/M^4\,,
\label{ghostmodel}
\end{equation}
where $M$ is a constant. A more general version of this model, called 
the dilatonic ghost condensate \cite{PiaTsu}, is 
\begin{equation}
P=-X+e^{\kappa \lambda \phi}X^2/M^4\,,
\label{ghostmodel2}
\end{equation}
which is motivated by a dilatonic higher-order correction to the 
tree-level action [as we have discussed in the case (A)].

\vspace{0.3cm}
\item (C) Tachyon

A tachyon field appears as an unstable mode of 
D-branes [non-Bogomol'nyi-Prasad-Sommerfield (non-BPS) branes]. 
The effective 4-dimensional Lagrangian is 
given by \cite{Garousi,Asen,Gibbons}
\begin{equation}
P=-V(\phi) \sqrt{1-2X}\,,
\label{tachyonmodel}
\end{equation}
where $V(\phi)$ is a potential of the tachyon field $\phi$.
While the tachyon model is difficult to be compatible 
with inflation in the early Universe because of the problem
for ending inflation \cite{Malcolm,Feinstein,Kofmanta}, 
one can use it for dark energy provided that the 
potential is shallower than 
$V(\phi)=V_0\phi^{-2}$ \cite{Paddyta,Abramo,Ruth,Garousi04,CGST,CalLiddle}. 

\vspace{0.3cm}
\item (D) Dirac-Born-Infeld (DBI) theory 

In the so-called ``D-cceleration'' mechanism in which a scalar field 
$\phi$ parametrizes a direction on the approximate Coulomb branch 
of the system in ${\cal N}=4$ supersymmetric Yang Mills theory, 
the field dynamics can be described by the DBI action for a probe
D3-brane moving in a radial direction of the Anti de Sitter (AdS) 
space-time \cite{Silver1,Silver2}. 
The Lagrangian density with the field potential 
$V(\phi)$ is given by 
\begin{equation}
P=-f(\phi)^{-1} \sqrt{1-2f(\phi)X}+f(\phi)^{-1}-V(\phi)\,,
\label{DBImodel}
\end{equation}
where $f(\phi)$ is a warped factor of the AdS throat.
In this theory one can realize the acceleration of the Universe 
even in the regime where $2f(\phi)X$ is close to 1.
The application of this theory to dark energy has been 
carried out in Refs.~\cite{Martin08,Ohta08,CoMizuno}.

\end{itemize}

\vspace{0.2cm}

For the theories with the action (\ref{kessence}),
the pressure $P_\phi$ and the energy density $\rho_{\phi}$ 
of the field are $P_{\phi}=P$ and 
$\rho_{\phi}=2XP_{,X}-P$, respectively. 
The equation of state of k-essence is given by  
\begin{equation}
w_{\phi}=\frac{P_{\phi}}{\rho_{\phi}}=
\frac{P}{2XP_{,X}-P}\,.\label{wphikes}
\end{equation}
As long as the condition $|2XP_{,X}|\ll|P|$ is satisfied, 
$w_{\phi}$ can be close to $-1$.
In the ghost condensate model (\ref{ghostmodel})
we have
\begin{equation}
w_{\phi}=\frac{1-X/M^{4}}{1-3X/M^{4}}\,,
\end{equation}
which gives $-1<w_{\phi}<-1/3$ for $1/2<X/M^{4}<2/3$. 
In particular the de Sitter solution ($w_{\phi}=-1$) is realized at $X/M^{4}=1/2$.
Since the field energy density is $\rho_{\phi}=M^{4}/4$ at the de
Sitter point, it is possible to explain the cosmic acceleration today
for $M\simeq 10^{-3}$~eV.

In order to discuss stability conditions of k-essence in the Ultra-Violet (UV) 
regime, we decompose the field into the homogenous and perturbed parts as 
$\phi (t, {\bf x})=\phi_0(t)+\delta \phi (t, {\bf x})$
in the Minkowski background and derive the Lagrangian
and the Hamiltonian for perturbations. 
The resulting second-order Hamiltonian reads \cite{PiaTsu} 
\begin{equation}
\delta H=\left(P_{,X}+2XP_{,XX}\right)
\frac{(\delta\dot{\phi})^{2}}{2}+P_{,X}\frac{(\nabla\delta\phi)^{2}}{2}
-P_{,\phi\phi}\frac{(\delta\phi)^{2}}{2}\,.
\label{Ldelpsi}
\end{equation}
The term $P_{,\phi \phi}$ is related with the effective mass of the field, 
which is unimportant in the UV regime as long as the field is
responsible for dark energy.
The positivity of the first two terms 
in Eq.~(\ref{Ldelpsi}) leads to the following stability conditions
\begin{eqnarray}
 P_{,X}+2XP_{,XX}\,\ge\,0\,,\qquad P_{,X}\,\ge\,0\,.
 \label{xi}
\end{eqnarray}

The phantom model with a negative kinetic energy $-X$ 
with a potential $V(\phi)$, i.e. $P=-X-V(\phi)$, do not satisfy 
the above conditions. 
Although the phantom model with $P=-X-V(\phi)$ can lead to 
the background cosmological dynamics allowed by SN Ia observations
($w_{\phi}<-1$) \cite{Trod,SinghSami,SamiTo}, 
it suffers from a catastrophic 
particle production of ghosts and normal fields because of
the instability of the vacuum \cite{Trod,Moore}
(see 
Refs.~\cite{Dabrowski,Emilio04,Feng05,Guo05o,Li05,Wei05,Arefeva05,Wei05d,Cai05,ZhaoXia,Zhang06,WZhao,Guo06,Nojiri06,YCai,Setare} for related works).
This problem is overcome in the ghost condensate model (\ref{ghostmodel})
in which the conditions (\ref{xi}) are satisfied for 
$X/M^4>1/2$ \footnote{It is possible that the dilatonic ghost condensate 
model crosses the cosmological constant boundary $w_{\phi}=-1$, but the
quantum instability problem is present in the region $w_{\phi}<-1$.
This crossing does not occur for the single-field k-essence Lagrangian 
with a linear function of $X$ \cite{Vikmank}.}.
Thus the successful k-essence models need to be constructed to 
be consistent with the conditions (\ref{xi}), while the field 
is responsible for dark energy under the condition 
$|2XP_{,X}|\ll|P|$.

The propagation speed $c_{s}$ of the field 
is given by \cite{Garriga} 
\begin{equation}
c_{s}^{2}=\frac{P_{\phi,X}}{\rho_{\phi,X}}=
\frac{P_{,X}}{P_{,X}+2XP_{,XX}}\,,
\label{soundkes}
\end{equation}
which is positive under the conditions (\ref{xi}). The speed $c_s$ remains
sub-luminal provided that 
\begin{equation}
P_{,XX}>0\,.
\label{PXX}
\end{equation}
This condition is ensured for the models (\ref{ghostmodel}), 
(\ref{ghostmodel2}), (\ref{tachyonmodel}), and (\ref{DBImodel}).

There are some k-essence models proposed to solve the coincidence problem 
of dark energy. One example is \cite{kes2,kes3}
\begin{equation}
P=\frac{1}{\phi^{2}}\left(-2.01+2\sqrt{1+X}+3\cdot10^{-17}X^{3}-10^{-24}X^{4}\right)\,.
\label{kesmodel1}
\end{equation}
In these models the solutions can finally approach the accelerating phase 
even if they start from relatively large values of the k-essence 
energy density $\Omega_{\phi}$ 
in the radiation era.
In such cases, however, there is a period in which the sound speed 
becomes superluminal before reaching the accelerated attractor \cite{Durrer}.
Moreover it was shown that the basins of attraction of a radiation scaling solution 
in such models are restricted to be very small \cite{Malquarti}.
We stress that these problems arise only for
the k-essence models constructed to solve the coincidence problem.

\subsection{Coupled dark energy}

Since the energy density of dark energy is the same order as that of 
dark matter in the present Universe, this implies that dark energy 
may have some relation with dark matter.
In this section we discuss the cosmological viability of 
coupled dark energy models and related topics such as 
scaling solutions, the chameleon mechanism, and 
varying $\alpha$.

%subsub
\subsubsection{The coupling between dark energy and dark matter}
\label{couplesec}
%subsub

In the flat FLRW cosmological background, a general coupling 
between dark energy (with energy density $\rho_{\rm DE}$
and equation of state $w_{\rm DE}$)
and dark matter (with energy density $\rho_m$) may be described 
by the following equations
\begin{eqnarray}
 \dot{\rho}_{\rm DE}+3H(1+w_{\rm DE})\rho_{\rm DE} &=& -\beta\,,
 \label{rhoDEeqco} \\
 \dot{\rho}_m+3H \rho_m &=& +\beta\,,
 \label{rhomeqco}
\end{eqnarray}
where $\beta$ is the rate of the energy exchange in the 
dark sector.

There are several forms of couplings proposed so far.
Two simple examples are given by 
\begin{eqnarray}
 & & {\rm (A)}~~\beta=\kappa Q \rho_m \dot{\phi}\,,
 \label{couplingA} 
 \\
 & & {\rm (B)}~~\beta=\alpha\, H \rho_m\,,
 \label{couplingB}
\end{eqnarray}
where $Q$ and $\alpha$ are dimensionless constants.
The coupling (A) arises in scalar-tensor theories after the conformal 
transformation to the Einstein frame \cite{Wetterich:1994bg,Amendola:1999qq,Lucacoupled,Holden:1999hm}.
In general the coupling $Q$ is field-dependent \cite{Huey,Das}, but 
Brans-Dicke theory \cite{Brans} (including $f(R)$ gravity) gives rise to 
a constant coupling \cite{TUMTY}. 
The coupling (B) is more phenomenological, 
but this form is useful to place observational bounds from the cosmic 
expansion history.
Several authors studied other couplings of the forms 
$\beta=(\alpha_m \rho_m+\alpha_{\rm DE}\rho_{\rm DE})H$ \cite{Zim1,Zim2,Chimento}, 
$\beta=\alpha \Omega_{\rm DE}H$ \cite{Dalal,Campos,Guo07}, 
and $\beta=\Gamma \rho_m$ \cite{Roy1,Jussi1,Roy2}.
See also Refs.~\cite{PavonSen,Cai05d,Guo05,Morita,Gumjudpai,Tomi05,delCampo,Lee06,Sadjadi,Bertolami,Mainini,Wei,Fukuyama,Urena,Caldera,Jussi2,Wei10} 
for related works.

\vspace{0.2cm}
{\bf (A) The coupling (A)}
\vspace{0.2cm}

Let us consider the coupling (A) in the presence of a coupled 
quintessence field with the exponential potential (\ref{exppo}).
We assume that the coupling $Q$ is constant.
Taking into account radiation uncoupled to dark energy 
($\rho_r \propto a^{-4}$)
the Friedmann equation is given by $3H^2=\kappa^2 
(\rho_{\rm DE}+\rho_m+\rho_r)$, where 
$\rho_{\rm DE}=\dot{\phi}^2/2+V(\phi)$.
Introducing the dimensionless variables 
$x_1=\kappa \dot{\phi}/(\sqrt{6}H)$, $x_2=\kappa \sqrt{V}/(\sqrt{3}H)$, 
and $x_3= \kappa \sqrt{\rho_r}/(\sqrt{3}H)$ as in Sec.~\ref{quinsec},  
we obtain
\begin{equation}
 x_1'=-3x_{1}+\frac{\sqrt{6}}{2}\lambda x_{2}^{2}+
 \frac{1}{2}x_{1}(3+3x_{1}^{2}-3x_{2}^{2}+x_{3}^{2})
 -\frac{\sqrt{6}}{2}Q (1-x_1^2-x_2^2-x_3^2)\,,
\end{equation}
and the same differential equations for $x_2$ and $x_3$ 
as given in Eqs.~(\ref{x2eq}) and (\ref{x3eq}).
For this dynamical system there is a scalar-field dominated 
fixed point given in Eq.~(\ref{fixedac}) as well as the 
radiation point $(x_1,x_2,x_3)=(0,0,1)$.
In the presence of the coupling $Q$, the standard matter era
is replaced by a
``$\phi$-matter-dominated epoch ($\phi$MDE)'' \cite{Lucacoupled}
characterized by 
\begin{equation}
(x_1, x_2, x_3)=(-\sqrt{6}Q/3,0,0)\,,\qquad
\Omega_{\phi}=2Q^2/3\,,\qquad
w_{\phi}=1\,.
\label{phiMDE}
\end{equation}
Defining the effective equation of state
\begin{equation}
w_{\rm eff}=-1-2\dot{H}/(3H^2)\,,
\end{equation}
one has $w_{\rm eff}=2Q^2/3$ for the $\phi$MDE, 
which is different from 0
in the uncoupled case. Provided that $2Q^2/3<1$, the $\phi$MDE is 
a saddle followed by the accelerated point (\ref{fixedac}) \cite{Lucacoupled}.

The evolution of the scale factor during the $\phi$MDE
is given by $a \propto t^{2/(3+2Q^2)}$, which is 
different from that in the uncoupled quintessence.
This leads to a change to the CMB shift parameter defined 
in Eq.~(\ref{CMBshift}). {}From the CMB likelihood analysis, 
the strength of the coupling is constrained to be 
$|Q|< 0.1$ \cite{Lucacoupled}. 
The evolution of matter density perturbations
is also subject to change by the effect of the coupling.
Under a quasi-static approximation on sub-horizon scales
the matter perturbation $\delta_m$ obeys the following 
equation \cite{Amenper,Amenper2}
\begin{equation}
\ddot{\delta}_m+(2H+Q\dot{\phi})\dot{\delta}_m-
4\pi G_{\rm eff} \rho_m \delta_m \simeq 0\,,
\label{delmeq}
\end{equation}
where the effective gravitational coupling is given by 
$G_{\rm eff}=(1+2Q^2)G$.
During the $\phi$MDE one can obtain the analytic solution to 
Eq.~(\ref{delmeq}), as $\delta_m \propto a^{1+2Q^2}$.
Hence the presence of the coupling $Q$ leads to a larger growth rate
relative to the uncoupled quintessence.
We can parameterize the growth rate of 
matter perturbations, as \cite{Peeblesbook}
\begin{equation}
f \equiv \frac{\dot{\delta}_m}{H \delta_m}
=(\Omega_m)^{\gamma}\,,
\label{gammadex}
\end{equation}
where $\Omega_m \equiv \kappa^2 \rho_m/(3H^2)$ is the density parameter 
of non-relativistic matter.
In the $\Lambda$CDM model the growth index $\gamma$ can be
approximately given by $\gamma \simeq 0.55$ \cite{Wang98,Linder05}.
In the coupled quintessence the growth rate can be fitted to the numerical 
solution by the formula $f=(\Omega_m)^{\gamma} (1+cQ^2)$, where 
$c=2.1$ and $\gamma=0.56$ are the best-fit values \cite{DiPorto}.
Using the galaxy and Lyman-$\alpha$ power spectra, 
the growth index $\gamma$ and the coupling $Q$ are constrained to be 
$\gamma=0.6^{+0.4}_{-0.3}$ and $|Q|<0.52$ (95 \% confidence level), 
respectively. This is weaker than the bound coming from
the CMB constraint \cite{DiPorto}. 
We also note that the equation for matter perturbations has been 
derived for the coupled k-essence scenario with a field-dependent 
coupling $Q(\phi)$ \cite{kesper1,kesper2}.
In principle it is possible to reconstruct the coupling from observations 
if the evolution of matter perturbations is 
known accurately \cite{Tsujirecon}.

There is another interesting coupled dark energy scenario called mass varying 
neutrino \cite{Hung00,Gu03,Fardon,Peccei,Takahashi06}
in which the mass $m_{\nu}$ of the neutrino depends on the quintessence
field $\phi$. The energy density $\rho_{\nu}$ and the pressure 
$P_{\nu}$ of neutrinos can be determined by assuming a Fermi-Dirac 
distribution with the neglect of the chemical potential. 
It then follows that the field $\phi$ obeys
the equation of motion \cite{Peccei}
\begin{equation}
\ddot{\phi}+3H\dot{\phi}+V_{,\phi}
=-\kappa Q(\phi) (\rho_{\nu}-3P_{\nu})\,,
\label{neueq}
\end{equation}
where $Q(\phi) \equiv \frac{{\rm d}\ln m_{\nu}(\phi)}{{\rm d} \phi}$.
In the relativistic regime in which the neutrino mass $m_{\nu}$ is much smaller
than the neutrino temperature $T_{\nu}$, the r.h.s. of Eq.~(\ref{neueq})
is suppressed because of the relation $\rho_{\nu} \simeq 3P_{\nu}$.
In the non-relativistic regime with $m_{\nu} \gg T_{\nu}$ the pressure
$P_{\nu}$ is much smaller than the energy density $\rho_{\nu}$, 
in which case the field equation (\ref{neueq}) mimics Eq.~(\ref{rhoDEeqco})
with the coupling (A).
In the mass varying neutrino scenario the field-dependent mass of neutrinos
determines the strength of the coupling $Q(\phi)$.
In the non-relativistic regime the neutrino energy density is approximately 
given by $\rho_{\nu} \simeq n_{\nu} m_{\nu} (\phi)$, where $n_{\nu}$ is the 
number density of neutrinos. Then the effective potential of the field is 
given by $V_{\rm eff}(\phi) \simeq V(\phi)+n_{\nu}m_{\nu} (\phi)$, 
which gives rise to a minimum for a runaway quintessence potential $V(\phi)$.
Since the field equation of state in this regime is $w_{\phi} \simeq 
-V(\phi)/[V(\phi)+n_{\nu}m_{\nu} (\phi)]$, one has $w_{\phi} \simeq -1$
for $n_{\nu}m_{\nu} \ll V(\phi)$. See 
Refs.~\cite{Kohri05,Brook1,Brook2,Amenneu,Ichiki08,Bambaneu} 
for a number of cosmological consequences of the mass varying neutrino scenario.

\vspace{0.2cm}
{\bf (B) The coupling (B)}
\vspace{0.2cm}

Let us consider the coupling (B) given in Eq.~(\ref{couplingB}).
For constant $\delta$ one can integrate Eq.~(\ref{rhomeqco}) as 
$\rho_m=\rho_m^{(0)}(1+z)^{3-\alpha}$, where 
$z$ is a redshift.
If the equation of state $w_{\rm DE}$ is constant, one obtains 
the following integrated solution for Eq.~(\ref{rhoDEeqco}):
\begin{equation}
\rho_{\rm DE}=\rho_{\rm DE}^{(0)}(1+z)^{3(1+w_{\rm DE})}
+\rho_m^{(0)} \frac{\alpha}{\alpha+3w_{\rm DE}}
\left[ (1+z)^{3(1+w_{\rm DE})}-(1+z)^{3-\alpha} \right]\,.
\end{equation}
The Friedmann equation, $3H^2=\kappa^2 (\rho_{\rm DE}+\rho_m)$, 
gives the parametrization for the normalized Hubble parameter
$E(z)=H(z)/H_0$, as 
\begin{equation}
E^2(z)=\Omega_{\rm DE}^{(0)}(1+z)^{3(1+w_{\rm DE})}
+\frac{1-\Omega_{\rm DE}^{(0)}}{\alpha+3w_{\rm DE}}
\left[ \alpha (1+z)^{3(1+w_{\rm DE})}+3w_{\rm DE}
(1+z)^{3-\alpha} \right]\,.
\end{equation}
This parametrization can be used to place observational constraints 
on the coupling $\alpha$.
The combined data analysis using  the observational data of the 5-year 
Supernova Legacy Survey (SNLS) \cite{SNLS}, 
the CMB shift parameter from the 3-year WMAP \cite{WMAP3}, 
and the BAO \cite{Eisenstein} shows that $\alpha$ and $w_{\rm DE}$
are constrained to be $-0.08<\alpha<0.03$ and 
$-1.16<w_{\rm DE}<-0.91$ (95 \% confidence level) \cite{Guo07}.
In Ref.~\cite{Jussi1} it was shown that, for constant $w_{\rm DE}$,
cosmological perturbations are subject to non-adiabatic 
instabilities in the early radiation era. This problem can be alleviated 
by considering the time-dependent $w_{\rm DE}$ satisfying the 
condition $w_{\rm DE}>-4/5$ at early times \cite{Jussi2}.

It is also possible to extend the analysis to the case in which 
the coupling $\alpha$ varies in time.
Dalal {\it et al.} \cite{Dalal} assumed the scaling relation, 
$\rho_{\rm DE}/\rho_m=(\rho_{\rm DE}^{(0)}/\rho_m^{(0)}) a^{\xi}$, 
where $\xi$ is a constant. For constant $w_{\rm DE}$ the coupling $\alpha$
is expressed in the form
\begin{equation}
\alpha (z)=\frac{\alpha_0}{\Omega_{\rm DE}^{(0)}+(1-\Omega_{\rm DE}^{(0)})
(1+z)^{\xi}}\,,
\end{equation}
where $\alpha_0=-(\xi+3w_{\rm DE})\Omega_{\rm DE}^{(0)}$.
If $\xi>0$, $\alpha (z)$ decreases for higher $z$.
In this case the Hubble parameter can be parametrized as
\begin{equation}
E^2(z)=(1+z)^3 \left[ 1-\Omega_{\rm DE}^{(0)}
+\Omega_{\rm DE}^{(0)}(1+z)^{-\xi} \right]^{-3w_{\rm DE}/\xi}\,.
\end{equation}
The combined data analysis using the SNLS, WMAP 3-year, and the BAO 
gives the bounds $-0.4<\alpha_0<0.1$ 
and $-1.18<w_{\rm DE}<-0.91$  (95 \% confidence level) \cite{Guo07}.
The $\Lambda$CDM model ($\alpha=0$, $w_{\rm DE}=-1$)
remains a good fit to the data.

%subsub
\subsubsection{Coupled dark energy and coincidence problem}
%subsub

In the coupled quintessence with an exponential potential 
($V(\phi)=V_0 e^{-\kappa \lambda \phi}$ with $\lambda>0$)
the $\phi$MDE scaling solution with $\Omega_{\phi}=2Q^2/3=$\,constant 
replaces the standard matter era. 
In this model there is another scaling solution given 
by \cite{Lucacoupled,Gumjudpai}
\begin{eqnarray}
& &\left( x_1, x_2, x_3 \right)=\left( \frac{\sqrt{6}}{2(Q+\lambda)}, 
\sqrt{\frac{2Q(Q+\lambda)+3}{2(Q+\lambda)^2}}, 0 \right)\,,
\qquad \Omega_{\phi}=\frac{Q(Q+\lambda)+3}{(Q+\lambda)^2}\,,
\nonumber \\
& &w_{\phi}=-\frac{Q (Q+\lambda)}{Q(Q+\lambda)+3}\,,\qquad
w_{\rm eff}=-\frac{Q}{Q+\lambda}\,.
\label{scaQ}
\end{eqnarray}
In the presence of the coupling $Q$ the condition 
for the late-time cosmic acceleration, $w_{\rm eff}<-1/3$,
is satisfied for $Q>\lambda/2>0$ or $Q<-\lambda<0$.
Then the scaling solution (\ref{scaQ}) can give rise to the global attractor 
with $\Omega_{\phi} \simeq 0.7$.
The $\phi$MDE solution (\ref{phiMDE}) followed by the accelerated scaling solution (\ref{scaQ})
may be used for alleviating the coincidence problem, because dark energy and
dark matter follow the same scaling relation from the end of the radiation era. 
In Ref.~\cite{Lucacoupled} it was shown, however, that the coupled quintessence with 
an exponential potential does not allow for such cosmological evolution
for the viable parameter space in the $(Q, \lambda)$ plane 
consistent with observational constraints.

It is possible to extend the analysis to coupled k-essence models
described by the action (\ref{kessence}) in the presence of the 
coupling (\ref{couplingA}).
{}From the requirement that the density parameter $\Omega_{\phi}$ ($\neq 0$)
and the equation of state $w_{\phi}$ are constants to realize scaling solutions, 
one can show that the Lagrangian density takes 
the following form \cite{PiaTsu,Tsujiscaling}
\begin{equation}
P=X\,g (Xe^{\kappa \lambda \phi})\,,
\label{scalinglag}
\end{equation}
where $\lambda$ is a constant, and $g$ is an arbitrary function in terms of 
$Y \equiv Xe^{\kappa \lambda \phi}$.
The result (\ref{scalinglag}) is valid for constant $Q$, but it can be 
generalized to a field-dependent coupling $Q(\phi)$ \cite{AQTW}.
The quintessence with an exponential potential 
($P=X-ce^{-\kappa \lambda \phi}$) corresponds to $g(Y)=1-c/Y$. 
Since the dilatonic ghost condensation model (\ref{ghostmodel2})
corresponds to $g(Y)=-1+Y/M^4$, this model has 
a scaling solution. We can also show that the tachyon field 
with the Lagrangian density $P=-V(\phi) \sqrt{1-2X}$
also has a scaling solution for the potential 
$V(\phi) \propto \phi^{-2}$ \cite{CGST}.
Even when the Hubble parameter squared $H^2$ is proportional to 
the energy density $\rho^n$, it is possible to obtain the Lagrangian 
density having scaling solutions in the form 
$P=X^{1/n}g(X e^{n \kappa \lambda \phi})$ \cite{Tsujiscaling}.
The cosmological dynamics of scaling solutions in such cases
(including the high-energy regime \cite{high1,high2}
in the Randall-Sundrum scenario \cite{RaS1,RaS2})
have been discussed by a number of 
authors \cite{Maedabrane,Mizuno1,Mizuno2}.

For the Lagrangian density (\ref{scalinglag}) there are two 
fixed points relevant to dark energy. 
Defining the dimensionless variables $x \equiv 
\dot{\phi}/(\sqrt{6}H)$ and 
$y \equiv e^{-\lambda \phi/2}/(\sqrt{3}H)$
(in the unit of $\kappa^2=1$), they are 
given by \cite{Tsujisingle}
\begin{itemize}
\item (A) Scalar-field dominated solution
\begin{equation}
x_A=\frac{\lambda}{\sqrt{6}P_{,X}}\,,\quad \Omega_{\phi}=1\,,\quad
w_{\rm eff}=w_{\phi}=-1+\frac{\lambda^2}{3P_{,X}}\,.
\end{equation}
\item (B) Scaling solution
\begin{eqnarray}
& & x_B=\frac{\sqrt{6}}{2(Q+\lambda)}\,,\quad 
\Omega_{\phi}=\frac{Q(Q+\lambda)+3P_{,X}}{(Q+\lambda)^2}\,.
\nonumber \\
& & w_{\rm eff}=-\frac{Q}{Q+\lambda}\,,\quad
w_{\phi}=-\frac{Q(Q+\lambda)}{Q(Q+\lambda)+3P_{,X}}\,.
\end{eqnarray}
\end{itemize}
The points (A) and (B) are responsible for the cosmic acceleration 
for (A) $\lambda^2/P_{,X}<2$ and (B) $Q>\lambda/2>0$ or $Q<-\lambda<0$, 
respectively. {}From the stability analysis about the fixed points
it follows that, when the point (A) is stable, the point (B) is not stable, 
and vice versa \cite{Tsujisingle}. 

The $\phi$MDE corresponds to a fixed point at which the kinetic
energy of the field dominates over the potential energy, 
i.e. $x \neq 0$ and $y=0$. 
Since the quantity $Y=Xe^{\kappa \lambda \phi}$ can be 
expressed as $Y=x^2/y^2$, the function $g(Y)$ cannot 
be singular at $y=0$ for the existence of the $\phi$MDE.
Then the function $g(Y)$ should be expanded in negative
powers of $Y$, i.e.
\begin{equation}
g(Y)=c_0+\sum_{n>0} c_n Y^{-n}
=c_0+\sum_{n>0} c_n (y^2/x^2)^n\,,
\label{gYform}
\end{equation}
which includes the quintessence with an exponential potential.
For this form of $g(Y)$, there is the following $\phi$MDE 
point (C):
\begin{equation}
(x_C, y_C)=\left(-\frac{\sqrt{6}Q}{3c_0}, 0 \right)\,,\quad 
\Omega_{\phi}=w_{\rm eff}=\frac{2Q^2}{3c_0}\,,\quad
w_{\phi}=1\,,
\end{equation}
together with the purely kinetic point $(x, y)=(\pm 1/\sqrt{c_0}, 0)$
and $\Omega_{\phi}=1$ for $c_0>0$.

An ideal cosmological trajectory that alleviates the coincidence 
problem of dark energy should be the $\phi$MDE (C) followed 
by the point (B).
However, it was shown in Ref.~\cite{AQTW} that such a trajectory 
is not allowed because the solutions cannot cross the singularity at 
$x=0$ as well as another singularity associated with the sound speed.
For example, when $c_0>0$, we find that $x_B>0$ and $x_C<0$ for
$Q>\lambda/2>0$, whereas $x_B<0$ and $x_C>0$ for
$Q<-\lambda<0$. These points are separated between the line 
$x=0$ at which the function (\ref{gYform}) diverges.
The $\phi$MDE solution chooses the accelerated point (A) 
as a final attractor. 
The above discussion shows that the coincidence problem is 
difficult to be solved even for the general Lagrangian density 
(\ref{scalinglag}) that has scaling solutions.
This problem mainly comes from the fact that 
a large coupling $Q$ required for the existence of 
a viable scaling solution (B) is not compatible 
with a small coupling $Q$
required for the existence of the $\phi$MDE.
We need a rapidly growing coupling to realize 
such a transition \cite{Lucastationary}.

%subsub
\subsubsection{Chameleon mechanism}
%subsub

If a scalar field $\phi$ is coupled to baryons as well as dark matter, this 
gives rise to a fifth force interaction that can be constrained experimentally.
A large coupling of the order of unity arises in modified gravity theories 
as well as superstring theories.
In such cases we need to suppress such a strong interaction with baryons
for the compatibility with local gravity experiments.
There is a way to screen the fifth force under the so-called chameleon 
mechanism \cite{chame1,chame2} in which the field mass is different depending 
on the matter density in the surrounding environment.
If the field is sufficiently heavy in the regions of high density, 
a spherically symmetric body can have a ``thin-shell''
around its surface such that the effective coupling between 
the field and matter is suppressed outside the body.

The action of a chameleon scalar field $\phi$ with a potential
$V(\phi)$ is given by 
\begin{equation}
S=\int{\rm d}^{4}x\sqrt{-g}\left[\frac{1}{2\kappa^2}R
-\frac{1}{2}g^{\mu\nu}
\partial_{\mu}\phi\partial_{\nu}\phi-V(\phi)\right]
+\int{\rm d}^{4}x\,{\cal L}_{M} 
(g_{\mu\nu}^{(i)},\Psi_{M}^{(i)})\,,
\label{actionchame}
\end{equation}
where $g$ is the determinant of the metric $g_{\mu\nu}$ 
(in the Einstein frame) and ${\cal L}_{M}$ is a matter Lagrangian
with matter fields $\Psi_{M}^{(i)}$ coupled to a metric
$g_{\mu\nu}^{(i)}$. 
The metric $g_{\mu\nu}^{(i)}$ is related to
the Einstein frame metric $g_{\mu\nu}$ via 
$g_{\mu\nu}^{(i)}=e^{2\kappa Q_{i}\phi}g_{\mu\nu}$,
where $Q_{i}$ are the strengths of the couplings for 
each matter component with the field $\phi$. 
The typical field potential is chosen to be of the runaway type 
(such as $V(\phi)=M^{4+n}\phi^{-n}$).
We also restrict the form of the potential such that 
$|V_{,\phi}| \to \infty$ as $\phi \to 0$.

Varying the action (\ref{actionchame}) with respect to $\phi$, 
we obtain the field equation
\begin{equation}
\square \phi-V_{,\phi}=-\sum_{i}
\kappa Q_i e^{4\kappa Q_i \phi} 
g_{(i)}^{\mu \nu} T_{\mu \nu}^{(i)}\,,
\label{squarephieq}
\end{equation}
where $T_{\mu \nu}^{(i)}=-(2/\sqrt{-g^{(i)}})\delta {\cal L}_M/\delta g^{\mu \nu}_{(i)}$
is the stress-energy tensor for the $i$-th form of matter.
For non-relativistic matter we have $g_{(i)}^{\mu \nu}T_{\mu \nu}^{(i)}=-\tilde{\rho}_i$, 
where $\tilde{\rho}_i$ is an energy density.
It is convenient to introduce the energy density $\rho_i \equiv \tilde{\rho}_i e^{3\kappa Q_i \phi}$, 
which is conserved in the Einstein frame.
In the following, let us consider the case in which the couplings 
$Q_i$ are the same for all species. i.e. $Q_i=Q$.
In a spherically symmetric space-time under the weak gravitational 
background (i.e. neglecting the backreaction of gravitational potentials), 
Eq.~(\ref{squarephieq}) reads
\begin{equation}
\frac{{\rm d}^2 \phi}{{\rm d} r^2}+
\frac{2}{r} \frac{{\rm d}\phi}{{\rm d} r}=
\frac{{\rm d}V_{\rm eff}}{{\rm d}\phi}\,,
\label{dreq}
\end{equation}
where $r$ is a distance from the center of symmetry,
and $V_{\rm eff}$ is the effective potential given by 
\begin{equation}
V_{\rm eff}(\phi)=V(\phi)+e^{\kappa Q \phi}\rho\,,
\label{Veff}
\end{equation}
and $\rho \equiv \sum_{i} \rho_i$.
For the runaway potential with $V_{,\phi}<0$ the positive coupling $Q$
leads to a minimum of the effective potential.
In $f(R)$ gravity the negative coupling ($Q=-1/\sqrt{6}$) gives
rise to a minimum for the potential with $V_{,\phi}>0$
(as we will see in Sec.~\ref{lgcsec}).

We assume that a spherically symmetric body has a constant
density $\rho=\rho_A$ inside the body ($r<r_c$)
and that the energy density outside the body ($r>r_c$)
is $\rho=\rho_B$. 
The mass $M_c$ of the body and the gravitational potential $\Phi_c$
at the radius $r_c$ are given by $M_c=(4\pi/3)r_c^3 \rho_A$
and $\Phi_c=GM_c/r_c$, respectively.
The effective potential $V_{\rm eff} (\phi)$ has two minima 
at the field values $\phi_A$ and $\phi_B$
satisfying $V_{\rm eff}' (\phi_A)=0$ and 
$V_{\rm eff}' (\phi_B)=0$, respectively. 
The former corresponds to the region with a high density that gives rise
to a heavy mass squared $m_A^2 \equiv V_{{\rm eff}}''(\phi_A)$,
whereas the latter to the lower density region with a lighter mass
squared $m_B^2 \equiv V_{{\rm eff}}''(\phi_B)$.
When we consider the ``dynamics'' of the field $\phi$
according to Eq.~(\ref{dreq}) we need to consider the inverted 
effective potential $(-V_{\rm eff})$ having two {\it maxima}
at $\phi=\phi_A$ and $\phi=\phi_B$.

The boundary conditions for the field are given by 
$\frac{{\rm d}\phi}{{\rm d}r}(r=0)=0$ and 
$\phi(r \to\infty)=\phi_{B}$.
The field $\phi$ is at rest at $r=0$ and begins to roll down the
potential when the matter-coupling term $\kappa Q\rho_{A}e^{\kappa Q\phi}$ 
becomes important at a radius $r_{1}$ in Eq.~(\ref{dreq}).
As long as $r_1$ is close to $r_c$ so that 
$\Delta r_c \equiv r_c-r_1 \ll r_c$, the body has 
a thin-shell inside the body.
Since the field acquires a sufficient kinetic energy in the thin-shell regime
($r_1<r<r_c$), it climbs up the potential 
hill outside the body ($r>r_c$).
The field profile can be obtained by matching the solutions of 
Eq.~(\ref{dreq}) at the radius $r=r_1$ and $r=r_c$. 
Neglecting the mass term $m_B$, we obtain the thin-shell field 
profile outside the body \cite{chame1,chame2,Tamaki} 
\begin{eqnarray}
\phi(r) \simeq \phi_{B}-\frac{2Q_{{\rm eff}}}{\kappa}
\frac{GM_{c}}{r}\,,\label{phisothin}
\end{eqnarray}
where
\begin{eqnarray}
Q_{\rm eff} = 3Q \epsilon_{\rm th}\,,\qquad
\epsilon_{\rm th} \equiv \frac{\kappa(\phi_B-\phi_A)}
{6Q\Phi_c}\,.
\label{epdef}
\end{eqnarray}
Here $\epsilon_{\rm th}$ is called the thin-shell 
parameter. Under the conditions $\Delta r_c/r_c \ll 1$ 
and $1/(m_A r_c) \ll 1$, the thin-shell parameter 
is approximately given by 
$\epsilon_{\rm th} \simeq \Delta r_c/r_c+1/(m_A r_c)$ \cite{Tamaki}.
As long as $\epsilon_{\rm th} \ll 1$ the amplitude of the 
effective coupling $Q_{\rm eff}$ can be much smaller than 1.
Hence it is possible for the large coupling models ($|Q|={\cal O}(1)$) 
to be consistent with local gravity experiments if the body has 
a thin-shell.

Let us study the constraint on the thin-shell 
parameter from the possible violation of the equivalence principle (EP). 
The tightest bound comes from the solar system tests of weak EP using
the free-fall acceleration of Moon ($a_{{\rm Moon}}$) 
and Earth ($a_{\oplus}$) toward Sun \cite{chame2}. 
The experimental bound on the difference
of two accelerations is given by \cite{Will01,Will05} 
\begin{equation}
\frac{|a_{{\rm Moon}}-a_{\oplus}|}
{(a_{{\rm Moon}}+a_{\oplus})/2}<10^{-13}\,.
\label{etamoon}
\end{equation}
If Earth, Sun, and Moon have thin-shells,
the field profiles outside the bodies are given by Eq.~(\ref{phisothin})
with the replacement of corresponding quantities. The
acceleration induced by a fifth force with the field profile $\phi(r)$
and the effective coupling $Q_{{\rm eff}}$ is $a^{{\rm fifth}}=|Q_{{\rm eff}}\nabla\phi(r)|$.
Using the thin-shell parameter $\epsilon_{{\rm th},\oplus}$ for Earth,
the accelerations $a_{\oplus}$ and $a_{{\rm Moon}}$ 
toward Sun (mass $M_{\odot}$) are \cite{chame2}
\begin{eqnarray}
a_{\oplus} & \simeq & \frac{GM_{\odot}}{r^{2}}\left[1+18Q^{2}
\epsilon_{{\rm th},\oplus}^{2}\frac{\Phi_{\oplus}}{\Phi_{\odot}}
\right]\,,\\
a_{{\rm Moon}} & \simeq & \frac{GM_{\odot}}{r^{2}}\left[1+18Q^{2}\epsilon_{{\rm th},\oplus}^{2}\frac{\Phi_{\oplus}^{2}}{\Phi_{\odot}\Phi_{{\rm Moon}}}\right]\,,
\end{eqnarray}
where $\Phi_{\odot}\simeq2.1\times10^{-6}$, $\Phi_{\oplus}\simeq7.0\times10^{-10}$,
and $\Phi_{{\rm Moon}} \simeq 3.1\times10^{-11}$ are the gravitational
potentials of Sun, Earth and Moon, respectively. Hence the condition
(\ref{etamoon}) translates into 
\begin{equation}
\epsilon_{{\rm th},\oplus}<8.8\times10^{-7}/|Q|\,.
\label{boep}
\end{equation}
Since the condition $|\phi_B| \gg |\phi_A|$ is satisfied for the field 
potentials under consideration, one has 
$\epsilon_{{\rm th},\oplus} \simeq \kappa \phi_B/(6Q \Phi_{\oplus})$ from 
Eq.~(\ref{epdef}). Then the condition (\ref{boep}) translates into 
\begin{equation}
|\kappa \phi_{B,\oplus}|<3.7 \times 10^{-15}\,.
\label{boep2}
\end{equation}

For example, let us consider the inverse power-law potential 
$V(\phi)=M^{4+n}\phi^{-n}$.
In this case we have 
$ \phi_{B,\oplus}=[(n/Q)(M_{\rm pl}^4/\rho_B)(M/M_{\rm pl})^{n+4}]^{1/(n+1)}
M_{\rm pl}$, where we recovered the reduced Planck mass
$M_{\rm pl}=1/\kappa$.
For $n$ and $Q$ of the order of unity, the constraint (\ref{boep2}) gives 
$M < 10^{-(15n+130)/(n+4)}M_{\rm pl}$.
When $n=1$, for example, one has $M < 10^{-2}$\,eV.
If the same potential is responsible for dark energy, the 
mass $M$ is constrained to be larger than this value \cite{Brax04}.
For the potential $V(\phi)=M^4 \exp (M^n/\phi^n)$, however, 
we have that $V(\phi) \approx M^4+M^{4+n}\phi^{-n}$ for $\phi>M$, 
which is responsible for dark energy 
for $M \approx 10^{-3}$ eV.
This can be compatible with the mass scale $M$
constrained by (\ref{boep2}) \cite{Brax04}.
See Refs.~\cite{Shaw,Brax07,Gubser,WeiCai,Shaw,Brax07,Brax07d,MotaShaw,Davis09,Hui09} 
for a number of cosmological 
and experimental aspects of the chameleon field.

%subsub
\subsubsection{Varying $\alpha$}
%subsub

So far we have discussed the coupling between dark energy 
and non-relativistic matter. In this section we discuss the case 
in which dark energy is coupled to an electromagnetic field.
In fact, a temporal variation of the effective fine structure
``constant'' $\alpha$ has been reported by a number of authors.
The variation of $\alpha$ constrained by the Oklo natural fission reactor
is given by $-0.9\times10^{-7}<\Delta\alpha/\alpha
\equiv(\alpha-\alpha_{0})/\alpha_{0}<1.2\times10^{-7}$ 
at the redshift $z \approx 0.16$, 
where $\alpha_{0}$ is the value of $\alpha$ today \cite{Oklo}.
{}From the absorption line spectra of distant quasars 
we have the constraints $\Delta\alpha/\alpha=(-0.574\pm0.102)\times10^{-5}$
over the redshift range $0.2<z<3.7$ \cite{dzuba,Webbetal} 
and $\Delta\alpha/\alpha=(-0.06\pm0.06)\times10^{-5}$
for $0.4<z<2.3$ \cite{chand}. Although the possibility of systematic
errors still remains \cite{Murphy}, this may provide important
implications for the existence of a light scalar field 
related with dark energy.

The Lagrangian density describing such a coupling between the field $\phi$
and the electromagnetic field $F_{\mu \nu}$ is given by 
\begin{equation}
{\cal L}_{F}(\phi)=-\frac{1}{4}B_{F}(\phi)F_{\mu\nu}F^{\mu\nu}.
\label{bclag}
\end{equation}
The coupling of the form $B_{F}(\phi)=e^{-\zeta\kappa(\phi-\phi_{0})}$
was originally introduced by Bekenstein \cite{Beken}, where 
$\zeta$ is a coupling constant and $\phi_{0}$ 
is the field value today.
There are also other choices of the coupling, see e.g., 
Refs.~\cite{SandvikBarrow,DvaliZa,ChibaKohri,Uzanreview,PBB04,Olive04,CNP04,MotaBarrow}.
The fine structure {}``constant'' $\alpha$ is inversely proportional to $B_{F}(\phi)$, 
so that this can be expressed as $\alpha=\alpha_{0}/B_{F}(\phi)$, where $\alpha_{0}$
is the present value. 
The exponential coupling $B_{F}(\phi)=e^{-\zeta\kappa(\phi-\phi_{0})}$
has a linear dependence $B_F (\phi) \simeq 1-\zeta \kappa (\phi-\phi_0)$
in the regime $|\zeta\kappa(\phi-\phi_{0})| \ll 1$, so that 
the variation of $\alpha$ is given by 
\begin{equation}
\frac{\Delta\alpha}{\alpha}=\frac{\alpha-\alpha_{0}}{\alpha_{0}}
 \simeq \zeta\kappa(\phi-\phi_{0})\,.\label{defDelta}
\end{equation}
Using the constraint $\Delta\alpha/\alpha\simeq-10^{-5}$ around
$z=3$ \cite{Webbetal} obtained from quasar absorption
lines, the coupling $\zeta$ can be expressed as 
\begin{equation}
\zeta\simeq-\frac{10^{-5}}{\kappa\phi(z=3)-\kappa\phi(z=0)}\,.\label{zetavalue}
\end{equation}

Let us consider the case in which the scalar field has a power-law dependence
in terms of the scale factor $a$, i.e.
\begin{equation}
\phi=\phi_{0}\,a^{p}=\phi_{0}(1+z)^{-p}\,.
\label{phiapz}
\end{equation}
In fact, in the so-called tracking regime of the matter-dominated 
era \cite{Paul99}, the inverse power-law potential 
$V(\phi)=M^{4+n}\phi^{-n}$ gives rise to a constant 
field equation of state: $w_{\phi}=-2/(n+2)$, which 
corresponds to the field evolution (\ref{phiapz}) with $p=3/(n+2)$.
Using Eq.~(\ref{phiapz}), the coupling $\zeta$ in Eq.~(\ref{zetavalue}) reads 
\begin{equation}
\zeta \simeq \frac{10^{-5}}{1-4^{-p}}
\left(\kappa\phi_{0}\right)^{-1}\,.
\label{zetaesti}
\end{equation}
Since $\kappa\phi_{0}$ is of the order of unity in order to
realize the present cosmic acceleration \cite{Paul99}, 
the coupling $\zeta$ is constrained to be $\zeta\approx10^{-5}$
for $p$ of the order of 1.

The above discussion is valid for the potentials having the 
solution (\ref{phiapz}) in the tracking regime.
The variation of $\alpha$ for other quintessence potentials was discussed
in Ref.~\cite{CNP04}. There is also a k-essence model
in which a tachyon field is coupled to electromagnetic 
fields \cite{Garousialpha}.

\subsection{Unified models of dark energy and dark matter}
\label{unifiedsec}

There are a number of works to explain the origin of dark energy 
and dark matter using a single fluid or a single scalar field.
Let us first discuss the generalized Chaplygin gas (GCG) model 
as an example of a single fluid model \cite{Kamen,Bento}.
In this model the pressure $P$ of the perfect fluid 
is related to its energy density $\rho$ via 
\begin{equation}
P=-A\rho^{-\alpha}\,,
\label{prhocha}
\end{equation}
where $A$ is a positive constant.
The original Chaplygin gas model corresponds to $\alpha=1$ \cite{Kamen}.

Plugging the relation (\ref{prhocha}) into the continuity equation 
$\dot{\rho}+3H(\rho+P)=0$, we obtain the integrated solution
\begin{equation}
\rho (t)=\left[ A+\frac{B}{a^{3(1+\alpha)}} \right]^{1/(1+\alpha)}\,,
\end{equation}
where $B$ is an integration constant.
In the early epoch ($a \ll 1$) the energy density evolves as $\rho \propto a^{-3}$, 
which means that the fluid behaves as dark matter.
In the late epoch ($a \gg 1$) the energy density approaches a constant 
value $A^{1/(a+\alpha)}$ and hence the fluid behaves as dark energy.
A fluid with the generalized Chaplygin gas therefore interpolates between 
dark matter and dark energy.

Although this model is attractive to provide unified description of 
two dark components, 
it is severely constrained by the matter power 
spectrum in large-scale structure.
The gauge-invariant matter perturbation $\delta_m$ with
a comoving wavenumber $k$ obeys the following equation 
of motion \cite{Waga} 
\begin{equation}
\ddot{\delta}_{m}+\left(2+3c_{s}^{2}-6w\right)H\dot{\delta}_{m}
-\left[\frac{3}{2}H^{2}(1-6c_{s}^{2}-3w^{2}+8w)
-\left(\frac{c_{s}k}{a}\right)^{2}\right]\delta_{m}=0\,,
\label{delmchap}
\end{equation}
where $w=P/\rho$ is the fluid equation of state, and  
$c_{s}$ is the sound speed given by 
\begin{equation}
c_{s}^{2}=\frac{{\rm d} P}{{\rm d} \rho}=-\alpha w\,.
\label{eq:chapcs}
\end{equation}
Since $w \to 0$ and $c_{s}^{2} \to 0$ in the limit $z\gg1$, the sound
speed is much smaller than unity in the deep matter era and starts
to grow around the end of it. Since $w$ is negative, 
$c_{s}^{2}$ is positive for $\alpha>0$ and negative for $\alpha<0$.

{}From Eq.~(\ref{delmchap}) the perturbations satisfying the following
condition grow via the gravitational instability
\begin{equation}
|c_{s}^{2}|<\frac{3}{2}\left(\frac{aH}{k}\right)^{2}\,.
\label{cs2}
\end{equation}
When $|c_{s}^{2}|>(3/2)\left( aH/k \right)^{2}$,
the perturbations exhibit either rapid growth or damped oscillations 
depending on the sign of $c_s^2$.
The violation of the condition (\ref{cs2}) mainly occurs around the
present epoch in which $|w|$ is of the order of unity and hence $|c_{s}^{2}|\sim|\alpha|$.
The smallest scale relevant to the galaxy matter power spectrum
in the linear regime corresponds to the wavenumber around 
$k=0.1\, h$~Mpc$^{-1}$.
Then the constraint (\ref{cs2}) gives the upper bound 
on the values of $|\alpha|$ \cite{Waga}:
\begin{equation}
|\alpha|~\lsim~10^{-5}\,.\label{alcon}
\end{equation}
Hence the generalized Chaplygin gas model is hardly distinguishable from 
the $\Lambda$CDM model. In particular the original Chaplygin gas model
($\alpha=1$) is excluded from the observations
of large-scale structure. Although nonlinear clustering may change
the evolution of perturbations in this model \cite{Avelino,Fabio}, 
it is unlikely that the constraint (\ref{alcon}) is relaxed significantly.

The above conclusion comes from the fact that in the Chaplygin gas 
model the sound speed is too large to match with observations.
There is a way to avoid this problem by adding 
a non-adiabatic contribution to Eq.~(\ref{delmchap}) to make $c_s$ 
vanish \cite{silent}. It is also possible to construct unified models 
of dark energy and dark matter using a purely kinetic scalar 
field \cite{Scherrer}. Let us consider k-essence models in which 
the Lagrangian density $P(X)$ has an extremum at some value $X=X_0$, 
e.g. \cite{Scherrer}
\begin{equation}
P=P_0+P_2 (X-X_0)^2\,.
\label{Peq}
\end{equation}
The pressure $P_{\phi}=P$ and the energy density $\rho_{\phi}=2XP_{,X}-P$
satisfy the continuity equation 
$\dot{\rho}_{\phi}+3H(\rho_{\phi}+P_{\phi})=0$, i.e.
\begin{equation}
\left( P_{,X}+2X P_{,XX} \right)\dot{X}+6HP_{,X}X=0\,.
\label{PXeq}
\end{equation}
The solution around $X=X_0$ can be derived by introducing a small 
parameter $\epsilon=(X-X_0)/X_0$. Plugging Eq.~(\ref{Peq}) into 
Eq.~(\ref{PXeq}), we find that $\epsilon$ satisfies the equation 
$\dot{\epsilon}=-3H \epsilon$ at linear order.
Hence we obtain the solution $X=X_0 \left[ 1+\epsilon_1 (a/a_1)^{-3} \right]$, 
where $\epsilon_1$ and $a_1$ are constants.
The validity of the above approximation demands that 
$\epsilon_1 (a/a_1)^{-3} \ll 1$.
Since $P_{\phi}\simeq P_0$ and $\rho_{\phi} \simeq -P_0+4P_2 X_0^2 \epsilon_1
(a/a_1)^{-3}$ in the regime where $X$ is close to $X_0$, 
the field equation of state is given by 
\begin{equation}
w_{\phi} \simeq -\left[ 1-\frac{4P_2}{P_0} X_0^2 \epsilon_1
\left( \frac{a}{a_1} \right)^{-3} \right]^{-1}\,.
\end{equation}
Since $w_{\phi} \to -1$ at late times it is possible to give rise to 
the cosmic acceleration.
One can also realize $w_{\phi} \approx 0$ during the matter era, 
provided that the condition $4P_2X_0^2/|P_0| \gg 1$ is satisfied.
The sound speed squared defined in Eq.~(\ref{soundkes}) is 
approximately given by 
\begin{equation}
c_s^2 \simeq \frac12 \epsilon_1
\left( \frac{a}{a_1} \right)^{-3}\,,
\end{equation}
which is much smaller than unity.
Hence the large sound speed problem can be evaded
in the model (\ref{Peq}).
In Ref.~\cite{Huk} it was shown that the above purely k-essence model is 
equivalent to a fluid with a closed-form barotropic equation of state plus 
a constant term that works as a cosmological constant 
to all orders in structure formation.
See Refs.~\cite{Lim,Bertacca,Bertacca2,Piatt,Urakawa} 
for generalized versions of the above model.

%%%%%%%%%%%%%%%%%%
\section{Modified gravity models}
\label{mosec}
%%%%%%%%%%%%%%%%%%

There is another class of dark energy models in which gravity is
modified from General Relativity (GR).
We review a number of cosmological and gravitational aspects 
of $f(R)$ gravity, Gauss-Bonnet gravity, scalar-tensor theories, 
and a braneworld model. 
We also discuss observational signatures 
of those models to distinguish them from 
other dark energy models.

\subsection{$f(R)$ gravity}

The simplest modification to GR is 
$f(R)$ gravity with the action
\begin{equation}
S=\frac{1}{2\kappa^{2}}\int{\rm d}^{4}x\sqrt{-g}f(R)
+\int {\rm d}^4 x\,{\cal L}_M (g_{\mu \nu}, \Psi_M)\,,
\label{fRaction}
\end{equation}
where $f$ is a function of the Ricci scalar $R$
and ${\cal L}_{M}$ is a matter Lagrangian for perfect fluids.
The Lagrangian ${\cal L}_M$ depends on the metric $g_{\mu \nu}$
and the matter fields $\Psi_M$.
We do not consider a direct coupling between the Ricci scalar 
and matter (such as $f_1(R){\cal L}_M$ 
studied in Refs.~\cite{Berto1,Berto2,Faracou}).

\subsubsection{Viable $f(R)$ dark energy models}
\label{viablesec}

In the standard variational approach called the metric formalism,
the affine connections $\Gamma^{\lambda}_{\mu \nu}$ are 
related with the metric $g_{\mu \nu}$ \cite{Weinbergbook}.
In this formalism the field equation can be derived by 
varying the action (\ref{fRaction}) with respect to $g_{\mu\nu}$: 
\begin{equation}
F(R)R_{\mu\nu}(g)-\frac{1}{2}f(R)g_{\mu\nu}
-\nabla_{\mu}\nabla_{\nu}F(R)+g_{\mu\nu}\square F(R)
=\kappa^{2}T_{\mu\nu}\,,
\label{fREin}
\end{equation}
where $F(R)\equiv\partial f/\partial R$, and
$T_{\mu \nu}=-(2/\sqrt{-g}) \delta {\cal L}_M/\delta g^{\mu \nu}$ is 
the energy-momentum tensor of matter.
Note that there is another way for the variation of the action 
called the Palatini formalism in which the metric and the connections
are treated as independent variables. In Sec.~\ref{Palasec} 
we shall briefly mention the application of 
Palatini $f(R)$ gravity to dark energy.
The trace of Eq.~(\ref{fREin}) is given by 
\begin{eqnarray}
3\,\square F(R)+F(R)R-2f(R)=\kappa^{2}T\,,
\label{trace}
\end{eqnarray}
where $T=g^{\mu\nu}T_{\mu\nu}=-\rho_M+3P_M$. Here $\rho_M$ 
and $P_M$ are the energy density and the pressure of matter, respectively.

The de Sitter point corresponds to a vacuum solution at which the
Ricci scalar is constant. Since $\square F(R)=0$ at this point,
we obtain 
\begin{equation}
F(R)R-2f(R)=0\,.
\label{fRdeSitter}
\end{equation}
The model $f(R)=\alpha R^{2}$ satisfies this condition and hence
it gives rise to an exact de Sitter solution.
In fact the first model of inflation proposed by Starobinsky \cite{Sta79}
corresponds to $f(R)=R+\alpha R^{2}$, in which 
the cosmic acceleration ends when the
term $\alpha R^{2}$ becomes smaller than $R$. 
Dark energy models based on $f(R)$ theories can be also constructed
to realize the late-time de Sitter solution satisfying 
the condition (\ref{fRdeSitter}).

The possibility of the late-time cosmic acceleration in $f(R)$
gravity was first suggested by Capozziello \cite{fR1} in 2002.
An $f(R)$ dark energy model of the form $f(R)=R-\mu^{2(n+1)}/R^n~(n>0)$
was proposed in Refs.~\cite{fR1d,fR2,fR3,Nojiri03} 
(see also Refs.~\cite{Soussa,Allem,Easson,Dick04,Allemandi:2004ca,Carloni}), 
but it became clear that this model suffers from a number of problems 
such as the matter instability \cite{Dolgov}, absence 
of the matter era \cite{APT,APT2}, and inability to satisfy local gravity 
constraints \cite{Chiba,OlmoPRL,Olmo,Navarro,Erick06,Chiba07}.
This problem arises from the fact that $f_{,RR}<0$ in this model.

In order to see why the models with negative values of $f_{,RR}$ are excluded, 
let us consider local fluctuations on a background characterized by
a curvature $R_{0}$ and a density $\rho_{0}$. We expand Eq.~(\ref{trace})
in powers of fluctuations under a weak field approximation. 
We decompose the quantities $F(R)$, $g_{\mu\nu}$, and $T_{\mu\nu}$
into the background part and the perturbed part: $R=R^{(0)}+\delta R$,
$F=F^{(0)}(1+\delta_{F})$, $g_{\mu\nu}=\eta_{\mu\nu}+h_{\mu\nu}$, 
and $T_{\mu\nu}=T_{\mu\nu}^{(0)}+\delta T_{\mu\nu}$,
where we have used the approximation that $g_{\mu\nu}^{(0)}$ 
corresponds the metric $\eta_{\mu \nu}$ in the Minkowski space-time.
Then the trace equation (\ref{trace}) reads \cite{Olmo,Navarro} 
\begin{equation}
\left(\frac{\partial^{2}}{\partial t^{2}}-\nabla^{2}\right)
\delta_{F}+M^{2}\,\delta_{F}=-\frac{\kappa^{2}}{3F^{(0)}}\delta T\,,
\label{delpsi}
\end{equation}
where $\delta T\equiv\eta^{\mu\nu}\delta T_{\mu\nu}$, and 
\begin{eqnarray}
M^{2}\equiv\frac{1}{3}\left[\frac{f_{,R}(R^{(0)})}{f_{,RR}(R^{(0)})}-R^{(0)}\right]
=\frac{R^{(0)}}{3}\left[\frac{1}{m(R^{(0)})}-1\right]\,.\label{Mpsi2}
\end{eqnarray}
Here the quantity $m=Rf_{,RR}/f_{,R}$ characterizes the deviation from 
the $\Lambda$CDM model ($f(R)=R-2\Lambda$).
In the homogeneous and isotropic cosmological background
(without a Hubble friction), $\delta_{F}$ is a function of the
cosmic time $t$ only and Eq.~(\ref{delpsi}) reduces to 
\begin{equation}
\ddot{\delta_{F}}+M^{2}\,\delta_{F}
=\frac{\kappa^{2}}{3F^{(0)}}\rho\,,
\label{perturfR}
\end{equation}
where $\rho\equiv-\delta T$. For the models where the deviation
from the $\Lambda$CDM model is small, 
we have $m(R^{(0)})\ll1$ so
that $|M^{2}|$ is much larger than $R^{(0)}$.
If $M^{2}<0$, the perturbation $\delta_{F}$ exhibits a violent instability. 
Then the condition $M^{2}\simeq f_{,R}(R^{(0)})/(3f_{,RR}(R^{(0)}))>0$
is needed for the stability of cosmological perturbations. 
We also require that $f_{,R}(R^{(0)})>0$  
to avoid anti-gravity (i.e. to avoid that the graviton becomes 
a ghost). Hence the condition $f_{,RR}(R^{(0)})>0$ needs to hold
for avoiding a tachyonic instability associated with the 
negative mass squared \cite{Carroll06,Song07,Bean07,Faulkner,Pogosian}.

For the consistency with local gravity constraints in solar system, 
the function $f(R)$ needs to be close to that in the $\Lambda$CDM 
model in the region of high density (in the region where the 
Ricci scalar $R$ is much larger than the cosmological 
Ricci scalar $R_0$ today).
We also require the existence of a stable late-time de Sitter point 
given in Eq.~(\ref{fRdeSitter}).
{}From the stability analysis about the de Sitter point, one can 
show that it is stable for 
$0<m=Rf_{,RR}/f_{,R}<1$ \cite{Muller,Faraonista,AGPT}.
Then we can summarize the conditions for the viability of 
$f(R)$ dark energy models:
\begin{itemize}
\item (i) $f_{,R}>0$ for $R \ge R_0$.
\item (ii) $f_{,RR}>0$ for $R \ge R_0$.
\item (iii) $f(R) \to R-2\Lambda$ for $R \gg R_0$.
\item (iv) $0<Rf_{,RR}/f_{,R}<1$ at the de Sitter point 
satisfying $Rf_{,R}=2f$.
\end{itemize}

The examples of viable models satisfying all these 
requirements are \cite{Hu,Star07,Tsuji07}
\begin{eqnarray}
 &  & {\rm (A)}~f(R)=R-\mu R_{c}\frac{(R/R_{c})^{2n}}{(R/R_{c})^{2n}+1}
 \qquad{\rm with}~~n,\mu,R_{c}>0\,,\label{Amodel}\\
 &  & {\rm (B)}~f(R)=R-\mu R_{c}\left[1-\left(1+R^{2}/R_{c}^{2}\right)^{-n}\right]
 \qquad{\rm with}~~n,\mu,R_{c}>0\,,\label{Bmodel}\\
 & &{\rm (C)}~f(R)=R-\mu R_{c}{\rm tanh}\,(R/R_{c})
\qquad{\rm with}~~\mu,R_{c}>0\,,
\label{tanh}
 \end{eqnarray}
where $\mu$, $R_c$, and $n$ are constants.
Models similar to (C) were proposed in Refs.~\cite{Appleby,Linder09}.
Note that $R_c$ is roughly of the order of the present 
cosmological Ricci scalar $R_0$.
If $R \gg R_c$ the models are close to the 
$\Lambda$CDM model ($f(R) \simeq R-\mu R_c$), 
so that GR is recovered in the region of high density. 
The models (A) and (B) have the following asymptotic behavior
\begin{equation}
f(R) \simeq R-\mu R_c \left[1- (R/R_c)^{-2n} \right]\,,\qquad
(R \gg R_c)\,,
\label{fRasy}
\end{equation}
which rapidly approaches the $\Lambda$CDM model 
for $n~\gsim~1$. The model (C) shows an even faster decrease of $m$
in the region $R \gg R_c$. 
The model $f(R)=R-\mu R_c (R/R_c)^n$ ($0<n<1$)
proposed in Refs.~\cite{AGPT,LiBarrow} is also viable, but 
it does not allow the rapid decrease of $m$ in the region of 
high density required for the consistency with local gravity tests.

For example, let us consider the model (B).
The de Sitter point given by the condition (\ref{fRdeSitter}) satisfies
\begin{equation}
\mu=\frac{x_1(1+x_1^2)^{n+1}}
{2[(1+x_1^2)^{n+1}-1-(n+1)x_1^2]}\,,
\label{Bmodellam}
\end{equation}
where $x_1 \equiv R_1/R_c$ and $R_1$ is the Ricci scalar
at the de Sitter point.
The stability condition ($0<m<1$) at this point gives \cite{Star07}
\begin{equation}
(1+x_1^2)^{n+2} > 1+(n+2)x_1^2+(n+1)(2n+1)x_1^4\,.
\label{Bmodelcon}
\end{equation}
The condition (\ref{Bmodelcon}) gives the lower bound
on the parameter $\mu$.
When $n=1$ one has $x_1>\sqrt{3}$ and 
$\mu>8\sqrt{3}/9$.
Under Eq.~(\ref{Bmodelcon}) one can show that 
the conditions $f_{,R}>0$ and $f_{,RR}>0$ 
are also satisfied for $R \ge R_1$.

\subsubsection{Observational signatures of $f(R)$ dark energy models}
\label{obsigsec}

In the flat FLRW space-time we obtain the following equations of 
motion from Eqs.~(\ref{fREin}) and (\ref{trace}):
\begin{eqnarray}
3FH^{2} & = & \kappa^{2} \rho_{m}+(FR-f)/2-3H\dot{F}\,,
\label{FRWfR1}\\
2F\dot{H} & = & -\kappa^{2} \rho_{m}
-\ddot{F}+H\dot{F}\,,
\label{FRWfR2}
\end{eqnarray}
where, for the perfect fluid, we have taken into account 
only the non-relativistic matter with energy density $\rho_m$.
In order to confront $f(R)$ dark energy models with SN Ia observations, 
we rewrite Eqs.~(\ref{FRWfR1}) and (\ref{FRWfR2}) as follows: 
\begin{eqnarray}
&  & 3AH^{2}=\kappa^{2} \left( \rho_{m}+\rho_{{\rm DE}} 
 \right)\,,\label{mofR1}\\
&  & -2A\dot{H}=\kappa^{2}\left( \rho_{m}
 +\rho_{{\rm DE}}+P_{{\rm DE}}\right)\,,
 \label{mofR2}
\end{eqnarray}
where $A$ is some constant and 
\begin{eqnarray}
\kappa^{2}\rho_{{\rm DE}} & \equiv & (1/2)(FR-f)-3H\dot{F}+3H^{2}
(A-F)\,,\label{fRrhode}\\
\kappa^{2}P_{{\rm DE}} & \equiv & \ddot{F}+2H\dot{F}
-(1/2)(FR-f)-(3H^{2}+2\dot{H})(A-F)\,.\label{fRPde}
\end{eqnarray}
By defining $\rho_{{\rm DE}}$ and $P_{{\rm DE}}$ in this
way, one can easily show that the following 
continuity equation holds 
\begin{eqnarray}
\dot{\rho}_{{\rm DE}}+3H(\rho_{{\rm DE}}+P_{{\rm DE}})=0\,.
\label{rhodecon}
\end{eqnarray}

We define the dark energy equation of state $w_{{\rm DE}}\equiv P_{{\rm DE}}/\rho_{{\rm DE}}$,
which is directly related to the one used in SN Ia observations. 
{}From Eqs.~(\ref{mofR1}) and (\ref{mofR2}) 
it is given by \cite{AmenTsuji,Moraes}
\begin{eqnarray}
w_{{\rm DE}}=-\frac{2A\dot{H}+3AH^2}
{3AH^2-\kappa^2 \rho_m}
=\frac{w_{{\rm eff}}}{1-(F/A)\tilde{\Omega}_m}\,,
\label{wDEfR}
\end{eqnarray}
where $\tilde{\Omega}_m \equiv \kappa^2 \rho_m/(3FH^2)$.
The viable $f(R)$ models approach the $\Lambda$CDM model in the past, 
i.e. $F \to 1$ as $R \to \infty$.
In order to reproduce the standard matter era in the high-redshift regime
we can choose $A=1$ in Eqs.~(\ref{mofR1}) and (\ref{mofR2}).
Another possible choice is $A=F_0$, where $F_0$ is the present value
of $F$. This choice is suitable if the deviation of $F_0$ from 1 is small
(as in scalar-tensor theory with a massless scalar field \cite{Torres,Boi00}).
In both cases the equation of state $w_{\rm DE}$ can be smaller than $-1$
before reaching the de Sitter 
attractor \cite{AmenTsuji,Hu,Tsuji07,Linder09,Motohashi10}.
This originates from the fact that the presence of non-relativistic matter makes 
the denominator in Eq.~(\ref{wDEfR}) smaller than 1 
(unlike Refs.~\cite{Bamba08,Bamba09} in which the authors did not
take into account the contribution of non-relativistic matter). 
Thus $f(R)$ dark energy models give rise to a phantom equation of state
without violating stability conditions of the system.
The models (A) and (B) are allowed from the SN Ia observations 
provided that $n$ is larger than the order of 
unity \cite{Dev,Mel09,Cardone09,Ali10}.

%%%%%%%%%%%%%%%%%%%%%%%%%%%%%
\begin{figure}
\begin{centering}
\includegraphics[width=3.2in,height=3.0in]{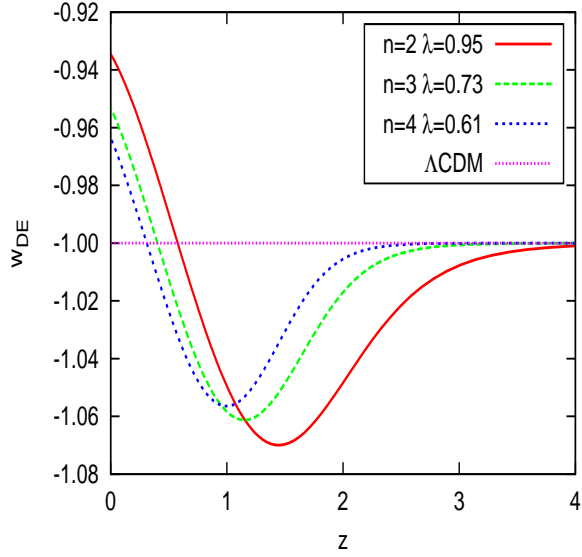} 
\par\end{centering}
\caption{Evolution of the dark energy equation of state $w_{\rm DE}$
for the model (B) with $A=1$ in Eqs.~(\ref{mofR1}) and (\ref{mofR2}).
The phantom equation of state and the cosmological constant 
boundary crossing are realized. {}From Ref.~\cite{Motohashi10}.}
\centering{}\label{frwde} 
\end{figure}
%%%%%%%%%%%%%%%%%%%%%%%%%%%%

The modification of gravity manifests itself in the effective gravitational 
coupling that appears in the equation of cosmological perturbations.
The full perturbation equations in $f(R)$ gravity are presented in 
Refs.~\cite{Hwang02,Hwang05,fRreview}.
When we confront $f(R)$ models with the observations of large-scale
structure, the wavenumbers $k$ of interest are sub-horizon modes
with $k/a \gg H$.
We can employ a so-called quasi-static approximation under 
which the dominant terms in perturbation equations correspond to 
those including $k^2/a^2$, $\delta \rho_m$, and $M^2$
\cite{Star98,Boi00,review,Tsujimatter}. 
Then the matter density perturbation $\delta_m$ approximately 
satisfies the following equation \cite{Tsujimatter,TsujiUddin}
\begin{equation}
\ddot{\delta}_m+2H \dot{\delta}_m-4\pi G_{\rm eff}
\rho_m \delta_m \simeq 0\,,
\end{equation}
where $\rho_m$ is the energy density of 
non-relativistic matter, and 
\begin{equation}
G_{\rm eff}=\frac{G}{f_{,R}}
\frac{1+4m\,k^2/(a^2 R)}{1+3m\,k^2/(a^2 R)}\,,
\end{equation}
where $m \equiv Rf_{,RR}/f_{,R}$.
This approximation is accurate for viable $f(R)$ dark energy models
as long as an oscillating mode of the scalar-field degree of freedom 
is suppressed relative to the matter-induced 
mode \cite{Star07,Tsuji07,Appleby2,delaCruz,delaCruz08,Motohashi09}.

In the regime where the deviation from the $\Lambda$CDM model
is small such that $m\,k^2/(a^2 R) \ll 1$, the effective gravitational 
coupling $G_{\rm eff}$ is very close to the gravitational constant $G$.
Then the matter perturbation evolves as $\delta_m \propto t^{2/3}$
during the matter dominance.
Meanwhile in the regime $m\,k^2/(a^2 R) \gg 1$ one has 
$G_{\rm eff} \simeq 4G/(3f_{,R})$, so that the evolution of $\delta_m$
during the matter era is given by 
$\delta_m \propto t^{(\sqrt{33}-1)/6}$ \cite{Star07,Tsuji07}.
The transition from the former regime to the latter regime occurs 
at the critical redshift \cite{Moraes09}
\begin{equation}
z_k \simeq \left[ \left( \frac{k}{a_0H_0} \right)^2 
\frac{2n(2n+1)}{\mu^{2n}} 
\frac{(2(1-\Omega_m^{(0)}))^{2n+1}}
{(\Omega_m^{(0)})^{2(n+1)}} \right]^{1/(6n+4)}-1\,,
\label{zk}
\end{equation}
where ``0'' represents the values today.
The time $t_k$ at the transition has a scale-dependence 
$t_k \propto k^{-3/(6n+4)}$, which means that the
transition occurs earlier for larger $k$.
The matter power spectrum $P_{\delta_m}=|\delta_m|^2$
at the onset of cosmic acceleration (at time $t_{\Lambda}$) 
shows a difference compared 
to the case of the $\Lambda$CDM model \cite{Star07}:
\begin{equation}
\frac{P_{\delta_m}}
{P_{\delta_m}{}^{\Lambda{\rm CDM}}}
=\left(\frac{t_\Lambda}{t_k}\right)
^{2\left(\frac{\sqrt{33}-1}{6}-\frac23\right)}
\propto k^{\frac{\sqrt{33}-5}{6n+4}}\,.
\label{Pratio}
\end{equation}
The ratio of the two power spectra today, i.e.
$P_{\delta_m}(t_0)/P_{\delta_m}{}^{\Lambda{\rm CDM}} (t_0)$
is in general different from Eq.~(\ref{Pratio}), but the difference 
is small for  $n$ of the order of unity \cite{Tsuji07}. 

The modified evolution of perturbations for the redshift $z<z_k$
gives rise to the integrated Sachs-Wolfe (ISW) effect
in CMB anisotropies \cite{Zhang05,SongHu1,LiBarrow,Peiris}, 
but this is limited to very large scales (low multipoles).
Since the CMB spectrum on the scales relevant to the 
large-scale structure ($k~\gsim~0.01\,h$\,Mpc$^{-1}$) is hardly 
affected by this modification, there is a difference between
the spectral indices of the CMB spectrum and 
the galaxy power spectrum: $\Delta n_s=(\sqrt{33}-5)/(6n+4)$.
Observationally we do not find any strong signature for
the difference of slopes of the two spectra.
If we take the mild bound $\Delta n_s<0.05$, 
we obtain the constraint $n>2$.

The growth index $\gamma$ defined in Eq.~(\ref{gammadex})
can be as close as 0.4 today in viable $f(R)$ models given 
in Eqs.~(\ref{Amodel})-(\ref{tanh}) \cite{Moraes}.
Depending on the epoch at which the perturbations cross the transition 
redshift $z_k$, the spatial dispersion of the growth index $\gamma_0 \equiv \gamma (z=0)$
appears in the range of values $0.40<\gamma<0.55$.
There are also regions in parameter space for which 
$\gamma_0$ converges to values around $0.40<\gamma<0.43$
\cite{Moraes09}.
These unusual dispersed or converged spectra will be useful to distinguish between 
$f(R)$ gravity models and the $\Lambda$CDM model in future high-precision 
observations. Since the modified evolution of matter perturbations directly 
affects the shear power spectrum in weak lensing, this is another important 
test for probing $f(R)$ gravity observationally \cite{Sapone08,TsujiTate,Schmidt,Borisov08,Narikawa}.
We also note that the nonlinear evolution of matter perturbations in $f(R)$ gravity 
(corresponding to the scales $k~\gsim~0.1\,h$\,Mpc$^{-1}$) has been 
studied in Refs.~\cite{Stabenau,Huparametrization,Oyaizu1,Oyaizu2,Oyaizu3,Laszlo,Tate,Koyama09,ClusterHu}.

\subsubsection{Local gravity constraints}
\label{lgcsec}

Let us discuss local gravity constraints on $f(R)$ dark energy models.
In the region of high density where gravitational experiments are carried out, 
the linear expansion of $R$ in terms of the cosmological value $R^{(0)}$
and the perturbation $\delta R$ is no longer valid because of the violation 
of the condition $\delta R \ll R^{(0)}$. In such a nonlinear regime, 
the chameleon mechanism \cite{chame1,chame2} can be at work to 
suppress the effective coupling between dark energy and non-relativistic matter.
In order to study how the chameleon mechanism works in $f(R)$ gravity, 
we transform the action (\ref{fRaction}) to the Einstein frame action 
under the conformal transformation 
$\tilde{g}_{\mu \nu}=F g_{\mu \nu}$: \cite{Maedacon}
\begin{equation}
S_E=\int {\rm d}^4 x \sqrt{-\tilde{g}} \left[ \frac{1}{2\kappa^2}
\tilde{R}-\frac12 \tilde{g}^{\mu \nu} \partial_{\mu} \phi
\partial_{\nu} \phi -V(\phi) \right]+\int {\rm d}^4 x\,
{\cal L}_M (g_{\mu \nu}, \Psi_m)\,,
\label{Einact}
\end{equation}
where $\kappa \phi \equiv \sqrt{3/2}\,\ln F$, 
$V(\phi)=(RF-f)/(2\kappa^2 F^2)$, and a tilde represents
quantities in the Einstein frame.

The action (\ref{Einact}) is the same as (\ref{actionchame})
with the correspondence that $g_{\mu \nu}$ in the Jordan frame
is equivalent to $g_{\mu \nu}^{(i)}$ in the action (\ref{actionchame}).
Since the quantity $F$ is given by $F=e^{-2\kappa Q \phi}$ with $Q=-1/\sqrt{6}$
in metric $f(R)$ gravity,
the field $\phi$ is coupled to non-relativistic matter (including baryons and 
dark matter) with a universal coupling $Q=-1/\sqrt{6}$.
Let us consider the models (\ref{Amodel}) and (\ref{Bmodel}), 
which behave as Eq.~(\ref{fRasy}) in the region of high density ($R \gg R_c$).
For the functional form (\ref{fRasy}) the effective potential defined 
in Eq.~(\ref{Veff}) is 
\begin{equation}
V_{{\rm eff}}(\phi) \simeq \frac{\mu R_{c}}{2\kappa^2}
e^{-4\kappa\phi/\sqrt{6}}
\left[1-(2n+1)\left(\frac{-\kappa\phi}{\sqrt{6}n\mu}
\right)^{2n/(2n+1)}\right]
+\rho e^{-\kappa\phi/\sqrt{6}}\,,
\label{Veff2}
\end{equation}
where 
\begin{equation}
F=e^{2\kappa \phi/\sqrt{6}}=1-2n\mu(R/R_{c})^{-(2n+1)}\,.
\label{Fre}
\end{equation}
Inside and outside a spherically symmetric body, the effective 
potential (\ref{Veff2}) has the following minima 
given, respectively, by 
\begin{eqnarray}
\kappa \phi_{A}\simeq-\sqrt{6}n\mu \left(\frac{R_{c}}{\kappa^2\rho_{A}} 
\right)^{2n+1}\,,\quad
\kappa \phi_{B}\simeq-\sqrt{6}n\mu \left(\frac{R_{c}}{\kappa^2\rho_{B}} 
\right)^{2n+1}\,.
\end{eqnarray}
One has $|\phi_B| \gg |\phi_A|$ provided that $\rho_A \gg \rho_B$.

The bound (\ref{boep2}) translates into 
\begin{equation}
\frac{n\mu}{x_{1}^{2n+1}}\left(\frac{R_{1}}{\rho_{B}}
\right)^{2n+1}<1.5\times10^{-15}\,,
\label{consmo1}
\end{equation}
where $x_{1}=R_1/R_c$ and $R_1$ is the Ricci scalar 
at the de Sitter point.
Let us consider the case in which the Lagrangian density 
is given by (\ref{fRasy}) for $R\ge R_{1}$. 
In the original models of Hu and Sawicki \cite{Hu} and 
Starobinsky \cite{Star07}  there are some modification to 
the estimation of $R_{1}$, but this change is not significant
when we place constraints on model parameters.
The de Sitter point for the model (\ref{fRasy}) corresponds to 
$\mu=x_{1}^{2n+1}/[2(x_{1}^{2n}-n-1)]$.
Substituting this relation into Eq.~(\ref{consmo1}), we find
\begin{equation}
\frac{n}{2(x_{1}^{2n}-n-1)}\left(\frac{R_{1}}{\rho_{B}}
\right)^{2n+1}<1.5\times10^{-15}\,.\label{cons2}
\end{equation}
The stability of the de Sitter point requires that $m(R_{1})<1$,
which translates into the condition $x_{1}^{2n}>2n^{2}+3n+1$. 
Then the term $n/[2(x_{1}^{2n}-n-1)]$ is smaller
than 0.25 for $n>0$.
Using the approximation that $R_{1}$ and $\rho_{B}$ are of 
the orders of the present cosmological density $10^{-29}$ g/cm$^{3}$ 
and the baryonic/dark matter density $10^{-24}$ g/cm$^{3}$ 
in our galaxy, respectively, we obtain the following 
constraint from (\ref{cons2}): \cite{CapoTsuji}
\begin{equation}
n>0.9\,.\label{bound3}
\end{equation}
Thus $n$ does not need to be much larger than unity. Under the condition
(\ref{bound3}), the deviation from the $\Lambda$CDM becomes 
important as $R$ decreases to the order of $R_{c}$.

{}From (\ref{Fre}) we find that there is a curvature singularity with $R \to \infty$
(and $M^2 \to \infty$) at $\phi=0$ for the models (\ref{Amodel}) and (\ref{Bmodel}).
At this singularity the field potential is finite, while its derivative goes to 
infinity. This singularity can be accessible as we go back to the 
past \cite{Frolov}, unless the oscillating mode of the scalar-field 
degree of freedom is suppressed.
This amounts to the fine-tuning of initial conditions for the 
field perturbation \cite{Star07}.
This past singularity can be cured by taking into account the $R^2$
term \cite{Appleby09}. 
The model of the type $f(R)=R-\alpha R_c \ln (1+R/R_c)$
was also proposed to address this problem \cite{Miranda}, 
but it satisfies neither local gravity 
constraints \cite{Thongkool} nor observational 
constraints of large-scale structure \cite{Maroto}.
Frolov \cite{Frolov} anticipated that the curvature singularity
may be accessed in a strong gravitational background 
such as neutron stars. Kobayashi and Maeda \cite{KM1,KM2} 
showed the difficulty of obtaining static spherically symmetric solutions
because of the presence of the singularity.
On the other hand, the choice of accurate boundary conditions 
confirms the existence of static solutions in 
a strong gravitational background with 
$\Phi_c~\lsim~0.3$ \cite{TTT09,Upadhye,Babi1,Babi2}.

\subsubsection{Palatini $f(R)$ gravity}
\label{Palasec}

In the so-called Palatini formalism of $f(R)$ gravity, 
the connections $\Gamma^{\lambda}_{\mu \nu}$ are
treated as independent variables when we vary the action (\ref{fRaction})
\cite{Palatini1919,Ferraris,Vollick,Vollick2,Flanagan0,Flanagan,Flanagan2}. 
Variation of the action (\ref{fRaction}) with respect to 
$g_{\mu \nu}$ gives 
\begin{equation}
\label{Pala1}
F(R)R_{\mu \nu}(\Gamma) -\frac12 f(R)g_{\mu \nu}
=\kappa^2 T_{\mu \nu}\,,
\end{equation}
where $F(R)=\partial f/\partial R$, 
$R_{\mu \nu}(\Gamma)$ is the Ricci tensor 
corresponding to the connections $\Gamma^{\lambda}_{\mu \nu}$, 
and $T_{\mu \nu}$ is the energy-momentum tensor of 
matter\footnote{There is another way for the variation of the action, known
as the metric-affine formalism \cite{Hehl}, in which  
the matter Lagrangian ${\cal L}_M$ depends 
not only on the metric $g_{\mu \nu}$ but also on the connection 
$\Gamma^{\lambda}_{\mu \nu}$. See Refs.~\cite{Liberati,Liberati2,Capome}
for the detail of such an approach.}.
$R_{\mu \nu}(\Gamma)$ is in general different from 
the Ricci tensor calculated in terms of metric connections $R_{\mu \nu}(g)$.
Taking the trace of Eq.~(\ref{Pala1}), we find
\begin{eqnarray}
\label{Pala2}
F(R)R-2f(R)=\kappa^2 T\,,
\end{eqnarray}
where $T=g^{\mu \nu}T_{\mu \nu}$.
The trace $T$ directly determines the Ricci scalar $R(T)$, 
which is related with the Ricci scalar 
$R(g)=g^{\mu \nu}R_{\mu \nu}(g)$ in the 
metric formalism via \cite{SotFaraoni}
\begin{equation}
R(T)=R(g)+\frac{3}{2(f'(R(T)))^2}(\nabla_{\mu}f'(R(T)))
(\nabla^{\mu}f'(R(T)))+\frac{3}{f'(R(T))} \square f'(R(T))\,,
\label{Rrela}
\end{equation}
where a prime represents a derivative in terms of $R(T)$.
Variation of the action (\ref{fRaction}) with respect to 
the connection leads to the following equation 
\begin{eqnarray}
\label{Pala3}
R_{\mu \nu}(g) -\frac12 g_{\mu \nu} R(g)
&=&\frac{\kappa^2 T_{\mu \nu}}{F}
-\frac{FR(T)-f}{2F}g_{\mu \nu}
+\frac{1}{F}(\nabla_{\mu} \nabla_{\nu}F
-g_{\mu \nu} \square F) \nonumber \\
& &-\frac{3}{2F^2} \left[ \partial_{\mu}F \partial_{\nu}F
-\frac12 g_{\mu \nu} (\nabla F)^2 \right]\,.
\end{eqnarray}

Unlike the trace equation (\ref{trace}) in the metric formalism, 
the kinetic term $\square F$ is not present in the corresponding 
equation (\ref{Pala2}) in the Palatini formalism.
Since the time-derivatives of the scalar-field degree of freedom 
do not appear in Palatini $f(R)$ gravity, cosmological solutions 
are not plagued by the dominance of the oscillating mode in the past.
In fact, the sequence of radiation, matter, and de Sitter epochs 
can be realized even for the model
$f(R)=R-\alpha/R^n$ ($n>0$) \cite{Meng,Meng2,Meng3,Sot,Sotinf,Motapala,FayTavakol}.
The combined data analysis of SN Ia, BAO, and the CMB 
shift parameter places the bound $n\in\left[-0.23,0.42\right]$
on the model $f(R)=R-\alpha/R^n$ ($n>-1$) \cite{FayTavakol}.

Although the background cosmology is well behaved in Palatini
$f(R)$ gravity, the evolution of non-relativistic
matter perturbations exhibits a distinguished feature relative to that in 
the $\Lambda$CDM model \cite{KoivistoPala2,LiPala0,LiPala,TsujiUddin}. 
Under the quasi-static approximation 
on sub-horizon scales, the equation of matter perturbations
is given by \cite{TsujiUddin}
\begin{equation}
\ddot{\delta}_m+2H \dot{\delta}_m-\frac{\kappa^2 \rho_m}
{2F} \left[ 1+\frac{m\,k^2/(a^2 R)}{1-m} \right] \delta_m \simeq 0\,.
\end{equation}
Although the matter perturbation evolves as $\delta_m \propto t^{2/3}$
in the regime $|m|\,k^2/(a^2 R) \ll 1$, the evolution of $\delta_m$
in the regime $|m|\,k^2/(a^2 R) \gg 1$ is completely different from 
that in GR. After the perturbations enter the regime 
$|m|\,k^2/(a^2 R)~\gsim~1$, 
they exhibit violent growth or damped oscillations depending on the 
signs of $m$ \cite{TsujiUddin}.
The $f(R)$ models are consistent with observations
of large-scale structure if the perturbations do not
enter the regime $|m|\,k^2/(a^2 R)~\gsim~1$ by today. 
This translates into the condition
\begin{equation}
\left| m(z=0) \right|~\lsim~(a_0H_0/k)^2\,.
\label{mcon}
\end{equation}
If we take the maximum wavenumber $k \approx 0.2\,h$\,Mpc$^{-1}$
(i.e., $k \approx 600 a_0H_0$), Eq.~(\ref{mcon}) gives the bound
$\left| m(z=0) \right|~\lsim~3 \times 10^{-6}$.
Hence the $f(R)$ models in the Palatini formalism are hardly 
distinguishable from the $\Lambda$CDM model.

There are also a number of problems in Palatini $f(R)$ dark energy models
associated with the non-dynamical nature of the scalar-field degree of freedom.
The dark energy model $f(R)=R-\mu^4/R$ is in conflict with 
the Standard Model of particle 
physics \cite{Flanagan0,Flanagan,Flanagan2,Kaloper,OlmoPRL2,Olmo08,Olmo09,Barausse1}
because of large non-perturbative corrections to the matter Lagrangian.
If we consider the models $f(R)=R-\mu^{2(n+1)}/R^n$, the only way to 
make such corrections small is to choose $n$ very close to 0 \cite{fRreview}.
Hence the deviation from the $\Lambda$CDM model needs to be 
very small. It was also shown that, for $f(R)$ dark energy models, 
a divergent behavior arises for the Ricci 
scalar at the surface of a static spherically symmetric star with 
a polytropic equation of state $P=c \rho_0^{\Gamma}$ ($3/2<\Gamma<2$), 
where $P$ is the pressure and $\rho_0$ is the rest-mass 
density \cite{Barausse1,Barausse2}.
These results show that Palatini $f(R)$ dark energy models
are difficult to be compatible with observational and experimental constraints, 
although this may not be the case for $f(R)$ models close 
to the Planck scale \cite{Singh,Barra09,Rei09}.

\subsection{Gauss-Bonnet dark energy models}

It is possible to extend $f(R)$ gravity to more general theories in which 
the Lagrangian density $f$ is an arbitrary function of $R$, 
$P\equiv R_{\mu\nu}R^{\mu\nu}$, and 
$Q\equiv R_{\mu\nu\alpha\beta}R^{\mu\nu\alpha\beta}$, where
$R_{\mu \nu}$ and  $R_{\mu\nu\alpha\beta}$ are Ricci tensor and 
Riemann tensor respectively \cite{CaDe,Mena}. 
The appearance of spurious spin-2 ghosts can be avoided by taking 
a Gauss-Bonnet (GB) combination \cite{ANunez,Calcagni05,Defelice1,Defelice2}
\begin{equation}
{\cal G}=R^2-4R_{\mu \nu}R^{\mu \nu}+
R_{\mu \nu \alpha \beta}R^{\mu \nu \alpha \beta}\,.
\end{equation}

A simple dark energy model motivated from low-energy 
effective string theory \cite{Gasreview} is given by \cite{NOS}
\begin{equation}
S=\int {\rm d}^4 x \sqrt{-g} \left[ \frac{1}{2\kappa^2} R
-\frac12 g^{\mu \nu} \partial_{\mu}\phi \partial_{\nu}\phi
-V(\phi)-f(\phi){\cal G} \right]+S_M\,,
\end{equation}
where $V(\phi)$ and $f(\phi)$ are functions of a scalar field $\phi$, 
and $S_M$ is a matter action.
The coupling $f(\phi){\cal G}$ allows the presence of 
a de Sitter solution even for a runaway field potential 
$V(\phi)$ \cite{Koivisto1,TsujiSami07,Koivisto2,Neupane}.
For the exponential potential $V(\phi)=V_0e^{-\kappa \lambda \phi}$ and 
the coupling $f(\phi)=(f_0/\mu)e^{\mu \kappa \phi}$, 
it was shown in Refs.~\cite{Koivisto1,TsujiSami07} that a scaling matter era 
can be followed by the late-time de Sitter solution for $\mu>\lambda$.

Koivisto and Mota \cite{Koivisto1} found that the parameter 
$\lambda$ in this model is constrained to be 
$3.5<\lambda<4.5$ (95 \% confidence level) from 
the observational data of SN Ia and WMAP 3-year.
The parameter $\lambda$ is constrained to be $3.5<\lambda<4.5$
at the 95\% confidence level. In the second paper \cite{Koivisto2},
they showed that the model is strongly disfavored from 
the combined data analysis including the constraints 
coming from BBN, LSS, BAO, and solar system data.
It was shown in Refs.~\cite{TsujiSami07,Ohta} that, when the 
GB term dominates the dynamics, 
tensor perturbations are subject to negative instabilities.
Amendola {\it et al.} \cite{AmenDavis} studied 
local gravity constraints on the above model and
showed that the energy contribution coming from the GB term needs
to be strongly suppressed for the consistency with solar-system experiments.
The above results imply that the GB term with the scalar-field
coupling $f(\phi){\cal G}$ can hardly be the source for dark energy.

The dark energy models based on the Lagrangian density 
${\cal L}=R/(2\kappa^2)+f({\cal G})$,
have been studied by a number of 
authors \cite{NO05,LBM,Cognola,DeHind,Davis,DeTsu,Cope09,Uddin,DeFelice09,DeFeliceMota}. 
In the presence of a perfect fluid with an energy density $\rho_M$, 
the Friedmann equation is given by \cite{NO05,LBM}
\begin{equation}
3H^2={\cal G}f_{,{\cal G}}-f-24 H^2 f_{,{\cal G}{\cal G}} \dot{{\cal G}}
+\rho_M\,.
\end{equation}
The Hubble parameter $H=H_1$ at the de Sitter point satisfies
$3H_1^2={\cal G}_1 f_{,{\cal G}} ({\cal G}_1)-f({\cal G}_1)$, 
where ${\cal G}_1=24 H_1^4$.
The stability of the de Sitter point requires the condition 
$0<H_1^6 f_{,{\cal G}{\cal G}} (H_1)<1/384$ \cite{DeTsu}.
In order to avoid the instability of solutions during radiation and 
matter eras, we also need the condition $f_{,{\cal G}{\cal G}}>0$.
In Ref.~\cite{DeTsu} the authors presented
a number of $f({\cal G})$ models that are cosmologically viable 
at the background level.
One of such viable models is given by 
\begin{equation}
f({\cal G})=\lambda \frac{{\cal G}}{\sqrt{{\cal G}_*}}
{\rm arctan} \left( \frac{{\cal G}}{{\cal G}_*} \right)
-\alpha \lambda \sqrt{{\cal G}_*}\,,
\end{equation}
where $\alpha$, $\lambda$ and ${\cal G}_*$ are constants.
This model can satisfy solar system constraints for a wide 
range of parameter space \cite{DeFelice09}.

If we consider cosmological perturbations, however, there is 
a Ultra-Violet (UV) instability in $f({\cal G})$ models associated with a negative 
propagation speed squared of a scalar-field degree of freedom \cite{LBM,DeFeliceMota}.
This growth of perturbations gets stronger on smaller scales, 
which is difficult to be compatible with the observed galaxy spectrum unless 
the deviation from GR is very small. 
Thus $f({\cal G})$ dark energy models are effectively ruled out as 
an alternative to the $\Lambda$CDM model.

\subsection{Scalar-tensor theories}

There is another class of modified gravity called scalar-tensor
theories in which the Ricci scalar $R$ is coupled to a scalar field $\varphi$.
One of the simplest examples is 
Brans-Dicke (BD) theory \cite{Brans} with the action 
\begin{equation}
S=\int{\rm d}^{4}x\sqrt{-g}\left[\frac{1}{2}\vp R
-\frac{\omega_{\rm BD}}{2\vp}(\nabla\vp)^{2}-U(\vp)\right]+S_{M}
(g_{\mu\nu},\Psi_{M})\,,
\label{BDaction}
\end{equation}
where $\omega_{\rm BD}$ is the BD parameter, $U(\vp)$
is the field potential, and $S_{M}$ is a matter action that depends 
on the metric $g_{\mu\nu}$ and matter fields $\Psi_{m}$. 
The original BD theory \cite{Brans} does not have the field 
potential $U(\varphi)$.

The general action for scalar-tensor theories can be written as 
\begin{equation}
S=\int{\rm d}^{4}x\sqrt{-g}\left[\frac{1}{2}f(\vp,R)
-\frac{1}{2}\omega(\vp)(\nabla\vp)^{2}\right]
+S_{M}(g_{\mu\nu},\Psi_{M})\,,
\label{stensoraction}
\end{equation}
where $f$ is a general function of the scalar field $\vp$ and the
Ricci scalar $R$, $\omega$ is a function of $\vp$.
We choose the unit $\kappa^{2}=1$.
We consider theories of the type 
\begin{eqnarray}
f(\vp,R)=F(\vp)R-2U(\vp)\,.
\label{stensor}
\end{eqnarray}
Under the conformal transformation $\tilde{g}_{\mu \nu}=F g_{\mu \nu}$,
the action in the Einstein frame is given by \cite{Maedacon}
\begin{equation}
S_{E}=\int{\rm d}^{4}x\sqrt{-\tilde{g}}\left[\frac{1}{2}\tilde{R}
-\frac{1}{2}(\tilde{\nabla}\phi)^{2}-V(\phi)\right]+
S_{M}(g_{\mu\nu},\Psi_{M})\,,
\label{SEframe}
\end{equation}
where $V=U/F^2$. We have introduced a new scalar field $\phi$ 
in order to make the field kinetic term canonical:
\begin{equation}
\phi \equiv \int{\rm d}\vp\,\sqrt{\frac{3}{2}
\left(\frac{F_{,\vp}}{F}\right)^{2}
+\frac{\omega}{F}}\,.\label{phire}
\end{equation}

We define the coupling between dark energy and non-relativistic 
matter, as
\begin{equation}
Q\equiv-\frac{F_{,\phi}}{2F}=-\frac{F_{,\vp}}{F}
\left[\frac{3}{2}\left(\frac{F_{,\vp}}{F}\right)^{2}+
\frac{\omega}{F}\right]^{-1/2}\,.\label{Q}
\end{equation}
In $f(R)$ gravity we have $\omega=0$ and hence $F=\exp(\sqrt{2/3}\phi)$ from 
Eq.~(\ref{phire}). Then the coupling is given by $Q=-1/\sqrt{6}$ from 
Eq.~(\ref{Q}). 
If $Q$ is constant as in $f(R)$ gravity, the following
relations hold from Eqs.~(\ref{phire}) and (\ref{Q}): 
\begin{equation}
F=e^{-2Q\phi}\,,\quad\omega=(1-6Q^{2})F\left(\frac{{\rm d}\phi}
{{\rm d}\vp}\right)^{2}\,.\label{conf_factor}
\end{equation}
In this case the action (\ref{stensoraction}) in the 
Jordan frame reads \cite{TUMTY} 
\begin{equation}
S=\int{\rm d}^{4}x\sqrt{-g}\Bigg[\frac{1}{2}F(\phi)R-
\frac{1}{2}(1-6Q^{2})F(\phi)(\nabla\phi)^{2}-U(\phi)\Bigg]+
S_{M}(g_{\mu\nu},\Psi_{m})\,.
\label{action2}
\end{equation}
In the limit that $Q\to0$ the action (\ref{action2}) reduces to the
one for a minimally coupled scalar field $\phi$ with the potential
$U(\phi)$. The transformation of the Jordan frame action (\ref{action2})
under the conformal transformation $\tilde{g}_{\mu\nu}=e^{-2Q \phi} g_{\mu\nu}$
gives rise to the Einstein frame action (\ref{SEframe}) with a constant
coupling $Q$. 

One can compare (\ref{action2}) with the action (\ref{BDaction})
in BD theory. Setting $\varphi=F=e^{-2Q\phi}$, one finds that two actions
are equivalent if the parameter $\omega_{{\rm BD}}$ is related to
$Q$ via the relation \cite{chame2,TUMTY} 
\begin{eqnarray}
3+2\omega_{{\rm BD}}=\frac{1}{2Q^{2}}\,.
\label{BDre}
\end{eqnarray}
Using this relation, we find that the General Relativistic limit ($\omega_{{\rm BD}}\to\infty$)
corresponds to the vanishing coupling ($Q \to 0$).
Since $Q=-1/\sqrt{6}$ in $f(R)$ gravity, this corresponds to 
the Brans-Dicke parameter $\omega_{{\rm BD}}=0$ \cite{ohanlon,teyssandier,Chiba}.
One can show that the field equation (\ref{Pala3}) in Palatini $f(R)$ gravity 
is equivalent to the one derived in BD theory with 
$\omega_{\rm BD}=-3/2$ \cite{SotFaraoni,fRreview}.
Hence Palatini $f(R)$ gravity corresponds to 
the infinite coupling ($Q^2 \to \infty$).

There are also other scalar-tensor theories that give rise to field-dependent
couplings $Q(\phi)$. For a nonminimally coupled scalar field 
with $F(\varphi)=1-\xi \varphi^2$ and 
$\omega (\varphi)=1$ in the action (\ref{stensoraction}) with (\ref{stensor}), 
the coupling is field-dependent, i.e. $Q(\varphi)=\xi \varphi/
[1-\xi \varphi^2 (1-6\xi)]^{1/2}$. The cosmological dynamics of dark energy 
models based on such theories have been studied by a number of 
authors \cite{Amendola:1999qq,Uzan,Chiba99,Bartolo,Mata,Bacci,Riazuelo}.

Let us consider BD theory with the action (\ref{action2}).
In the absence of the potential $U(\phi)$
the BD parameter $\omega_{\rm BD}$
is constrained to be $\omega_{\rm BD}>4.0 \times 10^4$
from solar-system experiments \cite{Will05}.
This bound also applies to the case
of a nearly massless field with the potential $U(\phi)$
in which the Yukawa correction $e^{-Mr}$ is close to unity
(where $M$ is the scalar field mass and $r$ is an interaction length).
Using the bound $\omega_{\rm BD}>4.0 \times 10^4$
in Eq.~(\ref{BDre}), we find
\begin{equation}
|Q|<2.5 \times 10^{-3}\,.
\end{equation}
In this case the cosmological evolution for such theories is 
hardly distinguishable from the $Q=0$ case.
Even for scalar-tensor theories with such small couplings, 
it was shown that the phantom equation state of dark energy 
can be realized without the appearance of a ghost 
state \cite{Peri,Peri2,Jerome,Gan}.

In the presence of the field potential it is possible for large coupling
models ($|Q|={\cal O}(1)$) to satisfy local gravity constraints under
the chameleon mechanism, 
provided that the mass $M$ of the field $\phi$ is sufficiently large in the
region of high density. 
In metric $f(R)$ gravity ($Q=-1/\sqrt{6}$) the field potential $U(\phi)$
in Eq.~(\ref{action2}) corresponds to $U=(FR-f)/2$ with 
$\phi=\sqrt{3/2}\,\ln F$. 
The viable $f(R)$ dark energy models (\ref{Amodel}) and (\ref{Bmodel})
have the asymptotic form (\ref{fRasy}), in which case 
the field potential is given by 
\begin{equation}
U(\phi)=\frac{\mu R_c}{2} \left[ 1-\frac{2n+1}{(2n \mu)^{2n/(2n+1)}}
\left( 1-e^{2\phi/\sqrt{6}} \right)^{2n/(2n+1)} \right]\,.
\label{Uphi}
\end{equation}
For BD theories with the constant coupling $Q$, one can generalize
the potential (\ref{Uphi}) to the form  
\begin{equation}
U(\phi)=U_{0}\left[1-C(1-e^{-2Q\phi})^{p}\right]
\qquad(U_{0}>0,~C>0,~0<p<1)\,.
\label{modelscalar}
\end{equation}

As $\phi \to 0$, the potential (\ref{modelscalar}) approaches
the finite value $U_0$ with a divergence of the field mass 
squared $M^2=U_{,\phi \phi} \to  \infty$. This model has a curvature singularity 
at $\phi=0$ as in the case of the $f(R)$ models (\ref{Amodel})
and (\ref{Bmodel}). The mass $M$ decreases as the field
evolves away from $\phi=0$. The late-time cosmic acceleration 
can be realized by the potential (\ref{modelscalar}) provided that 
$U_0$ is of the order of $H_0^2$.

Since the action (\ref{SEframe}) in the Einstein frame is equivalent to 
the action (\ref{actionchame}), the chameleon mechanism can be 
at work even for BD theories with large couplings ($|Q|={\cal O}(1)$).
Considering a spherically symmetric body with homogenous densities 
$\rho_A$ and $\rho_B$ inside and outside bodies respectively, 
the effective potential $V_{\rm eff}=V(\phi)+e^{Q\phi}\rho$ 
in the Einstein frame (where $V(\phi)=U(\phi)/F^2$)
has two minima characterized by 
\begin{equation}
\phi_A \simeq \frac{1}{2Q} 
\left( \frac{2U_0pC}{\rho_A} \right)^{1/(1-p)}\,,\qquad
\phi_B \simeq \frac{1}{2Q} 
\left( \frac{2U_0pC}{\rho_B} \right)^{1/(1-p)}\,.
\end{equation}
Using the experimental bound (\ref{boep2}) coming from the violation of
equivalence principle together with the condition for realizing 
the cosmic acceleration today, we obtain the constraint \cite{TUMTY}
\begin{equation}
p>1-\frac{5}{13.8-{\rm \log}_{10}\,|Q|}\,.
\end{equation}
When $|Q|=10^{-2}$ and $|Q|=10^{-1}$ we have $p>0.68$ and $p>0.66$, respectively. 
In $f(R)$ gravity the above bound corresponds to $p > 0.65$,
which translates into $n > 0.9$ for the model (\ref{fRasy}).

The evolution of cosmological perturbations in scalar-tensor theories has been 
discussed in Refs.~\cite{Boi00,Verde,Tsujimatter,TUMTY,Song10}.
Under the quasi-static approximation on sub-horizon scales,  
the matter perturbation $\delta_m$ for the theory (\ref{action2}) obeys 
the following equation of motion \cite{TUMTY,Song10} 
\begin{equation}
\ddot{\delta}_m+2H\dot{\delta}_m
-4\pi G_{{\rm eff}}\rho_{m}\delta_{m} \simeq 0\,,
\label{mattereqsca2a}
\end{equation}
where the effective (cosmological) gravitational coupling is 
\begin{equation}
G_{{\rm eff}}=\frac{G}{F}\frac{(k^{2}/a^{2})(1+2Q^{2})F
+M^{2}}{(k^{2}/a^{2})F+M^{2}}\,.
\label{eq:jfg}
\end{equation}
Here $M^2 \equiv U_{,\phi \phi}$ is the field mass squared. 
In the ``General Relativistic'' regime characterized by $M^2/F \gg k^2/a^2$, 
one has $G_{\rm eff} \simeq G/F$ and $\delta_m \propto t^{2/3}$.
In the ``scalar-tensor'' regime characterized by $M^2/F \ll k^2/a^2$, 
it follows that $G_{\rm eff} \simeq (1+2Q^2)G/F$ and 
$\delta_{m}\propto t^{(\sqrt{25+48Q^{2}}-1)/6}$.
If the transition from the former regime to the latter regime occurs 
during the matter era, this gives rise to a difference between the 
spectral indices of the matter power spectrum and of the CMB spectrum 
on the scales $0.01\,h$\,Mpc$^{-1}~\lsim~k~\lsim~0.2 \,h$\,Mpc$^{-1}$
\cite{TUMTY}:
\begin{equation}
\Delta n_s=\frac{(1-p)(\sqrt{25+48Q^{2}}-5)}{4-p}\,.
\label{deln}
\end{equation}
Under the criterion $\Delta n_s<0.05$,
we obtain the bounds $p>0.957$
for $Q=1$ and $p>0.855$ for $Q=0.5$.
As long as $p$ is close to 1, the model can be consistent with both 
cosmological and local gravity constraints.

For the perturbed metric ${\rm d}s^2=-(1+2\Psi){\rm d} t^2+a^2(t)(1-2\Phi)
\delta_{ij}{\rm d}x^i {\rm d} x^j$, the gravitational potentials obey the 
following equations under a quasi-static approximation on 
sub-horizon scales \cite{TUMTY}
\begin{eqnarray}
& &\frac{k^{2}}{a^{2}}\Psi\simeq-\frac{4\pi G}{F}
\frac{(k^{2}/a^{2})(1+2Q^{2})F+M^{2}}{(k^{2}/a^{2})F+M^{2}}
\rho_m \delta_m \,,
\label{PsiPhisca0}
\\
& &\frac{k^{2}}{a^{2}}\Phi\simeq-\frac{4\pi G}{F}
\frac{(k^{2}/a^{2})(1-2Q^{2})F+M^{2}}{(k^{2}/a^{2})F+M^{2}}
\rho_m \delta_m\,,
\label{PsiPhisca}
\end{eqnarray}
where we have recovered the gravitational constant $G$.
The results (\ref{PsiPhisca0}) and (\ref{PsiPhisca}) include those 
in $f(R)$ gravity by setting $Q=-1/\sqrt{6}$.
In the regime $M^2/F \ll k^2/a^2$ the evolution of $\Psi$ and $\Phi$
is subject to change compared to that in the GR regime characterized by 
$M^2/F \gg k^2/a^2$.
In general the difference from GR may be quantified by the parameters
$q$ and $\zeta$ \cite{Sapone08}:
\begin{equation}
\frac{k^2}{a^2}\Phi=-4 \pi G q \rho_m \delta_m\,,\qquad
\frac{\Phi-\Psi}{\Phi}=\zeta\,.
\label{qzetadef}
\end{equation}
In the regime $M^2/F \ll k^2/a^2$ of scalar-tensor theory (\ref{action2})
it follows that $q \simeq (1-2Q^2)/F$ and $\zeta \simeq -4Q^2/(1-2Q^2)$.

In order to confront dark energy models with the observations of weak
lensing, it may be convenient to introduce the following quantity \cite{Sapone08} 
\begin{equation}
\Sigma\equiv q(1-\zeta/2)\,.\label{eq:sigmamg}
\end{equation}
 {}From the definition (\ref{qzetadef}) we find that the
weak lensing potential $\psi=\Phi+\Psi$ can be expressed as 
\begin{equation}
\psi=-8\pi G\frac{a^{2}}{k^{2}}\rho_{m}\delta_{m}\Sigma\,.
\label{psisigma}
\end{equation}
In scalar-tensor theory (\ref{action2}) one has $\Sigma=1/F$.
The effect of modified gravity theories manifests itself in weak lensing
observations in at least two ways. One is the multiplication of the
term $\Sigma$ on the r.h.s. of Eq.~(\ref{psisigma}). Another is
the modification of the evolution of $\delta_{m}$. The latter depends
on two parameters $q$ and $\zeta$, or equivalently, $\Sigma$ and
$\zeta$. Thus two parameters ($\Sigma,\zeta$) will be useful to
detect signatures of modified gravity theories from future surveys
of weak lensing. See 
Refs.~\cite{Zhang07,Wang07,Jain08,Daniel,Berts,ZhaoPo,SongKoyama,SongDore,Guzik,Braxlensing,Daniel:2010ky,Bean:2010zq,Zhao3} 
for related works about testing 
gravitational theories in weak lensing observations.

\subsection{DGP model}

In the so-caled Dvali, Gabadadze, and Porrati (DGP) \cite{DGP} braneworld
it is possible to realize a ``self accelerating Universe'' even in the absence of dark energy.
In braneworlds standard model particles are confined on a 
3-dimensional (3D) brane embedded in the 5-dimensional bulk space-time 
with large extra dimensions.
In the DGP braneworld model \cite{DGP} the 3-brane is embedded 
in a Minkowski bulk space-time 
with infinitely large extra dimensions.
Newton gravity can be recovered by adding a 4D Einstein-Hilbert 
action sourced by the brane curvature to the 5D action.
Such a 4D term may be induced by quantum corrections 
coming from the bulk gravity and its coupling with matter on the brane. 
In the DGP model the standard 4D gravity is recovered for small distances,
whereas the effect from the 5D gravity manifests itself for large
distances. The late-time cosmic acceleration can be realized 
without introducing a dark energy component \cite{Deffayet1,Deffayet2}
(see also Ref.~\cite{SahniSh} for a generalized version of the DGP model).

The action for the DGP model is given by 
\begin{equation}
S=\frac{1}{2\kappa_{(5)}^{2}}\int {\rm d}^{5}X\sqrt{-\tilde{g}}\,
\tilde{R}+\frac{1}{2\kappa_{(4)}^{2}}\int {\rm d}^{4}X\sqrt{-g}R-
\int {\rm d}^{5}X\sqrt{-\tilde{g}}\,{\cal L}_{M}\,,\label{DGPaction}
\end{equation}
where $\tilde{g}_{AB}$ is the metric in the 5D bulk and 
$g_{\mu\nu}=\partial_{\mu}X^{A}\partial_{\nu}X^{B}\tilde{g}_{AB}$
is the induced metric on the brane with $X^{A}(x^{c})$ being the
coordinates of an event on the brane labelled by $x^{c}$.
The 5D and 4D (reduced) gravitational constants, $\kappa_{(5)}^{2}$ and 
$\kappa_{(4)}^{2}$, are related with the 5D and 4D Planck masses,
$M_{(5)}$ and $M_{(4)}$, via $\kappa_{(5)}^{2}=1/M_{(5)}^3$
and $\kappa_{(4)}^{2}=1/M_{(4)}^2$.
The first and second terms in Eq.~(\ref{DGPaction}) correspond to
Einstein-Hilbert actions in the 5D bulk and on the brane, respectively.
The matter action consists of a brane-localized matter whose action is given
by $\int {\rm d}^{4}x\sqrt{-g}\,(\sigma+{\cal L}_{M}^{{\rm brane}})$,
where $\sigma$ is the 3-brane tension and ${\cal L}_{M}^{{\rm brane}}$
is the Lagrangian density on the brane. Since the tension is not related
to the Ricci scalar $R$, it can be adjusted to be zero.

The Einstein equation in the 5D bulk is given by $G_{AB}^{(5)}=0$, 
where $G_{AB}^{(5)}$ is the 5D Einstein tensor.
Imposing the Israel junction conditions on the brane with a 
$Z_2$ symmetry, we obtain the 4D Einstein equation \cite{Porrati09}
\begin{equation}
G_{\mu \nu}-\frac{1}{r_c} (K_{\mu \nu}-K g_{\mu \nu})
=\kappa_{(4)}^2 T_{\mu \nu}\,,
\end{equation}
where $K_{\mu \nu}$ is the extrinsic curvature
on the brane and $T_{\mu \nu}$ is the energy-momentum
tensor of localized matter.
The cross-over scale $r_c$ is defined by $r_c \equiv 
\kappa_{(5)}^{2}/(2\kappa_{(4)}^{2})$.
The Friedmann equation on the flat FLRW brane takes a simple 
form \cite{Deffayet1,Deffayet2} 
\begin{equation}
H^{2}-\frac{\epsilon}{r_{c}}H=\frac{\kappa_{(4)}^{2}}{3}\rho_{M}\,,
\label{eqdgp}
\end{equation}
where $\epsilon=\pm 1$, and $\rho_M$ is the energy density of 
matter on the brane (with pressure $P_M$)
satisfying the continuity equation 
\begin{equation}
\dot{\rho}_M+3H(\rho_M+P_M)=0\,.
\end{equation}
If $r_c$ is much larger than the Hubble radius $H^{-1}$,
the first term in Eq.~(\ref{eqdgp})
dominates over the second one. In this case the standard Friedmann
equation, $H^{2}=\kappa_{(4)}^{2}\rho_{M}/3$, is recovered. Meanwhile,
in the regime $r_{c}< H^{-1}$, the presence of the second
term in Eq.~(\ref{eqdgp}) leads to a modification to the standard
Friedmann equation. In the Universe dominated by non-relativistic
matter ($\rho_{M}\propto a^{-3}$), the Universe approaches 
a de Sitter solution for $\epsilon=+1$: 
$H\to H_{{\rm dS}}=1/r_{c}$.
Hence it is possible to realize the present cosmic acceleration provided
that $r_{c}$ is of the order of the present Hubble radius $H_{0}^{-1}$.

%%%%%%%%%%%%%%%%%%%%%%%%%%%%%
\begin{figure}
\begin{centering}
\includegraphics[width=3.2in,height=3.0in]{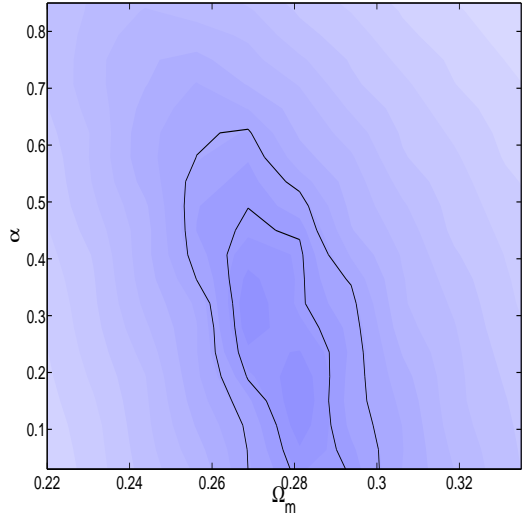} 
\par\end{centering}
\caption{Combined observational constraints on the modified DGP model
characterized by the Friedmann equation 
$H^2-H^{\alpha}/r_c^{2-\alpha}=\kappa_{(4)}^2 \rho_m/3$.
The two curves show $1\sigma$ and $2\sigma$ contours 
in the $(\Omega_m^{(0)}, \alpha)$ plane.
The original DGP model ($\alpha=1$) is incompatible with observations.
{}From Ref.~\cite{Xia}.}
\centering{}\label{dgp} 
\end{figure}
%%%%%%%%%%%%%%%%%%%%%%%%%%%%

Although the DGP braneworld is an attractive model allowing a self 
acceleration, the joint constraints from SNLS, BAO, and CMB 
data shows that this model is disfavored 
observationally \cite{Sawicki05,Fair,Maje,Alambrane,Song06,Xia}.
There is a modified version of the DGP model characterized by the Friedmann equation 
$H^2-H^{\alpha}/r_c^{2-\alpha}=\kappa_{(4)}^2 \rho_m/3$, where 
$\rho_m$ is the energy density of non-relativistic matter \cite{DvaliTurner}.
In Fig.~\ref{dgp} we show $1\sigma$ and $2\sigma$ contours in 
the $(\Omega_m^{(0)}, \alpha)$ plane constrained from the joint 
data analysis of SN Ia, BAO, CMB, gamma ray bursts, 
and the linear growth factor of matter perturbations \cite{Xia}. 
The parameter $\alpha$ is constrained to be 
$\alpha=0.254 \pm 0.153$ (68 \% confidence level) and hence
the flat DGP model ($\alpha=1$) is incompatible with observations.

The evolution of density perturbations in the DGP model has been 
studied in 
Refs.~\cite{Lue,Koyama,Silverstri06,Yamamoto06,Koyama2,HuteLinder,Wei08,Ishak09,Fu09}.
Under the quasi-static approximation on sub-horizon scales, the linear 
matter perturbation $\delta_m$ (with a homogenous density $\rho_m$) 
obeys the following equation
\begin{equation}
\ddot{\delta}_m+2H\dot{\delta}_m-4\pi G_{\rm eff} 
 \rho_{m}\delta_{m}=0\,,\qquad
 G_{\rm eff}=\left[1+1/(3\beta) \right]G\,,
\label{DGPeqdel}
\end{equation}
where $\beta \equiv 1-2Hr_c [1+\dot{H}/(3H^2)]$.
In the deep matter era ($Hr_{c} \gg 1$), $\beta$ is largely negative 
($|\beta|\gg1$). In this regime the matter perturbation 
evolves as $\delta_m \propto t^{2/3}$ as in GR.
Around the late-time de Sitter solution one has $\beta\simeq1-2Hr_{c}\simeq-1$ 
and $1+1/(3\beta)\simeq2/3$, so that the growth rate of $\delta_m$
gets smaller than that in the $\Lambda$CDM model.
The growth index $\gamma$ defined in Eq.~(\ref{gammadex}) is
given by $\gamma \approx 0.68$ \cite{Linder05}.

In the massless regime ($M^2/F \ll k^2/a^2$) in BD theory the effective 
gravitational coupling (\ref{eq:jfg}) is given by $G_{\rm eff}=(1+2Q^2)G/F
=(4+2\omega_{\rm BD})/(3+2\omega_{\rm BD})G/F$, where we 
used the relation (\ref{BDre}). Comparing this with the effective 
coupling (\ref{DGPeqdel}), we find that the DGP model 
is related to BD theory via $\omega_{\rm BD}=(3/2)(\beta-1)$ with $F=1$.
Since $\beta<0$ for the self-accelerating DGP solution, this implies 
that $\omega_{\rm BD}<-3/2$ and hence the DGP model 
contains a ghost mode.
It is however possible to construct a generalized DGP model free from 
the ghost problem by embedding our visible 3-brane with a 4-brane 
in a flat 6D bulk \cite{Rham}.

In the DGP model, a brane bending mode $\phi$ in the bulk corresponds to 
a scalar-field degree of freedom. In general such a field can mediate a
long range fifth force incompatible with local gravity experiments, but 
the presence of a self-interaction of $\phi$ allows the so-called 
Vainshtein mechanism \cite{Vainshtein} to work within a radius 
$r_*=(r_g r_c^2)^{1/3}$ ($r_g$ is the Schwarzschild radius of a source).
The DGP model can be consistent with local gravity constraints 
under some range of conditions on the energy-momentum 
tensor \cite{Deffayet:2001uk,Gruzinov:2001hp,Porrati:2002cp}.

The DGP model stimulated other approaches for constructing ghost-free
theories in the presence of nonlinear self-interactions of a scalar field $\phi$.
It is important to keep the field equations at second order 
in time derivatives to avoid that an extra degree of freedom gives 
rise to a ghost state. In particular Nicolis {\it et al.} \cite{Nicolis}
imposed a constant gradient-shift symmetry (``Galilean'' symmetry), 
$\partial_{\mu} \phi \to \partial_{\mu} \phi+b_{\mu}$, 
to restrict the equations of motion at second order, while keeping
a universal gravitational coupling with matter.
In the 4-dimensional Minkowski space-time they found five terms
${\cal L}_i$ ($i=1,\cdots 5$) giving rise to equations of motion 
satisfying the Galilean symmetry. The first three terms are 
given by ${\cal L}_1=\phi$, ${\cal L}_2=\nabla_{\mu} \phi \nabla^{\mu} \phi$, 
and ${\cal L}_3=\square \phi \nabla_{\mu} \phi \nabla^{\mu}\phi$.
The term ${\cal L}_3$ is the nonlinear field derivative that appears
in the DGP model, which allows the possibility for the consistency with 
solar system experiments through the Vainshtein mechanism.
Deffayet {\it et al.} \cite{DeffaGal,DeffaGalED} derived the covariant 
expression of the terms ${\cal L}_i$ ($i=1,\cdots 5$) 
by extending the analysis to the 
curved space-time.

Silva and Koyama \cite{KazuyaGal} considered BD theory (without a field potential)
in the presence of a nonlinear derivative term 
$\xi (\phi)\square \phi \nabla_{\mu} \phi \nabla^{\mu}\phi$, i.e.
\begin{equation}
S=\int {\rm d}^4 x \sqrt{-g} \left[ \frac12 \phi R
-\frac{\omega_{\rm BD}}{2\phi} (\nabla \phi)^2
+\xi (\phi)\square \phi \nabla_{\mu} \phi \nabla^{\mu}\phi \right]
+S_M\,,
\end{equation}
where $S_M$ is the matter action.
Although the nonlinear derivative term does not satisfy the Galilean invariance
in the FLRW cosmological background, the field equation of motion remains 
at second order (see also Ref.~\cite{JustinGal}). 
This term also arises as one of higher-order derivative corrections
to low-energy effective string theory.
Interestingly, for the function $\xi(\phi)=1/(M^2 \phi^2)$, there exists 
a de Sitter solution responsible for dark energy (provided that $M \approx H_0$).
Moreover, because of the presence of the nonlinear interaction, the problems 
of the appearance of ghosts and instabilities 
can be avoided for the BD parameter $\omega_{\rm BD}$ 
smaller than $-2$ \cite{KazuyaGal}. 
At early times General Relativity can be recovered by the 
cosmological Vainshtein mechanism.
A number of interesting observational signatures such as modified
growth of matter perturbations as well as a distinguished ISW effect 
have been studied in Refs.~\cite{KazuyaGal,KobayashiGal,KobayashiGal2}.

%%%%%%%%%%%%%%%%%%%%%%%%%%%%
\section{Cosmic acceleration without dark energy}
\label{withoutsec}
%%%%%%%%%%%%%%%%%%%%%%%%%%%%

There are attempts to explain the apparent cosmic acceleration 
by inhomogeneities in the distribution of matter
without recourse to a dark energy component
(see Ref.~\cite{Buchertreview} for review).

One of such approaches is the void model in which the presence of 
underdense bubbles leads to the faster expansion of 
the Universe compared to the outside.
In other words we live in the middle of a huge spherical region
and we interpret the evolution of this underdense region as
an apparent cosmic acceleration.
Originally, Tomita \cite{Tomita1,Tomita2} introduced a local homogenous 
void separated from the outside described by a homogenous 
FLRW space with a singular mass shell. 
The analysis was extended to the models with 
a continuous transition between the inside and outside 
the void \cite{Alnes}.
This can be described by a class of the Lema\^{i}tre-Tolman-Bondi (LTB) 
spherically symmetric models.
Theoretical and observational aspects of the LTB model have been 
extensively studied as an alternative to dark 
energy \cite{Celerier,Iguchi,Alnes,Vander,Paran,Moffat,Garfin,Chung,Alnes06,Alnes2,Alnes3,Kai,Enqvist,Biswas,Brouzakis,Tanimoto,Garcia,GarciaBellido2,Clifton,Alexander,Clarkson,Kainu09,Caldvoid,Quer09,Quartin,Romano,Cele09,Dunsby,Saito}.

The second approach is based on the backreaction of cosmological 
perturbations arising from perturbing the homogeneous 
Universe \cite{Rasanen,Kolb1,Kolb2}.
Unlike the void model, this tries to explain a {\it real} cosmic 
acceleration by arranging inhomogeneities  that come from 
the deviation of the FLRW metric.

There is another approach for explaining the apparent cosmic 
acceleration based on the ``Ultra Strong''  version of
equivalence principle \cite{Pia1,Pia2}. 
In this model the standard geometric description 
of space-time as a metric manifold holds as a small distance approximation
and hence General Relativity can be modified on large scales by 
a curvature-dependent subleading effect.
Although the original model proposed in Ref.~\cite{Pia3} do not 
explain the observational data of SN Ia very well, 
its modified version can be consistent with the SN Ia data
with a rather low value of the Hubble constant, 
$H_0 \approx 50$\,km\,sec$^{-1}$\,Mpc$^{-1}$.

In the following we shall briefly review the first two approaches.

\subsection{Inhomogeneous Lema\^{i}tre-Tolman-Bondi model}

In order to discuss a spherical inhomogeneity in local regions 
we take the LTB metric given by 
\begin{equation}
{\textrm{d}}s^{2}=-{\textrm{d}}t^{2}+X^{2}(t,r)\,{\textrm{d}}r^{2}
+R^{2}(t,r)\,{\textrm{d}}\Omega^{2}\,,
\label{eq:inh-met}
\end{equation}
where the expansion factor along the radial coordinate $r$ is different relative to the
surface line element $\rd\Omega^{2}=\rd\theta^{2}+\sin^{2}\theta\,\rd\phi^{2}$.
Solving the (0,1) component of the Einstein equation $G_{01}=0$ for the fluid at rest, 
it follows that $X(t,r)$ is separable as $X(t,r)=R'(t,r)/\sqrt{1+\beta(r)}$.
Here a prime represents a partial derivative with respect to $r$ and 
$\beta(r)$ is a function of $r$.
Then the metric (\ref{eq:inh-met}) is given by 
\begin{equation}
{\textrm{d}}s^{2}=-{\textrm{d}}t^{2}+\frac{\left[R'(t,r)\right]^{2}}
{1+\beta(r)}{\textrm{d}}r^{2}+R^{2}(t,r){\textrm{d}}\Omega^{2}\,.
\label{eq:LTB}
\end{equation}

The metric (\ref{eq:LTB}) recovers the one in the FLRW 
space-time by the choice $R=a(t)r$ and $\beta=-Kr^{2}$, 
where $K$ is a cosmic curvature.
In other cases the metric (\ref{eq:LTB}) describes a spherical 
inhomogeneity centered on the origin. 
We define the transverse Hubble function $H_{\perp}$
and the radial Hubble function $H_{||}$, as 
$H_{\perp} \equiv \dot{R}'/R'$ and $H_{\perp}=\dot{R}/R$.
The Einstein equations in the presence of non-relativistic 
matter (energy density $\rho_m$) give \cite{Alnes}
\begin{eqnarray}
H_{\perp}^{2}+2H_{||}H_{\perp}-\frac{\beta}{R^{2}}
-\frac{\beta'}{RR'} & = & 8\pi G\rho_{m}\,,
\label{eq:ltb-fri1}\\
6\frac{\ddot{R}}{R}+2H_{\perp}^{2}-2\frac{\beta}{R^{2}}
-2H_{||}H_{\perp}+\frac{\beta'}{RR'} & = & -8\pi G\rho_{m}\,.
\label{eq:ltb-fri2}
\end{eqnarray}
Eliminating the term $\rho_m$ from Eqs.~(\ref{eq:ltb-fri1}) and 
(\ref{eq:ltb-fri2}), we obtain the relation $2R \ddot{R}+\dot{R}^2=\beta (r)$.
Integrating this equation, it follows that 
\begin{equation}
H_{\perp}^{2}=\frac{\alpha(r)}{R^{3}}+\frac{\beta(r)}{R^{2}}\,,
\end{equation}
where $\alpha (r)$ is an arbitrary function of $r$.
{}From this we can introduce the today's effective density 
parameters of matter and the spatial curvature, respectively, as
\begin{equation}
\Omega_{m}^{(0)}(r)\equiv\frac{\alpha(r)}{R_{0}^{3}H_{\perp0}^{2}}\,,
\qquad
\Omega_{K}^{(0)}(r)=1-\Omega_{m}^{(0)}(r)=
\frac{\beta(r)}{R_{0}^{2}H_{\perp0}^{2}}\,.
\end{equation}

We define the time $t=0$ at the decoupling epoch (the redshift $z \simeq 1090$)
with $R(r, t=0)=0$. Introducing the conformal time $\eta$ as
${\rm d}\eta=(\sqrt{\beta}/R)\,{\rm d}t$, we obtain the following 
parametric solutions of Eqs.~(\ref{eq:ltb-fri1}) and (\ref{eq:ltb-fri2}) 
for $\beta>0$ \cite{Alnes}:
\begin{eqnarray}
& &R=\frac{\alpha(r)}{2\beta(r)}(\cosh\eta-1)=
\frac{R_{0}\Omega_{m}^{(0)}(r)}{2[1-\Omega_{m}^{(0)}(r)]}(\cosh\eta-1)\,,\label{eq:sol-R}\\
& &t=\frac{\alpha(r)}{2\beta^{3/2}(r)}(\sinh\eta-\eta)=
\frac{\Omega_{m}^{(0)}(r)}{2[1-\Omega_{m}^{(0)}(r)]^{3/2}
H_{\perp0}}(\sinh\eta-\eta)\,\,.\label{eq:sol-beta-t}
\end{eqnarray}
The structure of the void with an under density can be accommodated
by choosing $\Omega_m^{(0)}(r)$ and 
$h \equiv H_{\perp0}/(100$\,km\,sec$^{-1}$\,Mpc$^{-1}$)
in the following form \cite{Garcia} (see also Ref.~\cite{Alnes} for another choice)
\begin{eqnarray}
\Omega_{m}^{(0)}(r) & = & \Omega_{{\rm out}}+(\Omega_{{\rm in}}-
\Omega_{{\rm out}})f(r,r_{0},\Delta)\,,\\
h (r) & = & h_{{\rm out}}+(h_{{\rm in}}-h_{{\rm out}})f(r,r_{0},\Delta)\,,
\end{eqnarray}
where the function $f(r,r_0, \Delta)=[1-\tanh ((r-r_0)/2\Delta)]
/[1+\tanh (r_0/2\Delta)]$ describes a transition of a shell of 
radius $r_0$ and thickness $\Delta$
(``in'' and ``out'' represent quantities inside and outside 
the void, respectively).

The trajectory of photons arriving at $r=0$ today is characterized
by a path $t=\hat{t}(r)$ satisfying \cite{Alnes06}
\begin{equation}
\frac{{\rm d} \hat{t}}{{\rm d}r}=-\frac{R' (r, \hat{t})}
{\sqrt{1+\beta}}\,.
\end{equation}
Then one can show that the redshift $z=z(r)$ of photons
obeys the differential equation \cite{Iguchi}
\begin{equation}
\frac{{\rm d}z}{{\rm d}r}=(z+1) 
\frac{\dot{R}' (r, \hat{t})}{\sqrt{1+\beta}}\,,
\end{equation}
with $z(r=0)=0$.
The luminosity distance $d_L(z)$ is related to the diameter 
distance $d_A(z)=R(r, \hat{t})$ according to the usual 
duality relation 
\begin{equation}
d_L (z)=(1+z)^2 R(r, \hat{t})\,.
\end{equation}
Now we are ready to confront the inhomogeneous LTB model 
with the SN Ia observations.
{}From the requirement that the CMB acoustic peak is not 
spoiled we require that local density parameter 
$\Omega_{\rm in}$ is in the range 0.1-0.3, whereas 
$\Omega_{\rm out}=1$ as predicted by inflation.
The observed local value of $H$ is around $h_{\rm in} \approx 0.7$, 
whereas outside the void one requires $h \approx 0.5$
to be consistent with $\Omega_{\rm out}=1$.
Then the two parameters $r_0$ and $\Delta$ are constrained
by the SN Ia data. In Ref.~\cite{Garcia} it was shown that the inhomogeneous
LTB model can be consistent with the SN Ia data for $r_0=2.3 \pm 0.9$\,Gpc
and $\Delta/r_0>0.2$.

We note, however, that one can place other constraints on the void model.
If we do not live around the center of the void, the observed CMB dipole 
becomes much larger than that allowed by observations.
The maximum distance $r_c$ to the center is constrained 
to be smaller than 10-20 Mpc \cite{Alnes06,Quer09}.
Even if we happen to live very close to the center of the void, 
we observe distant off-centered galaxy clusters. 
Such off-centered clusters should see
a large CMB dipole  in their reference frame.
For us this manifests itself observationally as a kinematic 
Sunyaev-Zeldovich effect.
Using the observational data of only 9 clusters, the inhomogeneous 
LTB model with void sizes greater than 1.5 Gpc can be 
ruled out \cite{GarciaBellido2,Caldvoid}.
This is already in mild conflict with the constraint derived by 
the SN Ia data. It remains to see whether the void model 
can be ruled out or not in future observations.

\subsection{Backreaction of cosmological perturbations}

Let us finally discuss the possibility of realizing a real cosmic 
acceleration by the backreaction of inhomogeneities to the FLRW space-time.
In general averaging the inhomogeneities and then solving the Einstein 
equations (the standard approach) might not be the same as solving the full 
inhomogeneous Einstein equations first and then averaging them. 
In other words, the expected value of a nonlinear
function of $x$ is not the same as the nonlinear function of the
expected value of $x$. The argument is complicated and controversial, 
so we mention the basic ideas only briefly.
The readers who are interested in the detail of this line of research
may have a look at the original papers \cite{Rasanen,Kolb1,Kolb2,Hirata,Robert,Ishibashi,Buchert,Buchert2,Rasanen2,Marra,KasaiAsada,Kasai,Wiltshire,Brown,Paran2,Larena,Vene,Kolb09}.

Let us decompose the Einstein tensor $G_{\mu \nu}$ and the energy-momentum
tensor $T_{\mu \nu}$ into the background (0-th order FLRW Universe) and 
the perturbed parts, as $G_{\mu\nu}=G_{\mu\nu}^{(0)}+G_{\mu\nu}^{(1)}$ and 
$T_{\mu\nu}=T_{\mu\nu}^{(0)}+T_{\mu\nu}^{(1)}$, respectively.
Then the (00) component of the Einstein equation (\ref{Einsteineq}) gives 
\begin{equation}
G_{00}^{(0)}=8\pi G\,(T_{00}^{(0)}+T_{00}^{(1)})-G_{00}^{(1)}\,.
\label{G00eq}
\end{equation}
Identifying the average matter density at this order as 
$\langle\rho\rangle=T_{00}^{(0)}+T_{00}^{(1)}$
and averaging over Eq.~(\ref{G00eq}), we obtain
\begin{equation}
\langle G_{00}^{(0)}\rangle=8\pi G\langle\rho\rangle-\langle G_{00}^{(1)}\rangle\,,
\end{equation}
where $G_{00}^{(0)}=3H^2$ in the FLRW background.
This shows that $3H^2 \neq 8\pi G \langle \rho \rangle$ in general because 
of the presence of the term $\langle G_{00}^{(1)}\rangle$.
If we first average the metric as $\langle g_{\mu \nu}^{(1)} \rangle=0$, it then follows 
that $G_{00}^{(1)}=G_{00}(\langle g_{\mu\nu}^{(1)}\rangle)=0$ and hence 
$3H^2=8\pi G \langle \rho \rangle$.
The above argument can be extended to the second order, i.e.
\begin{equation}
\langle G_{00}^{(0)}\rangle=8\pi G\langle\rho\rangle-\langle G_{00}^{(1)}
+G_{00}^{(2)}\rangle\,.
\end{equation}

The cosmological evolution depends on the averaging procedure.
For example, if we take the average of the function $f(t, x)$
as \cite{Kolb1}
\begin{equation}
\langle f\rangle(t)=\frac{\int\rd^{3}x\sqrt{\gamma(t,x)}f(t,x)}
{\int\rd^{3}x\sqrt{\gamma(t,x)}}\,,
\end{equation}
where $\gamma$ corresponds to the determinant of the perturbed 
metric of spatial constant-time hypersurfaces, then
the second-order term $G_{00}^{(2)}$ contributes to the
expansion rate of the Universe (which is typically of the order of $10^{-5}$). 
If we use other ways of averaging, the amplitude of such a term 
is subject to change \cite{Ishibashi}.
Also the results are affected by adding the contributions higher than 
the second order. In fact Ref.~\cite{Hirata} pointed out the danger of arbitrarily 
stopping at some order by showing several examples in which 
many contributions cancel each other.

The backreaction scenario is very attractive if it really works, because
it is the most economical way of explaining the cosmic acceleration 
without using dark energy.
We hope that further progress will be expected in this direction.

%%%%%%%%%%%%%%%%%%
\section{Conclusions}
\label{consec}
%%%%%%%%%%%%%%%%%%

We summarize the results presented in this review.
\begin{itemize}
\item The cosmological constant ($w_{\rm DE}=-1$) is favored 
by a number of observations, but theoretically it is still challenging 
to explain why its energy scale is very small.
\item Quintessence leads to the variation of the field equation
of state in the region $w_{\phi}>-1$, but the current 
observations are not sufficient to distinguish between quintessence potentials.
\item In k-essence it is possible to realize the cosmic acceleration 
by a field kinetic energy, while avoiding the instability problem
associated with a phantom field. The k-essence models that aim to 
solve the coincidence problem inevitably leads to the superluminal 
propagation of the sound speed.
\item In coupled dark energy models there is an upper bound on 
the strength of the coupling from the observations of CMB, large-scale
structure and SN Ia.
\item The generalized Chaplygin gas model allows the unified description 
of dark energy and dark matter, but it needs to be very close to the 
$\Lambda$CDM model to explain the observed matter power spectrum.
There is a class of viable unified models of dark energy and dark matter using 
a purely k-essence field.
\item In $f(R)$ gravity and scalar-tensor theories it is possible to 
construct viable models that satisfy both cosmological and local 
gravity constraints. These models leave several interesting observational 
signatures such as the modifications to the matter power spectrum 
and to the weak lensing spectrum.
\item The dark energy models based on the Gauss-Bonnet term
are in conflict with a number of observations and experiments in general 
and hence they are excluded as an alternative to the $\Lambda$CDM model.
\item The DGP model allows the self-acceleration of the Universe, but it
is effectively ruled out from observational constraints and 
the ghost problem. However, some of the extension of works such as 
Galileon gravity allow the possibility for avoiding the ghost problem, 
while satisfying cosmological and local gravity constraints.
\item The models based on the inhomogeneities in the distribution of 
matter allow the possibility for explaining the apparent accelerated expansion of
the Universe. The void model can be consistent with the SN Ia data, 
but it is still challenging to satisfy all other constraints coming 
from the CMB and the kinematic Sunyaev-Zeldovich effect.
\end{itemize}

When the author submitted a review article \cite{review} on dark energy 
to International Journal of Modern Physics D in March 2006, we wrote 
in concluding section  that ``over 900 papers
with the words `dark energy' in the title have appeared
on the archives since 1998, and nearly 800 with the words
`cosmological constant' have appeared''.
Now in April 2010, I need to change the sentence to 
``over 2250 papers with the words `dark energy' in the title have appeared
on the archives since 1998, and nearly 1750 with the words
`cosmological constant' have appeared''.
This means that over 4000 papers about dark energy and cosmological 
constant have been already written, with more than 2300 papers
over the past 4 years. Many cosmologists, astrophysicists, and particle 
physicists have extensively worked on this new field of research
after the first discovery of the cosmic acceleration in 1998. 
We hope that the future progress of both theory and observations
will provide some exciting clue to reveal the origin of dark energy.

\section*{Acknowledgments}

I thank Sabino Matarrese to invite me to write this article in
a chapter ``dark energy: investigation and modeling'' of a book
published in Springer.
I am also grateful to my all collaborators and colleagues with whom 
I discussed a lot about dark energy.
I also thank Eiichiro Komatsu, Marek Kowalski, Elic Linder, 
Jun-Qing Xia, and Jun'ichi Yokoyama for permission to 
include figures from their papers.
This work was supported by Grant-in-Aid for Scientific Research 
Fund of the JSPS (No.\, 30318802) and 
Grant-in-Aid for Scientific Research on Innovative 
Areas (No.\, 21111006).

%%%%%%%%%%%%%%%%

\end{document}